%% file: Final_version.tex
\documentclass[journal]{IEEEtran}

\usepackage{rotating}
\usepackage{tikz}
\usepackage{bbding}
\usepackage{pifont}
\usepackage{amssymb}
\usepackage{diagbox}
\usepackage{amsmath,amsfonts}
\usepackage{multicol,multirow,makecell,booktabs}
\usepackage{upgreek}
\usepackage{mathtools}
\usepackage{array}  
\usepackage{enumitem}
\usepackage{picins}
\allowdisplaybreaks[4]
\ifCLASSOPTIONcompsoc
  \usepackage[nocompress]{cite}
\else
  \usepackage{cite}
\fi

   \usepackage{graphicx}

\makeatother

\usepackage[switch]{lineno}

\title{Towards Quantum-Native Communication Systems: State-of-the-Art, Trends, and Challenges}
\author{Xiaolin Zhou, \textit{Senior Member, IEEE,} Anqi Shen, Shuyan Hu, \textit{Member, IEEE,} Wei Ni, \textit{Fellow, IEEE,} \\ Xin Wang, \textit{Fellow, IEEE,} and Ekram Hossain, \textit{Fellow, IEEE}

\thanks{This work was supported in part by the Natural Science Foundation of Shanghai under Grant 24ZR1407100, and in part by the National Natural Science Foundation of China under Grant 62231010, and Grant 61571135. \textit{(Corresponding author: Xin Wang.)}}
\thanks{X.~Zhou, A. Shen, and X. Wang are with the Key Laboratory for Information Science of Electromagnetic Waves, School of Information Science and Technology, Fudan University, Shanghai 200433, China (E-mail: zhouxiaolin@fudan.edu.cn; 21110720067@m.fudan.edu.cn; xwang11@fudan.edu.cn).}
\thanks{S. Hu is with the College of Electronics and Information Engineering, Tongji University, Shanghai 201804, China. She was with Fudan University, Shanghai 200433, China
(E-mail: syhu@tongji.edu.cn).}
\thanks{W.~Ni is with Data61, Commonwealth Scientific and Industrial Research Organization (CSIRO), Sydney, NSW 2122, Australia (E-mail: wei.ni@csiro.au).}
\thanks{E. Hossain is with the University of Manitoba, Winnipeg, MB R3T 2N2, Canada (E-mail: ekram.hossain@umanitoba.ca).}
}
\date{August 2021}

\begin{document}

\maketitle

\input{ch00_intro}

\input{ch0_preliminary}

 \input{Chap_Survey_of_Surveys}

 \input{Chapter_1}

\input{Chapter_2}

\input{Chapter_3}

 \input{Chapter_4}

\input{Chapter_5}

\input{Chapter_6}

\input{Chapter_7}

\bibliographystyle{IEEEtran}
\bibliography{reference}

\begin{IEEEbiography}[{\includegraphics[width=1in,height=1.25in,clip,keepaspectratio]{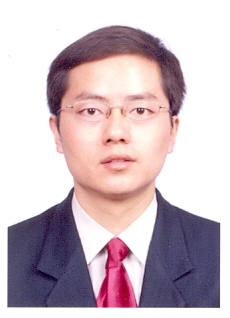}}]{Xiaolin Zhou} (Senior Member, IEEE) received the B.S. degree from Xidian University, China, in 1996, and the Ph.D. degree in communications and information systems from Shanghai Jiaotong University, Shanghai, China, in 2003. From 2005 to 2006, he was a Visiting Researcher with Monash University, Australia. He is currently an Associate Professor with the Department of Communication Science and Engineering, Fudan University, China. His research interests include wireless quantum communications, photon-counting communications, wireless multi-user communications, and iterative detection.
\end{IEEEbiography}

\begin{IEEEbiography}[{\includegraphics[width=1in,height=1.25in,clip,keepaspectratio]{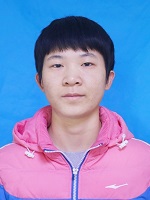}}]{Anqi Shen} received the B.S. and M.S. degrees in information and communication engineering from Inner Mongolia University, Hohhot, China, in 2018 and 2021. She is currently working toward the Ph.D. degree in electromagnetic field and microwave technology in Fudan University, Shanghai, China. Her research interests include quantum communications and optical wireless communications.
\end{IEEEbiography}

\begin{IEEEbiography}[{\includegraphics[width=1in,height=1.25in,clip,keepaspectratio]{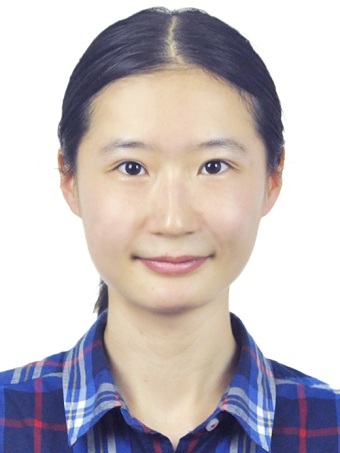}}]{Shuyan Hu} (M'20) received the B. Eng. degree in Electrical Engineering from Tongji University, Shanghai, China, in 2014, and the Ph.D. degree in
Electronic Science and Technology from Fudan University, Shanghai, China, in 2019.
She is currently an Associate Professor at Tongji University.
She was a Postdoctoral Research Fellow with Fudan University from 2019 to 2024.
Her research interests include machine learning and convex optimization and their applications to unmanned aerial vehicles, wireless communication, and intelligent systems.
\end{IEEEbiography}

\begin{IEEEbiography}[{\includegraphics[width=1in,height=1.25in,clip,keepaspectratio]{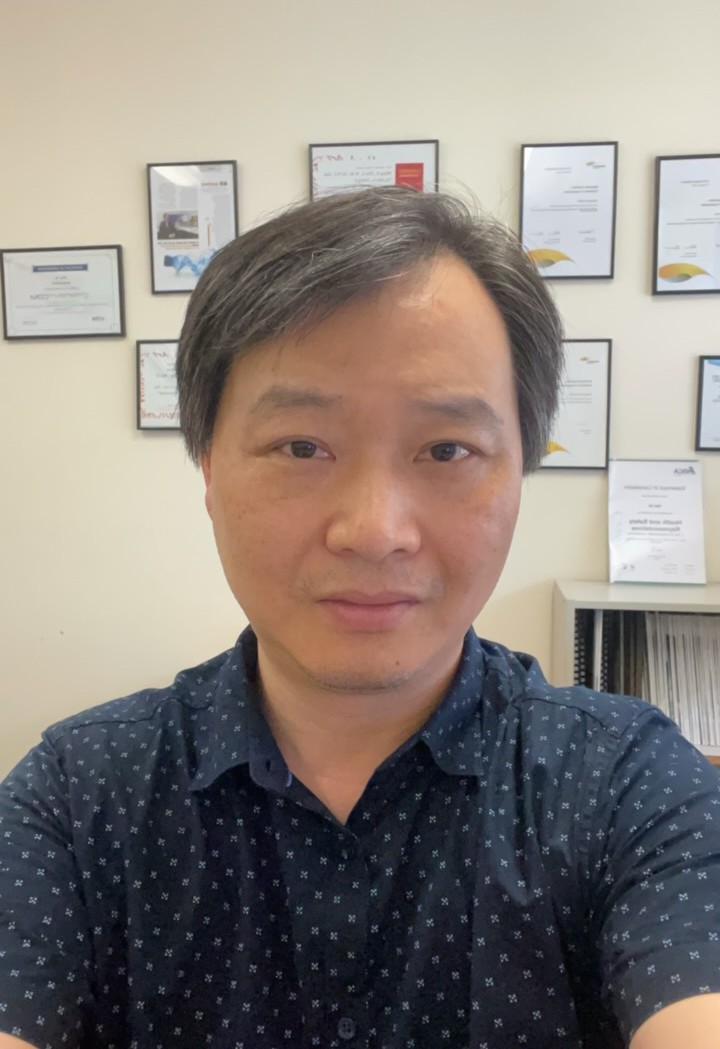}}]{Wei Ni}  (M'09-SM'15-F'24) received the B.E. and Ph.D. degrees in Electronic Engineering from Fudan University, Shanghai, China, in 2000 and 2005, respectively. He is a Senior Principal Research Scientist at CSIRO and a Conjoint Professor at the University of New South Wales. He was a Postdoctoral Research Fellow at Shanghai Jiaotong University from 2005 -- 2008; Deputy Project Manager at the Bell Labs, Alcatel/Alcatel-Lucent from 2005 to 2008; and Senior Researcher at Devices R\&D, Nokia from 2008 to 2009. He has authored five book chapters, more than 300 journal papers, more than 100 conference papers, 27 patents, and ten standard proposals accepted by IEEE. His research interests include machine learning, online learning, stochastic optimization, and their applications to system efficiency and integrity.  
 Dr. Ni has been an Editor for IEEE Transactions on Wireless Communications since 2018, an Editor for IEEE Transactions on Vehicular Technology since 2022, an Editor for IEEE Transactions on Information Forensics and Security and IEEE Communications Surveys and Tutorials since 2025, and an Editor for IEEE Transactions on Network Science and Engineering since 2025. He served first as the Secretary, then the Vice-Chair and Chair of the IEEE VTS NSW Chapter from  2015 to 2022, Track Chair for VTC-Spring 2017, Track Co-chair for IEEE VTC-Spring 2016, Publication Chair for BodyNet 2015, and Student Travel Grant Chair for WPMC 2014.
\end{IEEEbiography}

\begin{IEEEbiography}[{\includegraphics[width=1in,height=1.25in,clip,keepaspectratio]{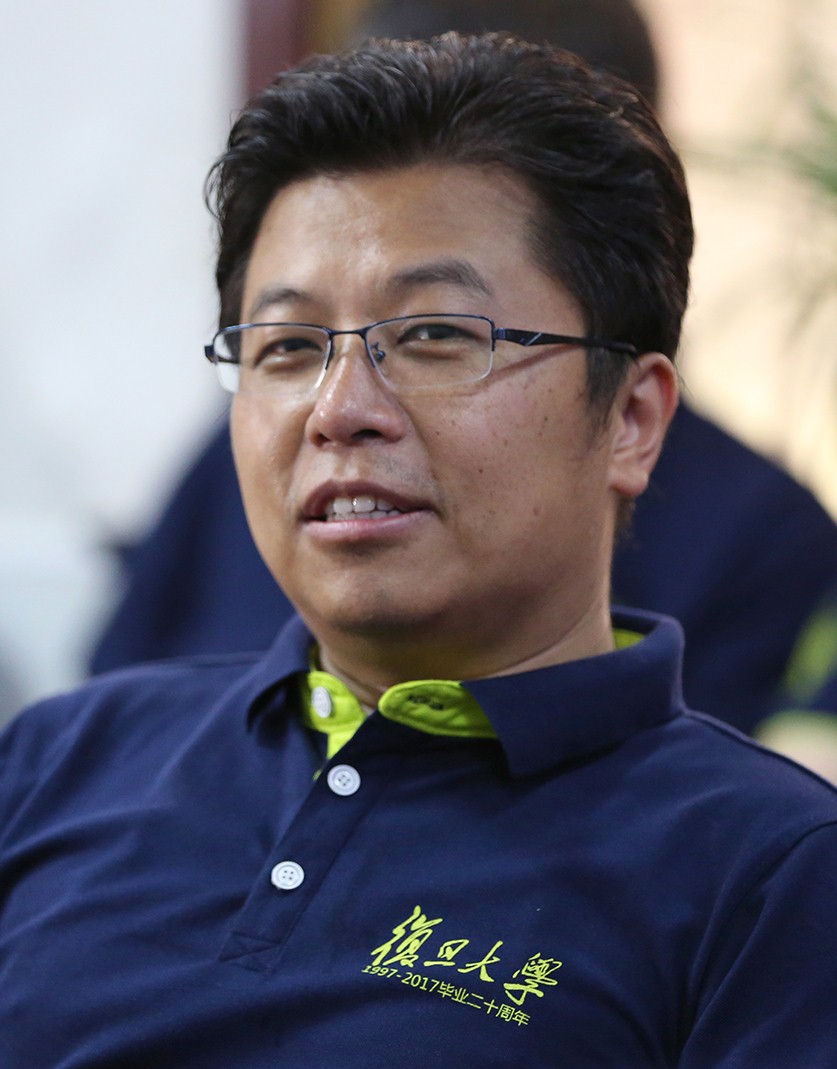}}]{Xin Wang} (SM'09-F'23) received the B. Sc. and M. Sc. degrees from Fudan University, Shanghai, China, in 1997 and 2000, respectively, and the Ph.D. degree from Auburn University, Auburn, AL, USA, in 2004, all in electrical engineering.
From September 2004 to August 2006, he was a Postdoctoral Research Associate with the Department of Electrical and Computer Engineering, University of Minnesota, Minneapolis. In August 2006, he joined the Department of Electrical Engineering, Florida Atlantic University, Boca Raton, FL, USA, as an Assistant Professor, and was later promoted to a tenured Associate Professor in 2010. He is currently a Distinguished Professor and the Chair of the Department of Communication Science and Engineering, Fudan University, Shanghai, China. His research interests include stochastic network optimization, energy-efficient communications, cross-layer design, and signal processing for communications. 
Dr. Wang served as a Senior Area Editor for the IEEE Transactions on Signal Processing, as an Associate Editor for the IEEE Transactions on Signal Processing, as an Editor for the IEEE Transactions on Wireless Communications, as an Editor for the IEEE Transactions on Vehicular Technology, and as an Associate Editor for the IEEE Signal Processing Letters. He is a member of the Signal Processing for Communications and Networking Technical Committee of IEEE Signal Processing Society, and a Distinguished Speaker of the IEEE Vehicular Technology Society.
\end{IEEEbiography}

\begin{IEEEbiography}[{\includegraphics[width=1in,height=1.25in,clip,keepaspectratio]{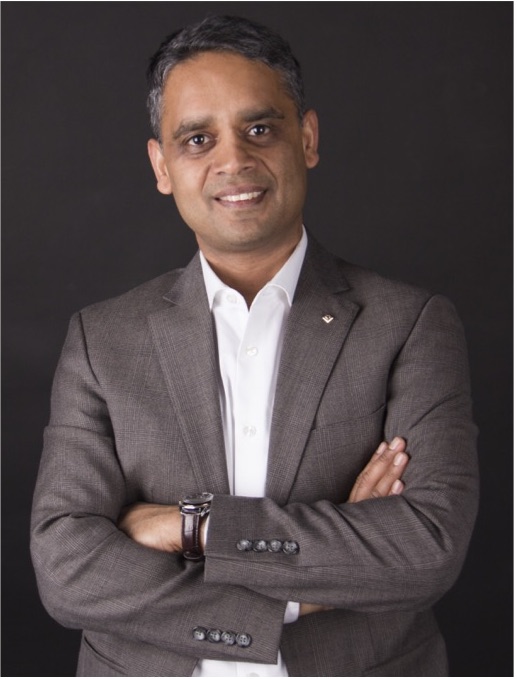}}]{Ekram Hossain}(Fellow, IEEE) is a Professor and the Associate Head (Graduate Studies) of the Department of Electrical and Computer Engineering, University of Manitoba, Canada. He is a Member (Class of 2016) of the College of the Royal Society of Canada. He is also a Fellow of the Canadian Academy of Engineering and the Engineering Institute of Canada. He has won several research awards, including the 2017 IEEE Communications Society Best Survey Paper Award and the 2011 IEEE Communications Society Fred Ellersick Prize Paper Award. He was listed as a Clarivate Analytics Highly Cited Researcher in Computer Science from 2017-2024. Previously, he served as the Editor-in-Chief (EiC) for the IEEE Press (2018–2021) and the IEEE Communications Surveys and Tutorials (2012–2016). He was a Distinguished Lecturer of the IEEE Communications Society and the IEEE Vehicular Technology Society. He served as the Director of Magazines (2020-2021) and the Director of Online Content (2022-2023) for the IEEE Communications Society.
\end{IEEEbiography}
\vspace{4\baselineskip}
\end{document}

%% file: ch00_intro.tex
\begin{abstract}

The potential synergy between quantum communications and future wireless communication systems is explored. By proposing a quantum-native or quantum-by-design philosophy, the survey examines technologies such as quantum-domain (QD) multi-input multi-output, QD non-orthogonal multiple access, quantum secure direct communication, QD  resource allocation, QD routing, and QD artificial intelligence. The recent research advances in these areas are summarized. Given the behavior of photonic and particle-like Terahertz (THz) systems, a comprehensive system-oriented perspective is adopted to assess the feasibility of using quantum communications in future systems. This survey also reviews quantum optimization algorithms and quantum neural networks to explore the potential integration of quantum communication and quantum computing in future systems. Additionally, the current status of quantum sensing, quantum radar, and quantum timing is briefly reviewed in support of future applications.
The associated research gaps and future directions are identified, including extending the entanglement coherence time, developing THz quantum communications devices, addressing challenges in channel estimation and tracking, and establishing the theoretical bounds and performance trade-offs of quantum communication, computing, and sensing. This survey offers a unique perspective on the potential for quantum communications to revolutionize future systems and pave the way for even more advanced technologies.

\end{abstract}


\begin{table}
\caption{Acronyms and Abbreviations} 
\begin{tabular}{ll}
\toprule
\textbf{Acronym} & \textbf{Full form} \\ \midrule
AI & Artificial Intelligence\\
{\color{black} BB84 } &  Bennett-Brassard-84\\
{\color{black}BER } &  Bit Error Rate\\
{\color{black}CD } &  Classical-Domain\\
{\color{black}CDMA } &  Code Division Multiple Access\\
CVQKD &  Continuous-Variable Quantum Key Distribution\\
DSQC  & Deterministic Secure Quantum Communication \\
{\color{black}EPR } &  Einstein-Podolsky-Rosen \\
FSO  &  Free-Space Optical\\
{\color{black}GEO }  &  Geostationary-Earth-Orbit\\
{\color{black} GHZ }  &  Greenberger-Horne-Zeilinger\\
HAP &  High-Altitude Platform\\
HL  &  Helstrom Limit\\
IDMA  &  Interleave Division Multiple Access\\
IoT   &   Internet of Things \\
ISAC &  Integrated Sensing and Communications \\
LEO &  Low-Earth-Orbit\\
{\color{black} LLR } &  Log-Likelihood Ratio\\
{\color{black} MAC } &  Media Access Control\\
MCF  & Multi-Core Fiber\\
{\color{black} MEC } &  Mobile Edge Computing\\
MIMO   &  Multi-Input Multi-Output\\
ML  &  Machine Learning   \\
mmWave  &  Millimeter Wave\\
NFV   &  Network Functions Virtualization\\
NG     & Next-Generation \\
NISQ &  Noisy Intermediate-Scale Quantum\\
NOMA  &  Non-Orthogonal Multiple Access\\
{\color{black} OAM } &   Orbital Angular Momentum\\
{\color{black}OFDMA } &  Orthogonal Frequency Division Multiple Access\\
O-RAN &  Open RAN  \\
{\color{black}PIC } &  Parallel Interference Cancellation\\
POVM  &  Positive Operator-Valued Measure\\
PSO &  Particle Swarm Optimization  \\
QBER   &  Quantum Bit Error Rate\\
QCL   & Quantum Cascade Laser\\
{\color{black}QD}  & Quantum-Domain \\ 
QECC  & Quantum Error Correction Code\\
QEM  & Quantum Error Mitigation \\
Qinternet  &  Quantum Internet  \\
QI  & Quantum Illumination  \\
QIT  &  Quantum Information Technology  \\
QKD  &  Quantum Key Distribution\\
QKP  &  Quantum Key Pool\\
QNN &     Quantum Neural Network  \\
QoS  &  Quality-of-Service\\
QRNG  & Quantum Random Number Generators \\
QSDC  & Quantum Secure Direct Communication\\
QSIC &  Quantum Secure Indirect Communication\\
QTC &  Quantum Turbo Code\\
Qubit  &  Quantum Bit\\
{\color{black} RAN } &  Radio Access Network\\
RCWA   &  Routing, Core, and Wavelength Allocation\\
RL   &   Reinforcement Learning  \\
{\color{black} RWTA } &  Routing, Wavelength, and Timeslot Allocation\\
{\color{black} SDM } &  Spatial Division Multiplexing\\
SDN  &  Software-Defined Networking\\
{\color{black}SNR}  &  Signal-to-Noise Ratio\\
SQL  &  Standard Quantum Limit\\
{\color{black} SVM } &  Support Vector Machine\\
{\color{black}TDMA } & Time Division Multiple Access\\
THz  &  Terahertz\\
UAV  & Unmanned Aerial Vehicle\\
VLC &  Visible Light Communication\\
VNF  &  Virtual Network Function\\
{\color{black}WDM } &  Wavelength Division Multiplexing\\
{\color{black}WDMA } &  Wavelength Division Multiple Access\\
{\color{black} WSDM } &  Wavelength-Space Division Multiplexing\\
\bottomrule
\end{tabular}
\label{tab.nomenclature}
\end{table}

\section{Introduction}

\subsection{Background}

The growing demand for high-quality, confidential information services drives the rapid development of wireless communication systems and inspires transformative innovation in information technologies. Following the global roll-out of the fifth-generation (5G) wireless systems, the community's research efforts are becoming focused on future wireless systems, including the sixth-generation (6G) systems and beyond~\cite{CuiQimei2025}. 
For example, the Third Generation Partnership Project will start standardizing 6G technology in the second half of 2025~\cite{3GPP}, and the technology is expected to be commercially available around 2030. The capabilities of 6G communications will facilitate a quantum leap from the connected things of the 5G era to connected intelligence~\cite{0Quantum } and realizing the vision of ``intelligent connection of everything and digital twin"~\cite{Zhaoyajun20196G, WOS:000754250700024,6Gvision_Yazar,2022Vision}.

\begin{figure*}
\centering
\includegraphics[width=7.5 in]{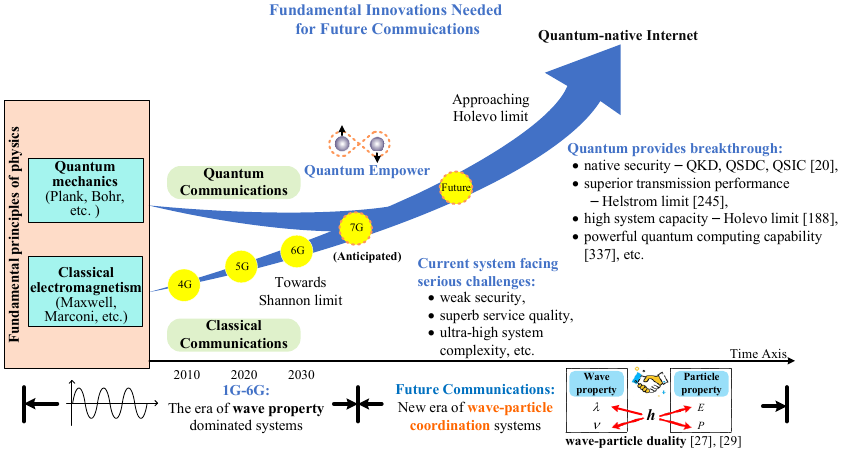}
\caption{Quantum communication boosts communication development and evolves to the future. In future communication systems, quantum communication can play an increasingly important role.  In the future, the communication network can evolve into a quantum network to achieve a high degree of unity of the wave-particle duality physical law in the communication system.  And with the development of classical and quantum communications, the deep integration of these two systems will be realized in the future, and finally, the Qinternet. In this figure, $E = h\nu $, $P = {h \mathord{\left/
 {\vphantom {h \lambda }} \right.
c \kern-\nulldelimiterspace} \lambda }$, where $E$ represents energy, $P$ represents momentum, $h$ is the Planck's constant, $\nu$ denotes frequency, and $\lambda$ denotes wavelength.}
\label{Quantum_communication_boosts_7G_development_and_evolves_to_the_future}
\end{figure*}

It is anticipated that future communication systems are expected to pay more attention to communication security and to the protection of user privacy. Potential key technologies, such as quantum communication, will provide improved security in future communication networks. Furthermore, quantum technologies can also play an important role in improving transmission data rate and reliability.
Some important quantum-related research topics are included as:  
\begin{itemize}
\item
Quantum communications is a field of engineering that involves the exchange of quantum states representing quantum keys, e.g.,~\cite{8439931}.  
This involves the use of quantum mechanical phenomena such as superposition and entanglement to transport quantum states from one place to another.
To fully exploit these resources, the transferred state must remain unobserved during transmission, prohibiting copying or amplifying during transmission~\cite{6576337}.

\item
Quantum computing relies on quantum bits (Qubits) as the basic unit of information, rather than classical bits. Quantum computers operate by harnessing the laws of quantum mechanics~\cite{6198345} and are able to perform specific types of calculations much faster than classical computers. The potential for quantum computers to solve specific problems that are intractable for classical computers has led to considerable research~\cite{GYONGYOSI201951}.

\item
Quantum sensing relies on the principles of quantum mechanics to measure and transform physical quantities. These sensors take advantage of the unique phenomena that occur at the quantum level, such as quantum entanglement~\cite{8750548}, quantum coherence~\cite{9542465}, and quantum state squeezing~\cite{9908325}. Quantum sensors can be much more sensitive and accurate than classical sensors~\cite{0Quantum,2019Entanglement}.

\item
Quantum circuitry comprises interconnected quantum gates that manipulate qubits, which are arranged in a specific pattern for mitigating the avalanche-line proliferation of qubit errors imposed by the deleterious effects of quantum decoherence~\cite{9965678}.

\item
Quantum teleportation is a process that enables the transfer of quantum information, in the form of qubits, from one location to another, while relying on classical-domain (CD) assistance. This is made possible through the use of quantum entanglement, a phenomenon in which two or more particles become connected in such a way that the quantum state of one particle can influence that of the other, even if they are far apart. Quantum entanglement allows the qubits to be instantaneously transferred from one location to another as if they were teleported~\cite{9023997}. But again, because of the reliance on CD operations, the speed of light cannot be exceeded.

\item
Quantum communication protocols are methods conceived for transmitting information carried by quantum states by exploiting the principles of quantum mechanics~\cite{2017Quantum,962022,bookchina}. 
Some examples are quantum key distribution (QKD) and quantum secure direct communication (QSDC).
Quantum communication protocols provide secure communication with applications in areas, such as cryptography, networking, and computing. However, QKD only represents the negotiation of secret keys, which are then used in a similar way to its classical encryption counterpart by taking the module-two connection of the data and the key. By contrast, QSDC constitutes a fully-fledged quantum communication protocol originally proposed in~\cite{WOS:000174548500044}. This protocol has rapidly evolved in recent years~\cite{9130765}.
\end{itemize}

\subsection{Synergy of Quantum and Classical Systems} 

As shown in Fig.~\ref{Quantum_communication_boosts_7G_development_and_evolves_to_the_future}, quantum communications and classical communications will co-evolve~\cite{8840309,9623536}.
Quantum communications will empower classical communications. For example, the natural security of quantum communications can support the demanding security requirements of future classical communications (e.g., financial stock trading)~\cite{Zhang18oe}.
In general, the nano-scale world may be viewed as being governed by quantum theory. In the quantum world, the states are probabilistic, and the associated uncertainty has nothing to do with experimental technology. This feature is embodied in the principle of superposition of quantum states. By contrast, classical Newtonian theory may describe the physical phenomena of the classical world. The classical world is characterized by the fact that the physical quantities and states of objects are completely determined at any given moment. 

In future communication systems, quantum communication is expected to play an increasingly important role by exploiting the wave-particle duality of physics~\cite{Greiner2005Quantum,2008THE} in practical communication systems. As illustrated in Fig.~\ref{Quantum_communication_boosts_7G_development_and_evolves_to_the_future}, the confluence of classical and quantum communications is expected to illuminate the conception of the quantum Internet (Qinternet).
Compared to the operational 4G and 5G communication systems, future systems are expected to adopt both wireless optical communications and visible light communications, which are conducive to the employment of quantum communications in higher frequency bands~\cite{6191306}, where the particle-like characteristics of electromagnetic waves are more pronounced according to quantum physics~\cite{7451959}. For example, the Terahertz (THz) band, ranging from 0.3 THz to 3 THz, exhibits strong quantum-like behaviors and particle characteristics. 

There are increasing discussions in the quantum research community about exploiting the THz band for supporting quantum communications~\cite{ 9739032 }. The quantum behavior of THz signals in silicon has been investigated~\cite{ PhysRevB.95.125424}.
Existing experiments and engineering implementations of quantum communications have typically been dedicated to optical fibers~\cite{9836755}, wireless laser channels~\cite{9571213}, and the THz band~\cite{ 9739032 }.  
Another reason for considering quantum communications in future systems is the benefit of the relative maturity and cost efficiency of optical quantum devices~\cite{0Integrated}.

\begin{figure}[t]
\centering
\includegraphics[width=3.5 in]{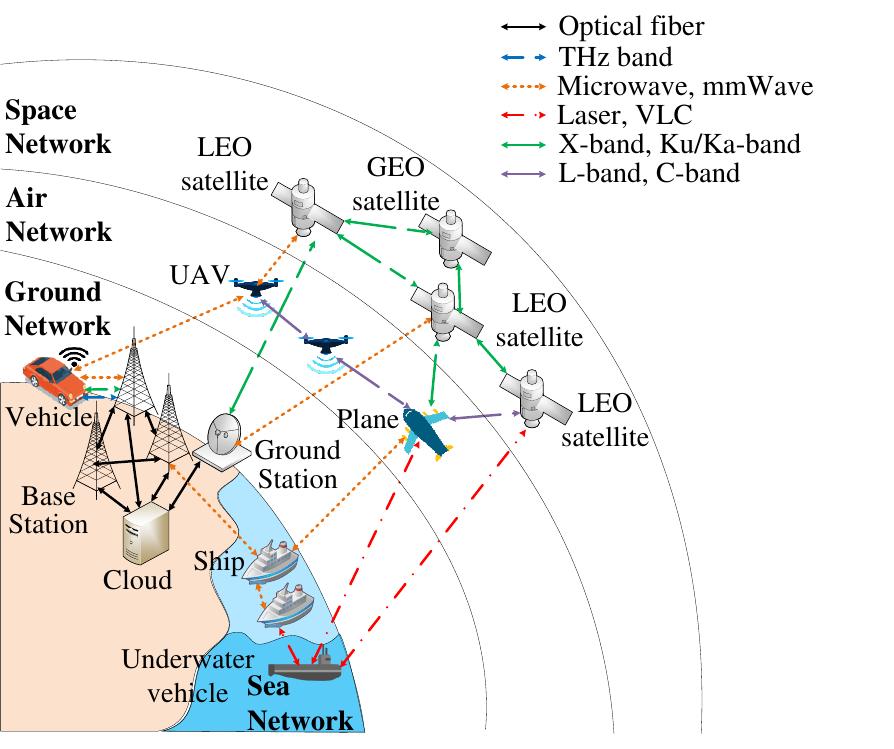}
\caption{Future space-air-ground-sea integrated secure communication network model. The future full-band communication network will cover different interface media, such as optical fiber, THz, microwave and mmWave, laser and VLC, X-band and Ku/Ka-band, L-band and C-band. They will provide the security, connectivity, reliability, and throughput needed for future communication systems~\cite{9524814,8760401,9247451}.}
\label{Space-air-ground-sea_integrated_network}
\end{figure}

Fig.~\ref{Space-air-ground-sea_integrated_network} depicts a promising future space-air-ground-sea integrated network that can achieve fast, reliable, and secure connectivity across broad coverage ranges~\cite{9628162}. Aerial facilities, including low-Earth-orbit (LEO) and geostationary-Earth-orbit (GEO) satellites, planes and unmanned aerial vehicles (UAVs), ground stations and vehicles, as well as ships and underwater vehicles, will be integrated. 
Different propagation media, including optical fiber, THz, microwave and millimeter wave (mmWave), laser, and visible light communication (VLC), X-band and Ku/Ka-band, L-band, and C-band, will have to be critically appraised in support of seamless connectivity, reliability, and high throughput~\cite{9524814}. 

For example, optical fibers are extensively considered for connecting network entities in the core networks~\cite{9607182}. Lasers are used for connecting different satellites and individual satellites with ground stations~\cite{9354784}. The X-band, Ku/Ka-band, and THz bands have the potential in inter-satellite connections. The THz band has also been considered for providing local coverage on the ground~\cite{9569475}. 
These propagation media can potentially support quantum communications in the future.
Moreover, quantum systems have been reported to harness highly directional antennas in the THz, infrared, and optical bands due to recent breakthroughs in nanotechnology~\cite{2020Quantum }.

\begin{table*}
\caption{Stages of quantum-empowered communication systems~\cite{UpdateofSRAandroadmap, FGQIT4N2}}
\centering
\begin{tabular}{|m{1.75cm}<{ \flushleft} | m{4.25cm}<{ \flushleft} | m{5.125cm}<{ \flushleft}|m{5.125cm}<{ \flushleft}|}
 \hline
 \makecell*[c]{\textbf{Involved Field}}  &  \makecell*[c]{\textbf{Key Technologies}}  &\makecell*[c]{\textbf{Near-Term Objectives}} &\makecell*[c]{\textbf {Long-Term Objectives}} \\ \hline 
\textbf{\begin{tabular}[c]{@{}l@{}} Quantum \\communication \end{tabular}}
& 
$\bullet$ Quantum cryptography~\cite{WOS:000911965000001}; 

$\bullet$ QRNG~\cite{RevModPhys.92.025002};

$\bullet$ QKD;

$\bullet$ Photon detection technologies;

$\bullet$ Quantum repeaters;

$\bullet$ Quantum memory~\cite{8083123};

$\bullet$ Quantum entanglement distribution devices;

$\bullet$ Distributed quantum processors.

&  $\bullet$  Advances in QKD, QRNG, and quantum-secure authentication systems~\cite{8118926}; 

$\bullet$ Develop trusted node network functionality~\cite{8872467};

$\bullet$ Develop global secret key distribution~\cite{9749727};

$\bullet$ Demonstrate a basic link as a building block for future quantum repeater;

$\bullet$ Develop network performance, application, protocol, and software test suites.

& $\bullet$ Demonstrate a chain of quantum repeaters~\cite{9192923} at physical distances;

$\bullet$ Demonstrate a quantum network node of at least 20 qubits connected to a quantum network;

$\bullet$ Demonstrate device-independent-inspired QRNG and QKD;

$\bullet$ Demonstrate entanglement generation using satellite-based links~\cite{10089128};

 \\  \hline
 
\textbf{\begin{tabular}[c]{@{}l@{}} Quantum \\computing \end{tabular}}
&

$\bullet$ Scalable quantum computing~\cite{7738703};

$\bullet$ Quantum error correction and fault-tolerant quantum computing~\cite{8436751};

$\bullet$ Quantum computing chip or device;

$\bullet$ Quantum algorithm;

$\bullet$ Classical networking enhancement for quantum computing;

$\bullet$ Quantum networking interconnection of quantum computers~\cite{1635967}.

&  $\bullet$ Demonstrate practical strategies for fault-tolerant universal quantum computer~\cite{6620283};

$\bullet$ Identify algorithms and use cases with quantum advantages;

$\bullet$ Enhance the noisy intermediate-scale quantum processing mechanism with error mitigation~\cite{9504796};

$\bullet$ Initiate research in quantum device physics, qubits, and gate control;

$\bullet$ Liaise with standards bodies.

& $\bullet$ Demonstrate quantum processors equipped with quantum error correction~\cite{9570857} and robust qubits with a general gate set;

$\bullet$ Demonstrate quantum algorithms with quantum advantages;

$\bullet$ Coordinate research, development and integration in quantum device physics, qubit and gate control, quantum memory;

$\bullet$ Extend the suite of quantum algorithms, such as digital error correction systems, optimization compilers~\cite{8980363} and libraries.
 \\  \hline
 
\textbf{\begin{tabular}[c]{@{}l@{}} Quantum \\sensing \end{tabular}}
& 
$\bullet$ Quantum clocks~\cite{8699653};

$\bullet$ Improved optical sensing/imaging;

$\bullet$ New space-based time and frequency transfer techniques;

$\bullet$ Improved single-photon detectors and arrays~\cite{6843851};

$\bullet$ Spin- and entanglement-based sensors~\cite{8750548};

$\bullet$ Quantum-enhanced radio frequency, microwave, and optical signal processing and detection;

&

$\bullet$ Laboratory demonstrations of engineering quantum states in real-world applications.

$\bullet$ Develop prototypes of quantum-enhanced, super-resolved, and sub-shot noise microscopy, spectroscopy~\cite{8083316}, interferometry, quantum lidar and radar~\cite{9477766};

$\bullet$  Develop chips that integrate photonics, electronics and atoms, miniaturized lasers;

$\bullet$ Establish standardization, calibration, and traceability for new sensor technologies;

& $\bullet$ Integrate quantum measurement~\cite{8872312} standard for self-calibration of instrumentation;

$\bullet$ Develop magnetometers for improved magnetic resonance imaging, high-performance optical clocks, and atomic interferometers~\cite{8872784};

$\bullet$ Develop quantum sensor networks as well as space-borne quantum-enhanced sensors, including optical clocks atomic and optical inertial sensors~\cite{7935677}.

 \\  \hline

\end{tabular}
\label{tab.Evolutionofquantum-empowered6G_B6Gsystems}
\end{table*}

\subsection{Emerging Quantum Use Cases and Standardization}

The International Telecommunication Union's Telecommunication  Standardization Sector formed a Focus Group on Quantum Information Technology (QIT) for Networks~\cite{ FGQIT4N }. 
In November 2021, the group delivered nine technical reports to specify the general QIT terminologies~\cite{FGQIT4N1}, use cases~\cite{ FGQIT4N2 }, and standardization outlook~\cite{ FGQIT4N3 }. The group also considered QKD networks~\cite{ FGQIT4N4 } and their user cases~\cite{ FGQIT4N5 }, protocols~\cite{ FGQIT4N6 }, control and management~\cite{ FGQIT4N7 }, network transport technologies~\cite{ FGQIT4N8 }, and standardization prospects~\cite{ FGQIT4N9 }.
A roadmap for quantum-empowered future systems is provided in Tab.~\ref{tab.Evolutionofquantum-empowered6G_B6Gsystems}, with some real-world examples of the use cases in the following.

In 2019, researchers from Singapore's Nanyang Technological University reported in~\cite{Quantumcommunicationchips} the fabrication of the first millimeter-scale quantum communication chip, which can provide outstanding quantum security to compact devices, e.g., smartphones, tablets, and smartwatches.
On February 17, 2022, JPMorgan Chase announced the joint launch of a QKD network with Toshiba and the U.S. telecom system provider Ciena, demonstrating the feasibility of a metropolitan QKD network for securing its payment network link in the face of steganographic attacks~\cite{JPMorganChase}. 

In 2007, the Defense Advanced Research Projects Agency initiated a research program to examine the benefits of optical quantum sensors~\cite{QuantumSensorsProgram}. These sensors can improve the accuracy of lidar and other optical sensors through techniques, such as quantum lithography. 
Apart from defense~\cite{9904994}, navigation~\cite{7550613}, and mapping of contour~\cite{5156503}, quantum sensing is the mature quantum technology for gas-sensing~\cite{9932599} and deep space exploration~\cite{9717834}.

Some other use cases are authentication, digital signatures, and secure access control keys, which can be generated using its quantum random number generators (QRNG), e.g., by ID Quantique~\cite{idquantique}.
ID Quantique provides long-term data protection from future quantum computer attacks when large-scale quantum computers become available. 
As opposed to magnetic stripes or embedded computer chips, quantum-secured authentication uses nanoparticles 
to create a unique pattern with the aid of a laser beam directed at the nanoparticles, which cannot be replicated.
Anecdotally, during the election process in Geneva, ID Quantique implemented an Ethernet encryption system with QKD, ensuring the integrity and security of voting processes.
Moreover, the randomness of online gaming must be ensured to secure a uniform winning probability~\cite{2019Quantum}.

\begin{figure*}
\centering
\includegraphics[width=7 in]{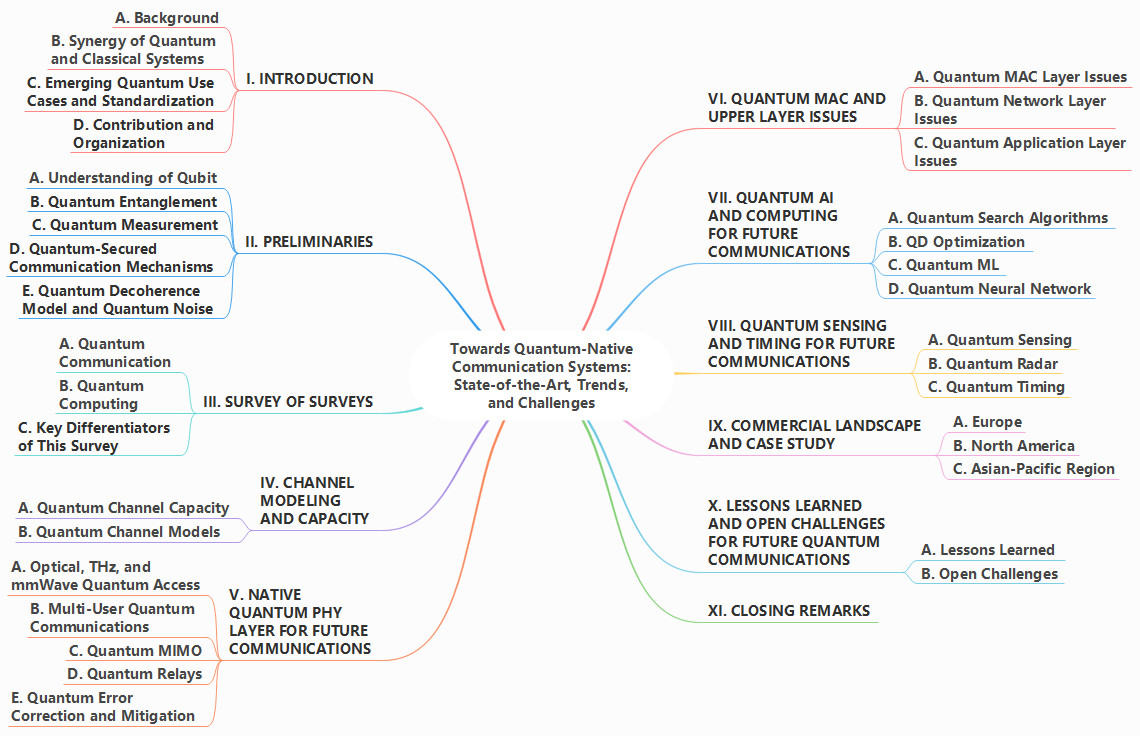}
\caption{The outline of this survey.}
\label{Outline_of_this_survey_paper}
\end{figure*}

\subsection{Contribution and Organization}

This survey advocates radical frontier research and the notion of quantum-by-design, where the potential synergy between quantum communications and future systems is discussed. The family of candidates quantum communication technologies, such as quantum multi-input multi-output (MIMO), quantum non-orthogonal multiple access (NOMA), QSDC/quantum secure indirect communication (QSIC), quantum artificial intelligence (AI), quantum sensing and timing for future systems, are explored in the light of the relevant recent research breakthroughs. 

The key contributions of the survey are as follows. 

\begin{itemize}
\item We highlight a wide range of quantum technologies in line with the future communication vision and application scenarios. In contrast to existing surveys heavily focused on QKD, we also provide a broader perspective on QSDC and QSIC. (Sections I, II and VI)

\item We explore a new perspective, namely for future high-frequency communications, e.g., mmWave/THz, where the particle effect is stronger and the feasibility of quantum communications is improved primarily because of hardware maturity, as well as the lower vacuum noise, which is inversely proportional to the frequency. 
    It is promising to exploit the wave-particle duality of quantum systems to improve the communication performance in future systems. (Section I)

\item A new networking perspective is adopted for assessing the synergy between quantum techniques and different segments of future communication systems, including multi-band quantum access, quantum relays, quantum resource allocation, and quantum routing in support of future quantum-native systems. (Sections V and VI)

\item We review quantum optimization algorithms and neural networks to explore the new research direction for the potential integration of quantum technologies and AI. (Section VII)

\item We highlight the status quo of quantum sensing, quantum radar, and quantum timing, which are important research areas and can potentially extend the sensing capability for integrated sensing and communications (ISAC). (Section~VIII)

\item Research gaps and future directions are identified in each network segment of future systems in support of quantum-native design. In particular,  (Section X)
\begin{itemize}
    \item The entanglement coherence time severely limits the effective generation and distribution of entangled qubits. Therefore, it is urgent to develop new technologies
to extend the entanglement coherence time.

    \item The design of THz quantum communications devices is in its infancy. Challenges, including noise suppression techniques, high-gain antennas, and accurate parameter estimation algorithms, are yet to be addressed.
    
    \item The high mobility of quantum communication nodes, such as satellites and UAVs, can lead to rapid propagation channel fluctuations, leading to new challenges in channel estimation and tracking. The Doppler shift will also affect the measurement of quantum states.  
    
    \item A joint design of quantum communication and quantum computing based on quantum superposition, entanglement, and coherence constitutes a critical step toward quantum-native designs. Yet, little to no effort has been documented in the open literature.

        \item 
    Quantum error correction~\cite{MichaelJBiercuk2022QuantumEC} will be critical in a potential joint design of quantum communications and computing. However, its technical maturity is low.

    \item The fundamental relationships, theoretical bounds, and performance trade-offs of quantum communication, quantum computing, and quantum sensing have to be established in support of quantum-native designs. 

\end{itemize}

\end{itemize} 

The rest of this survey is structured as depicted in Fig.~\ref{Outline_of_this_survey_paper}.
Section II introduces the fundamental concepts of quantum mechanics and quantum-secured communication schemes. Section III reviews the existing surveys on quantum communication and computing. 
Section IV reviews quantum channel models. 
Section V summarizes the latest research on communication technologies in the quantum physical (PHY) layer, including some facets of multi-user, MIMO, and relaying techniques in the quantum context. 
Section VI analyzes resource allocation, routing protocols/algorithms, multiplexing techniques, and quantum-secured network services in quantum media access control (MAC) and the upper layer. 
Quantum AI and computing are discussed in Section VII. In Section VIII, quantum sensing and quantum timing are appraised in the context of next-generation (NG) applications. 
The commercialization efforts of quantum technologies are summarized in Section IX, and the open challenges are discussed in Section X, followed by concluding remarks in Section XI. Tab.~\ref{tab.nomenclature} lists the acronyms and abbreviations used.

%% file: ch0_preliminary.tex
\section{Preliminaries}
A quantum communication system uses quantum states as the information carrier. The propagation of quantum states in a quantum channel follows the law of quantum mechanics~\cite{9542199}.
This section provides some rudimentary preliminaries on quantum mechanics, including the concepts of qubits, quantum entanglement, and quantum measurement. Three popular quantum communications techniques, namely, QKD, QSDC and QSIC, are briefly introduced and compared.

\begin{table*}
\caption{A comparison of QSDC, QSIC and QKD}
\centering
\begin{tabular}{| m{3cm}<{ \centering} | m{5cm}<{ \flushleft} | m{4cm}<{ \flushleft}|m{4cm}<{ \flushleft}|}
 \hline
 &  \makecell*[c]{\textbf{QSDC}}  &\makecell*[c]{\textbf{QSIC}} &\makecell*[c]{\textbf {QKD}} \\ \hline

\textbf{Information carrier}
& 
$\bullet$ Quantum state~\cite{bookchina}
&  $\bullet$  Einstein-Podolsky-Rosen photons~\cite{2015Advances }
& $\bullet$ Quantum state~\cite{Zhang18oe}
 \\  \hline 
\textbf{Basic principles}
& 
\makecell*[l] { $\bullet$ Basic principles of quantum \\mechanics and properties ~\cite{9130765}.}
&  $\bullet$ Remote association of quantum entanglement pairs~\cite{5235303}.
& $\bullet$ Heisenberg uncertainty principle~\cite{0Heisenberg} and quantum no-cloning theorem~\cite{Masato1998No}.
 \\  \hline 
\textbf{Whether to generate a secret key}
& 
$\bullet$ No need to generate a secret key and do not depend on the encryption and decryption algorithms~\cite{bookchina}.
&  $\bullet$  No need to generate a secret key~\cite{Flamini2018,bookchina}.
& $\bullet$ Used to generate and distribute secret keys without transmitting any actual messages~\cite{2012Comparison}.
 \\  \hline 
\textbf{Advantages}
& 
$\bullet$ Directly and safely transmit secret information in the quantum channel.
$\bullet$ Has higher security requirements than QKD (requires the discovery of eavesdropping and guarantees the security of the data transmitted before the discovery of eavesdropping)~\cite{9129730}.
&  $\bullet$  Quantum teleportation is the simplest form of quantum communication~\cite{7784726}. 
$\bullet$ Does not transmit any matter or energy.
& $\bullet$ The research is the earliest, fastest, and most mature theory, and has been applied in practice~\cite{8745822}.
 \\  \hline
\textbf{Shortcomings}
& $\bullet$ Not conducive to the implementation of eavesdropping detection, error correction, or privacy amplification.
& $\bullet$ Not applicable in faster scenarios than optical communications.
& $\bullet$ The problem of the trusted relay security.\\  \hline
\textbf{Transmission channel}
&  \multicolumn{3}{c|}{$\bullet$ Optical fiber, free-space channel}
\\  \hline
\end{tabular}
\label{tab.quantumsecure}
\end{table*}

\subsection{Understanding of Qubit}

In classical communications, a classical bit is the basic unit of transmission and storage. Each bit can represent one of the binary symbols from the set $\left\{ {0,1} \right\}$. In the quantum communication field, information is represented by a qubit~\cite{9023997}.
{\color{black} Dirac's bra-ket notation is often used to label quantum states (i.e., qubit), quantum operations, and quantum measurements. For a complex linear space ${\mathbb{C}^{2 \times 1}}$ composed of column vectors, a common set of orthogonal normalized bases can be expressed as $\left( {\begin{array}{*{20}{c}}
1\\
0
\end{array}} \right)$ and $\left( {\begin{array}{*{20}{c}}
0\\
1
\end{array}} \right)$.  
Using Dirac notation, these column vectors are represented as right vectors: $\left| 0 \right\rangle  = \left( {\begin{array}{*{20}{c}}
1\\
0
\end{array}} \right)$ and $\left| 1 \right\rangle  = \left( {\begin{array}{*{20}{c}}
0\\
1
\end{array}} \right)$, where $\left|  \cdot  \right\rangle $ is the ket operator originating from the word ``bra-ket''. The set of right vectors can be represented by $\left\{ {\left| 0 \right\rangle ,\left| 1 \right\rangle } \right\}$. Any vector in a linear space ${C^{2 \times 1}}$ can be represented as a right vector $\left| \psi  \right\rangle $, which is a linear combination of the basis vectors $ \left| 0 \right\rangle$ and $\left| 1 \right\rangle $, as given by
\begin{align}
\left| \psi  \right\rangle  = \alpha \left| 0 \right\rangle  + \beta \left| 1 \right\rangle ,\forall \alpha ,\beta  \in \mathbb{C}.
\end{align}}A qubit may be in the state $\left| 0 \right\rangle $, or in the state $\left| 1 \right\rangle $, or in the superposition state $\alpha \left| 0 \right\rangle  + \beta \left| 1 \right\rangle $, where $\alpha  = \cos \frac{\theta }{2}$, and $\beta  = {e^{i\gamma  }}\sin \frac{\theta }{2}$. Here, $i = \sqrt{-1}$. 

\begin{figure}
\centering
\includegraphics[width=3 in]{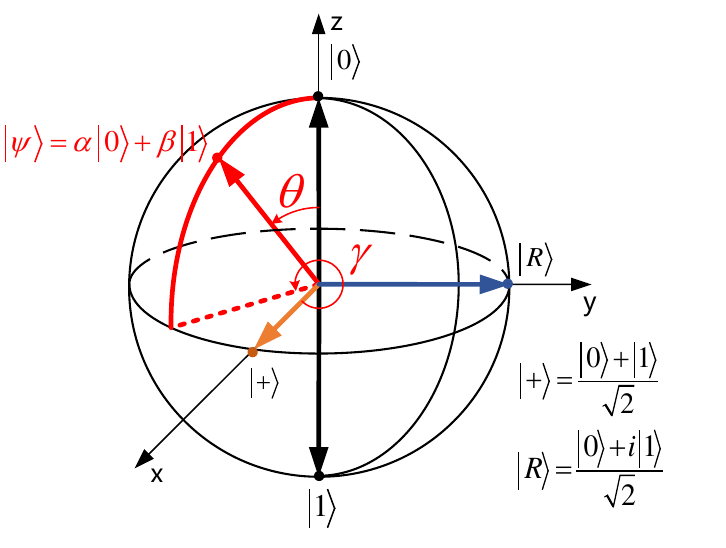}
\caption{Bloch sphere. Any point on the Bloch sphere represents a qubit, which can be represented by the angle parameters $\theta $ and $\gamma $. A qubit can be expressed as $\left| \psi  \right\rangle   = \alpha \left| 0 \right\rangle  + \beta \left| 1 \right\rangle$, where $\alpha  = \cos \frac{\theta }{2}$, and $\beta  = {e^{i\gamma }}\sin \frac{\theta }{2}$~\cite{nielsen2002quantum}.}
\label{Bloch_sphere}
\end{figure}

As shown in Fig.~\ref{Bloch_sphere}, any point on the  Bloch sphere represents a qubit~\cite{nielsen2002quantum}, which can be uniquely identified by the angle parameters $\theta $ and $\gamma$. As it transpires from this figure, there can be an infinite number of qubits on the Bloch sphere.

Using a so-called density operator constitutes an efficient method for describing the quantum system by a series of probabilistic superpositions based on their states~\cite{bookchina}, i.e., ${\rho }{\rm{ = }}\sum\limits_j {{p_{_j}}\left| {{\psi _j}} \right\rangle \langle {\psi _j}|}$, where $\left| {{\psi _j}} \right\rangle $ indicates a legitimate state of the system, and ${p_j}$ is the probability of the state. The density operator contains all state information of a quantum system.

\subsection{Quantum Entanglement }
Schrödinger introduced the concept of quantum entanglement into quantum mechanics in 1935, which became a unique resource for quantum communications~\cite{schrodinger1935}.
In a two-qubit composite system, if the quantum state cannot be expressed in the form of a direct
tensor product $ |\psi {\rangle _{AB}} = |\psi {\rangle _A} \otimes |\psi {\rangle _B},$ the composite state is said to be a two-qubit entangled state. The Bell state is the simplest type of 
two-qubit entangled state,
comprising four maximally entangled two-qubit states~\cite{ razaviintroduction }:
\begin{align}
{{{\left| {{\psi ^ + }} \right\rangle }_{AB}}}&{ = \frac{1}{{\sqrt 2 }}\left( {|0{\rangle _A}|1{\rangle _B} + |1{\rangle _A}|0{\rangle _B}} \right);}\\
{{{\left| {{\psi  ^ - }} \right\rangle }_{AB}}}&{ = \frac{1}{{\sqrt 2 }}\left( {|0{\rangle _A}|1{\rangle _B} - |1{\rangle _A}|0{\rangle _B}} \right);}\\
{{{\left| {{\phi ^ + }} \right\rangle }_{AB}}}&{ = \frac{1}{{\sqrt 2 }}\left( {|0{\rangle _A}|0{\rangle _B} + |1{\rangle _A}|1{\rangle _B}} \right);}\\
{{{\left| {{\phi ^ - }} \right\rangle }_{AB}}}&{ = \frac{1}{{\sqrt 2 }}\left( {|0{\rangle _A}|0{\rangle _B} - |1{\rangle _A}|1{\rangle _B}} \right).}
\end{align}
A specific property of the Bell states are that $A$ and $B$ are uncertain before observing or synonymously measuring them. After the measurement of one of the states, the state of the other also becomes known. For ${{{\left| {{\phi ^ + }} \right\rangle }_{AB}}}$, if the observed state of $A$ is $\left| 0 \right\rangle $, then the state of $B$ must be $\left| 0 \right\rangle $. If the observed state of $A$ is  $\left| 1 \right\rangle $, then the state of $B$ must be  $\left| 1 \right\rangle $. 
\subsection{Quantum Measurement }
Quantum measurement is a key concept in quantum mechanics, and one of the salient operations in quantum communication and quantum computing. These techniques have no direct counterpart in classical systems. 
The receiver observes the quantum states using a quantum measurement operator, selects relevant observations from a finite set, and restores classical information. This process is vital for the utilization of quantum technologies~\cite{1973Statistical}.

In quantum mechanics, the measurement of an arbitrary quantum state $\left| \psi \right\rangle $ in the Hilbert space ${\cal H}$ is performed using a set of measurement operators $\left\{ {{{\Pi }_m}} \right\}$. These operators must satisfy the completeness condition of $\sum\limits_m {\Pi_m^\dag } {\Pi_m} = {I_{\cal H}}$, where $(\cdot)^\dag$ denotes conjugate transpose, and $I_{\cal H}$ is the identity operator in the Hilbert space. The index $m$ denotes the measurement result obtained after the measurement of the quantum state $\left| \psi \right\rangle $ with the aid of the measurement operator $\Pi_m$. The probability of obtaining the specific result $m$ is formulated as:
\begin{align}
p\left( m \right) = \langle \psi | \Pi _m^\dag \Pi _m^{}\left| \psi \right\rangle,
\end{align}
which must satisfy $\sum\limits_{\forall m} {p\left( m \right)}=1$ since all probabilities must sum to unity. After the measurement, the quantum state collapses onto the state $\frac{{{{\Pi }_m}\left| \psi \right\rangle }}{{\sqrt {\langle \psi | \Pi _m^\dag \Pi _m^{}\left| \psi \right\rangle } }}$.

For measurements of non-orthogonal quantum states having known probability distributions, the positive operator-valued measure (POVM) technique~\cite{6760055} may be used. 
Assuming the quantum state is a pure state $\left| {{\psi}} \right\rangle $, it has a density operator $ \rho = \left| \psi \right\rangle \left\langle \psi \right|$, and the measurement operator is ${ \Pi _m} = \left| {{\mu _m}} \right\rangle \langle {\mu _m}|$, the probability of correctly detecting this quantum state is ${P_c} = {\sum\limits_m {\left| {\left\langle {{{\mu _m}}}
\mathrel{\left | {\vphantom {{{\mu _m}} \psi }}
\right. \kern-\nulldelimiterspace}
{\psi } \right\rangle } \right|} ^2}$, where $\left| {{\mu _m}} \right\rangle $ is a vector of measurement operator. 
Some other measurements, such as the projective measurement~\cite{4539425}, are special cases of POVM measurement.

\subsection{Quantum-Secured Communication Mechanisms}

Quantum cryptography offers a high level of security in communication systems by exploiting the principles of quantum physics and statistical theory. One of its most significant advantages is its ability to detect eavesdropping using quantum technology~\cite{WOS:000911965000001}. 
Compared to classical cryptography, quantum cryptography significantly reduces the likelihood of data interception and tampering, making it a superior option for secure communication~\cite{WOS:000911965000001}.

The family of quantum-secured communications protocols includes QSDC~\cite{WOS:000174548500044, 9130765}, QSIC~\cite{5235303}, and QKD~\cite{Quantumcommunicationchips, Zhang18oe}, as summarized in Tab.~\ref{tab.quantumsecure}.
In addition, there are quantum secret sharing~\cite{6225432}, quantum private key encryption~\cite{1594855}, quantum public key encryption~\cite{8832134}, quantum authentication~\cite{9076167}, and quantum signature~\cite{9271729}.

\begin{figure}[t]
\centering
\includegraphics[width= 3.5 in]{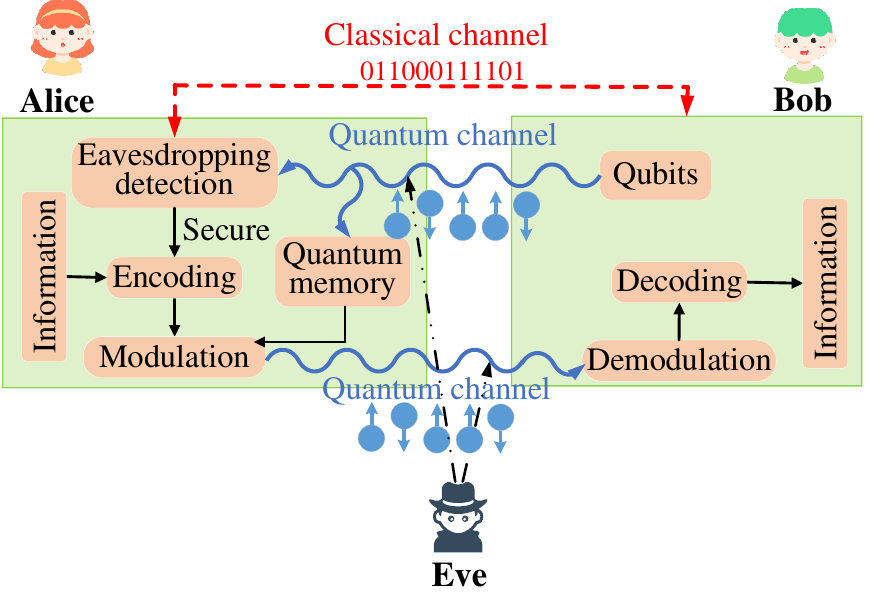}
\caption{Schematic diagram of QSDC. Bob prepares a sequence of qubits and sends it to Alice through the quantum channel. Alice randomly selects a sequence (control qubits) from the received sequence for eavesdropping detection. After that, Alice sends the selected sequence position, measurement basis and measurement results to Bob through the classical channel, and Bob performs security evaluation~\cite{bookchina}.  }
\label{Schematic_diagram_of_quantum_secure_direct_communication}
\end{figure}

\subsubsection{QSDC}

QSDC is a branch of quantum communication that enables the direct transmission of secret messages over a quantum channel using entangled photon-pair or at the single-photon level~\cite{9129730}. It is characterized by theoretically perfect information security.
This technique allows for the direct transmission of a secret message from Alice to Bob without the need for classical ciphertext communication as an intermediary~\cite{bookchina}.
As shown in Fig.~\ref{Schematic_diagram_of_quantum_secure_direct_communication}, Bob generates and sends a series of qubits to Alice through a quantum channel. Alice selects a subset of these qubits (known as control qubits) at random to detect potential eavesdropping, which is essentially based on estimating the quantum bit error rate (QBER). {\color{black}If an eavesdropper observes a qubit, it collapses back into the CD. This causes a qubit error. Then, Alice sends the positions of the selected qubits, the measurement basis used, and the measurement results to Bob through a classical channel. Bob performs a security evaluation based on this information, again by estimating the QBER.}
If the channel is considered safe without eavesdropping, Alice encodes the information sequence through the agreed encoding method, maps it to the remaining qubits and sends it to Bob through the quantum channel. Bob obtains the information sequence after demodulation and decoding. As a benefit, QSDC does not have to generate any secret key, nor does it need any decryption algorithm. Even if there are eavesdroppers, it will not lead to information leakage~\cite{9129730}.

\begin{figure}
\centering
\includegraphics[width=3.5 in]{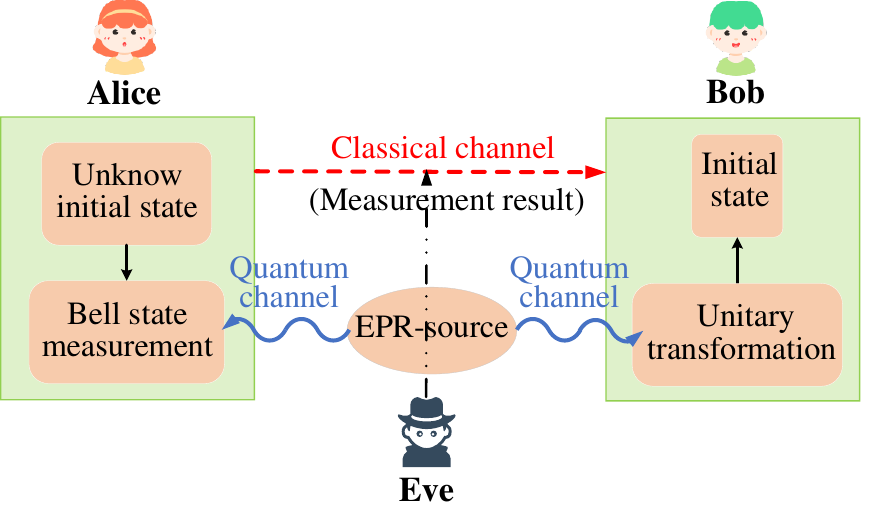}
\caption{Schematic diagram of QSIC. A pair of EPR photons are first distributed to Alice and Bob, and then Alice performs a joint Bell state measurement on the unknown initial quantum state to be transmitted with the EPR photons.  Then Alice transmits the measured classical results to Bob through classical channels. Bob uses the EPR photon and the received classical information to perform the corresponding unitary transformation. Thus, Bob turns the EPR photon into the initial quantum state, which Alice wants to send~\cite{bookchina}. 
}\label{Schematic_diagram_of_quantum_indirect_communication}
\end{figure}

\subsubsection{QSIC}
In contrast to other quantum-secured communication methods, QSIC does not have to transmit quantum states directly. Instead, it uses state entanglement to transmit information~\cite{Flamini2018,bookchina}. A popular QSIC technique is quantum teleportation~\cite{7784726}.
{\color{black} Einstein-Podolsky-Rosen (EPR) entanglement is a unique form of quantum entanglement~\cite{2015Advances }. It usually involves photon polarization. Consider a pair of entangled photons in a quantum superposition state, where one photon is polarized horizontally and the other vertically. When a photon is detected, its polarization is determined. This measurement produces a direct effect on the polarization state of the other photons, even though they are far away. In other words, when the polarization of the first photon is measured, the polarization of the second photon can be predicted without direct observation~\cite{bookchina}. This unusual correlation of EPR entanglement has been demonstrated experimentally.}

The QSIC mode based on EPR photon pairs is shown in Fig.~\ref{Schematic_diagram_of_quantum_indirect_communication}: a pair of EPR photons are first provided both for the transmitter Alice as well as for the receiver Bob. Then, Alice applies joint Bell state measurement to the unknown initial quantum state (including the information encoded in the quantum state) to be transmitted with the aid of the EPR photons. Then Alice transmits the measured classical results to Bob through classical channels. Bob uses the EPR photon and the received classical information to perform the corresponding unitary transformation~\cite{ANDERSON1993157}. Thus, Bob turns the EPR photon into the initial quantum state that Alice wants to send. This method does not directly transmit quantum states but nonetheless completes the transmission of quantum states, known as quantum teleportation.

Another approach to QSIC is constituted by quantum dense coding~\cite{BAN200197}, where the transmitter applies a unitary transformation to one of two entangled particles and then sends both particles to the receiver~\cite{BAN200197}. The receiver jointly measures the pair of particles and determines the type of transformation conducted by the transmitter based on the measurement results.

\begin{figure}[t]
\centering
\includegraphics[width=3.5 in]{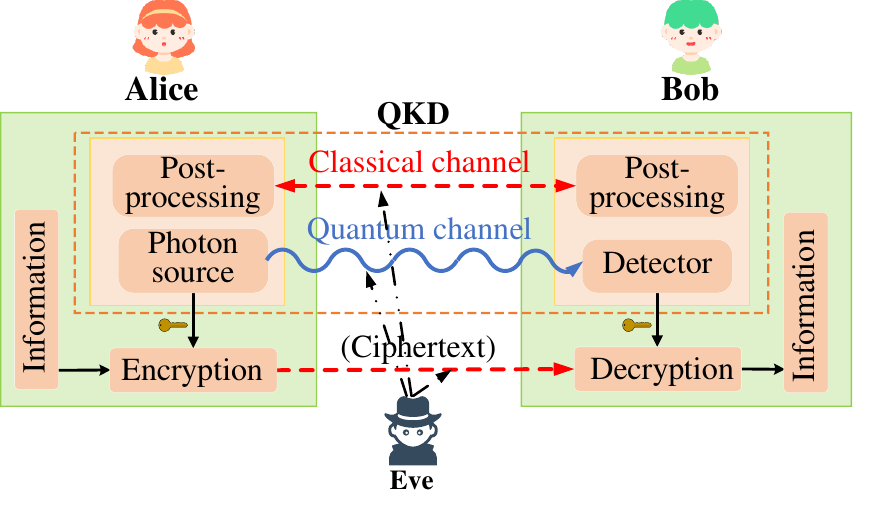}
\caption{Schematic diagram of QKD. The transmitter, Alice, uses the secret key allocated by the QKD system to encrypt the classical information sequence and sends it to the receiver, Bob, through the classical channel. After receiving the ciphertext, Bob decrypts it with his own key to get the classical information. The third-party eavesdropper Eve is unable to crack the ciphertext because it is impossible to obtain the secret key generated by the QKD system~\cite{bookchina}.}\label{Schematic_diagram_of_quantum_secure_communication_system_based_on_QKD}
\end{figure}

\subsubsection{QKD}

This is a specific key-negotiation protocol of quantum-secured communication relying on secret keys provided for both parties, which are then used for encrypting and transmitting data via a classical communication system. 
QKD distributes unconditionally secure symmetric secret keys between Alice and Bob~\cite{9684555}.
It provides a means of ensuring unconditional security in key exchange.
A QKD system is also capable of detecting the presence of an eavesdropper (Eve). It generates an absolutely secure quantum key, thanks to the no-cloning theorem, which implies that qubits cannot be copied because, upon observation, they fall back to the CD~\cite{2020Quantum}.
Any unauthorized tampering with legitimate communication on a QKD link directly increases the error rate, alerting both the transmitter and receiver~\cite{WOS:000730601600001}.

The basic structure of a QKD-based quantum-secured communication system is shown in Fig.~\ref{Schematic_diagram_of_quantum_secure_communication_system_based_on_QKD}. 
{\color{black}In this system, both the transmitter (Alice) and the receiver (Bob) are constituted by a classical communication module and a QKD module. }
The system generates and stores a secret key, and has two communication channels: a quantum channel for transmitting photons used in QKD (assuming the use of optical quantum communication) and a classical channel for transmitting encrypted data and auxiliary information generated during the QKD process~\cite{9684555}. Alice uses the secret key provided by the QKD system to encrypt a classical information sequence and send it to Bob via the classical channel. Upon receipt of the ciphertext, Bob uses its own secret key to decrypt it and retrieve the original classical information~\cite{2012Comparison}. 

\begin{figure}
\centering
\includegraphics[width=3.5 in]{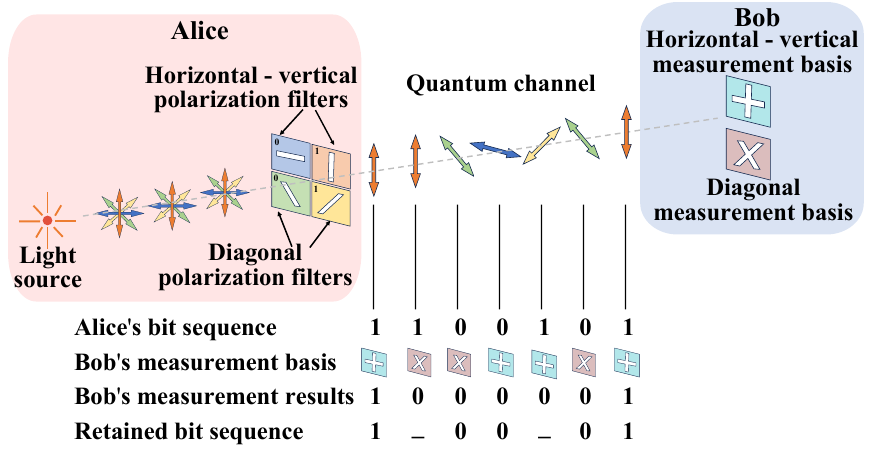}
\caption{Quantum communication process. Alice prepares a set of qubits and encodes the information onto the quantum states of the qubits. Then, Alice transmits qubits to Bob through a quantum channel. Bob measures the quantum states after receiving the transmitted qubits and decodes the information in the qubits by measuring the results~\cite{bennett2014quantum}.}
\label{Quantum_communication_process}
\end{figure}

The Bennett-Brassard-84 (BB84) protocol~\cite{Publickeydistributionandcointossing,bennett2014quantum} is the first QKD protocol in the world. As shown in Fig.~\ref{Quantum_communication_process}, the protocol procedure can be summarized in the following steps:
\begin{itemize}
\item \textit{Preparation of qubits:} Alice prepares qubits typically carried by photons or atoms. Qubits are prepared onto specific quantum states using experimental procedures, e.g., manipulating photons with laser pulses.

\item \textit{Encoding of quantum states:} Alice encodes the information to be sent into the quantum states of the qubits. This is typically done by quantum gate operations, such as changing a photon's polarization or an atom's spin state to represent different information.

\item \textit{Transmission of quantum states:} After the encoding is complete, Alice transmits the qubits to Bob. This can be carried out through optical fiber, free space transmission, or other quantum channels. 
To ensure the stability and security of transmission, it is usually necessary to adopt some technical means, e.g., quantum entanglement.

\item \textit{Reception of quantum states:} Bob measures the quantum state after receiving the qubits. The measurement method usually corresponds to the encoding process, measuring the specific properties of the qubits to restore the encoded information. When Bob selects an incorrect measurement basis at random, he discards relevant bits.

\item \textit{Decoding of quantum states:} Bob decodes the information in the qubits by measuring the result. The decoding process must be done according to Alice's encoding mode to recover the original information.
\end{itemize}

\subsection{Quantum Decoherence Model and Quantum Noise}

Quantum systems can be susceptible to environmental disturbances, resulting in noise and decoherence and destroying quantum information~\cite{ Decoherence2016, WOS:000385644900001}. 
Quantum noise is the random fluctuations caused by the interaction of a quantum system with the environment~\cite{ WOS:001325277500001}.
Quantum gates operate on qubits and manipulate quantum states through unitary transformations that are sensitive to quantum noise~\cite{ QuantumComputationandQuantumInformation, WOS:000345871100004}. In quantum-native applications (e.g., gate-based quantum computing), quantum noise can significantly affect the fidelity and reliability of quantum gates and, in turn, the overall performance of quantum systems~\cite{ WOS:000435544200004}.

The loss of coherence has an impact on the dependability of quantum states, such as entanglement, which is crucial for quantum communication protocols, such as QKD and quantum teleportation. Markov decoherence~\cite{Mirrahimi2021Dynamics} is the most commonly studied form of decoherence in quantum systems, in which the system interacts with the environment without any memory of previous interactions. 

In the presence of decoherence, the evolution of quantum systems is often modeled using the Lindblad master equation, which has the general form~\cite{ WOS:000519594600001,breuer2002theory}:

\begin{align}
\frac{{d\rho }}{{dt}} =  - \frac{i}{{\hbar}}\left[ {{H_s},\rho } \right] + \sum\limits_k {\left( {{L_k}\rho L_k^\dag  - \frac{1}{2}\left\{ {{L_k^\dag}L_k ,\rho } \right\}} \right)},
\end{align}
where $\rho $ is the density matrix describing the state of the system; ${{H_s}}$ is the Hamiltonian of the system, representing its nondissipative evolution; $\hbar$ is reduced Planck constant; $L_k $ are the Lindblad operators, which represent different types of interactions between the system and environment; ${L_k^\dag }$ is conjugate transpose of ${{L_k}}$. The dissipative term $\sum\limits_k {\Big( {{L_k}\rho L_k^\dag  - \frac{1}{2}\big\{ {{L_k^\dag}L_k ,\rho } \big\}} \Big)} $ accounts for irreversible processes that introduce decoherence in the system.

Depending on the nature of the interaction between the system and environment, different types of decoherence can occur:
\subsubsection{Dephasing Decoherence}
Dephasing is the most commonly studied type of decoherence that describes loss of phase information in a system. It serves as the main form of decoherence.
For a single qubit, a simple dephasing process can be represented by a Lindblad operator ${L_k} = {\sigma _z}$, where ${\sigma _z}$ is the Pauli-Z operator~\cite{WOS:000404144400006}.

\subsubsection{Energy Relaxation}
Energy relaxation describes the relaxation process caused by the energy level transition of the system. This is relevant to quantum computing, where a qubit may transition from an excited state to a ground state due to interaction with the environment~\cite{WOS:000443408500002}.
The Lindblad operators $L_k $ are related to the lowering and raising operators of the system,  ${\sigma _ - }$ and  ${\sigma _ + }$, which correspond to transitions from the excited state to the ground state and the other way around, respectively.

\subsection{Summary}

{\color{black}This section touches upon the fundamentals of quantum communication, including qubits, quantum entanglement, and quantum measurement, as well as various quantum-secured communication protocols, such as QKD, QSDC, and QSIC. While much of the research has focused on QKD, this survey covers QKD, QSDC, and QSIC. We compare these methods and highlight their advantages. For example, QSDC allows for direct data communication without the need for key distribution, eliminating potential security vulnerabilities related to key storage and ciphertext attacks~\cite{xjzz20198183}.  
Moreover, QSDC uses quantum mechanical principles to quantify the amount of data that an eavesdropper could potentially steal, offering unparalleled practical security~\cite{9130765}. Quantum teleportation, a type of QSIC, allows for the ``transmission” of qubits without the physical transfer of the particles storing the qubits~\cite{9023997}.

%% file: Chap_Survey_of_Surveys.tex
\section{Survey of Surveys}

\begin{table*}
\caption{Comparison of this survey with other related surveys, where \checkmark indicates that the survey has the corresponding topics, and a blank space indicates otherwise}
\centering
\begin{tabular}{|c|c|c|c|c|c|c|c|}
\hline

\textbf{\makecell*[c]{Paper}}  & \textbf{\makecell*[c]{Year}} &\textbf{\makecell*[c]{Quantum\\ Channels}} & \textbf{\makecell*[c]{Quantum  \\ PHY Layer}} &\textbf{\makecell*[c]{Quantum \\ MAC and Upper Layer}} &\textbf{\makecell*[c]{Quantum AI\\ and Computing}} &\textbf{\makecell*[c]{Quantum Sensing\\ and Timing}}  & \textbf{\makecell*[c]{Commercial\\ Landscape}}\\ \hline
This survey &  2025 &  \checkmark&   \checkmark &  \checkmark & \checkmark  &  \checkmark&  \checkmark\\ 
{\color{black}~\cite{ 10440135}} & 2024  &  &  \checkmark & \checkmark  &   &   &  \\ 
{\color{black}~\cite{ WOS:001137396000001}} & 2024  &  &  & & \checkmark &   & \\ 
{\color{black}~\cite{ WOS:001141997100001}} & 2024  &  & \checkmark & &  & & \\ 
~\cite{ arXivHanzo} & 2024  &  &  \checkmark &   &  \checkmark &  \checkmark &  \\ 
{\color{black}~\cite{ WOS:001087470200001}} & 2023 & &   & \checkmark & \checkmark  & \checkmark & \\ 
{\color{black}~\cite{ WOS:001054330900001}} & 2023  &  & & &  \checkmark & & \\ 
{\color{black}~\cite{ WOS:001029956200002}} & 2023  &  & \checkmark & &  & & \\ 
{\color{black}~\cite{ WOS:001034626400001}} & 2023  &  & & &  \checkmark & & \\ 
{\color{black}~\cite{ WOS:001114162000001}} & 2023  &  &  & & \checkmark & \checkmark & \\ 
{\color{black}~\cite{ WOS:001045556300001}} & 2023  &  &  & & \checkmark &   & \\ 
{\color{black}~\cite{ WOS:000939178900001}} & 2023  &  &  & & \checkmark &   & \\ 
{\color{black}\cite{10064036}} &  2023  & & \checkmark  &  \checkmark & \checkmark & & \\ 
{\color{black}\cite{ CHAWLA20232191}} &  2023  & & \checkmark  &  \checkmark & & & \\ 
{\color{black}\cite{ WOS:000958092100001}} &  2023  & &  &  & &  \checkmark & \\ 
{\color{black}\cite{ WOS:000937115100001}} &  2023  &\checkmark & \checkmark  &  \checkmark & & & \\ 
{\color{black}\cite{ WOS:001062432100002}} &  2023  & & \checkmark  &  \checkmark &\checkmark & & \\ 
{\color{black}\cite{ WOS:001083998400012} }&  2023  & & \checkmark &  \checkmark &  &  & \checkmark \\ 
{\color{black}\cite{ WOS:000977361500001}} &  2023  & & \checkmark &  \checkmark &  &  &  \\ 
{\color{black}\cite{ WOS:001062144100001}}  &  2023  & & \checkmark &  \checkmark &  &  &  \\ 
{\color{black}~\cite{ 9870532}} & 2022  &  &  & & \checkmark &   & \\ 
{\color{black}\cite{ WOS:000923685400002}} &  2022  & \checkmark & &    &  &  &  \\ 
{\color{black}\cite{ ILLIANO2022109092}} &  2022  & & \checkmark  &  \checkmark & & & \\ 
~\cite{9684555}& 2022 & \checkmark&  \checkmark &\checkmark &  & & \\ 
~\cite{9211474} & 2022 & &  \checkmark &\checkmark &  & & \\ 
~\cite{HeLiangHuang2022NearTermQC} &2022  & &   &  &  \checkmark & & \\ 
~\cite{9528843} & 2021 & \checkmark&  \checkmark & \checkmark & \checkmark &\checkmark & \\ 
~\cite{9384317} & 2021  & &   &  &\checkmark  & & \\ 
{\color{black}~\cite{9420742}} & 2021  & &   &  &\checkmark  & & \\ 
{\color{black}~\cite{20202Towards}} & 2020 & &   &  &\checkmark  & & \\ 
~\cite{8972916} & 2020 & &   &  &\checkmark  & & \\ 
~\cite{Standards4Quantum} & 2020 & & \checkmark  &  & \checkmark & \checkmark& \\ 
~\cite{RevModPhys.92.025002} & 2020 & \checkmark& \checkmark  &  \checkmark &   & & \\ 
~\cite{9023997} & 2020  & \checkmark &   &  &  & & \\ 
~\cite{2019Survey} & 2019 & & \checkmark  & \checkmark &   & & \\ 
~\cite{8540839} & 2019 & &   &  & \checkmark  & & \\ 
~\cite{GYONGYOSI201951} & 2019  & &   &  & \checkmark  & & \\ 
~\cite{2019Quantum} & 2019  & & \checkmark   &  &   \checkmark & & \\ 
~\cite{cozzolino2019high} & 2019 & \checkmark&   &  &   & & \\ 
~\cite{8439931}  & 2019 &\checkmark &  \checkmark &  &   & & \\ 
~\cite{Dunjko2018} & 2018 & &   &  &  \checkmark & & \\ 
~\cite{wehner2018quantum} & 2018 & \checkmark &  \checkmark & \checkmark &  \checkmark & & \\ 
~\cite{gyongyosi2018survey} &2018 & \checkmark &   &  &   & & \\ 
~\cite{Zhang18oe} & 2018 & & \checkmark  &  &   & & \\ 
~\cite{Flamini2018} & 2018 & & \checkmark  &  &   & & \\ \hline
\end{tabular}
\label{tab.surveyofsurveys}
\end{table*}

The existing surveys on QIT can be divided into quantum communication and quantum computing. 
They typically focus on a limited number of aspects related to quantum communication networks, such as QKD, channel modeling,
optimization algorithms, quantum circuits, and coding without a systematic, holistic picture of techniques and architectures of quantum-aided NG networks. 
By contrast, this survey endeavors to deliver a comprehensive perspective of the potential applications of quantum communication techniques for critically appraising their pros and cons in NG systems. Tab.~\ref{tab.surveyofsurveys} compares this survey against the existing surveys.

\subsection{{\color{black}Quantum Communication}}

\subsubsection{{\color{black}Quantum Security Protocols}}
{\color{black}
Quantum security protocols are cryptographic protocols designed to leverage the principles of quantum mechanics to provide security for communication and computation. QKD, as one of the fastest developing quantum security protocols, has been widely discussed from different aspects, such as networking~\cite{Zhang18oe,9684555}, security~\cite{RevModPhys.92.025002,WOS:000977361500001}, applications to smart grids~\cite{9211474}, and designs in different frequency bands \cite{ WOS:001141997100001}.

Cao \textit{et al.}~\cite{9684555} discussed short-range, metropolitan, and long-haul QKD networks from an engineering perspective and described the architecture, interfaces, protocols, and various enabling technologies of QKD. 
Zhang \textit{et al.}~\cite{Zhang18oe} introduced salient experimental studies of large-scale QKD, including fiber-based QKD in metro networks, trusted relay-based backbone networks, and satellite-borne QKD in free space.
Xu \textit{et al.}~\cite{RevModPhys.92.025002} focused on the security of practical QKD, including security proof, implementation, and vulnerabilities.
Moreover, the application, testing, and verification practices of QKD were discussed in~\cite{WOS:000977361500001}.
Some researchers studied the application of QKD protocols in smart grid communications~\cite{9211474} and microwave continuous-variable QKD (CVQKD) and THz CVQKD systems \cite{ WOS:001141997100001}.

QSDC and quantum teleportation have also attracted researchers' attention. Pan \textit{et al.} \cite{ 10440135} highlighted the motivation and current status of QSDC research, with a focus on the theoretical basis and experimental verification.
Cacciapuoti \textit{et al.}~\cite{9023997} reviewed the quantum teleportation process by providing the mathematical model and some representative schemes conceived for practical entanglement generation and distribution.
Some other branches of quantum cryptography, such as quantum secret sharing, quantum signature, and quantum private query, were briefly introduced in \cite{2019Survey}.

\subsubsection{Quantum Systems and Networks}

Leveraging the principles of quantum mechanics, a myriad of devices can constitute a robust quantum network to ensure secure, reliable, and high-speed communications across different users, thus paving the way for Qinternet~\cite{QuantumInternetintheSky}.
Hasan \textit{et al.}~\cite{ WOS:000937115100001} surveyed the vision, design goals, information processing, and architectures of quantum communication. They also introduced a quantum communication system model and discussed a suite of prospective quantum technology applications.
Paudel \textit{et al.}~\cite{WOS:001083998400012} presented a comprehensive review of the latest advancements in quantum networks within the energy industry, encompassing platforms, devices, and protocols of QKD and quantum teleportation. 

One of the most striking advantages of CV systems over DV technology is their intrinsic compatibility with the latest optical technologies.
Hosseinidehaj \textit{et al.}~\cite{8439931} examined recent advances in CV quantum communication designed for LEO satellites. With their cost effectiveness, scalability, and stability, emerging quantum photonic devices are flourishing and opening up new possibilities in smaller footprints.
Luo \textit{et al.} \cite{ WOS:001029956200002} reviewed the development of quantum photon chips for quantum communication and outlined the popular photonic integrated manufacturing platforms and the key components of integrated quantum communications.

\subsubsection{Qinternet}
This is a cutting-edge paradigm that uses the unique characteristics of quantum technology to influence communication networks radically.
Rozenman \textit{et al.} \cite{ WOS:001087470200001} described communication activities that can be carried out via a quantum network and correlated these communication activities with experimental needs and phases of the Qinternet. 
The functionalities, potential applications, enabling technologies, and components of the Qinternet were discussed in~\cite{9528843}.
In addition, there are some researches on Qinternet protocols. For example, Wehner \textit{et al.}~\cite{wehner2018quantum} provided examples of known application protocols for each phase of development towards a fully functional Qinternet, and  
Illiano \textit{et al.}~\cite{ ILLIANO2022109092} proposed a major paradigm shift in the design of Qinternet protocol stacks from the perspective of harnessing the properties of quantum entanglement and quantum information.

\subsubsection{Quantum Channel}
A quantum channel is a communication channel that transmits quantum information. Unlike classical communication channels that transmit classical bits, quantum channels transmit quantum states, i.e., qubits, which can be in a superposition or entangled state.
A profound property of quantum communications has recently been revealed, that is, the channel’s capability of delivering information is also influenced by the superadditivity, superactivation and causal activation phenomena. In this context,
Koudia \textit{et al.}~\cite{ WOS:000923685400002} offered an informal description of the three unconventional phenomena and summarized the differences and similarities between these three phenomena. Additionally, Gyongyosi \textit{et al.}~\cite{gyongyosi2018survey} provided an overview of quantum channel capacities, including the properties of quantum communication channels, the fundamental differences between the classical and quantum channels, the capacity measures and practical implementations of quantum channels.

\subsubsection{Standardization of Quantum Technology}
Research on standardization of quantum technology and quantum ecology has been discussed in~\cite{Standards4Quantum,WOS:001062144100001}.
Jenet \textit{et al.}~\cite{ Standards4Quantum } provided a comprehensive report on the standardization of quantum technology. Their report introduced how standardization can support the deployment of quantum technologies in the areas of quantum security, confidential communications, measurement, and sensing. 
Purohit \textit{et al.}~\cite{WOS:001062144100001} emphasized the need for a quantum-ready ecosystem. They formulated standard quantum technology readiness levels and defined innovative models and tools for it. 

\subsubsection{Quantum Communication and Computing}
Some authors carried out relevant research on quantum computing and quantum communication.
Ali \textit{et al.}~\cite{ WOS:001062432100002} delved into the subjects of quantum security, quantum repeaters, and quantum interconnection, offering insightful discussions on these topics. Their study also outlined the challenges of quantum-enabled 6G networks, complemented by potential future research ideas.
Yang \textit{et al.}~\cite{10064036} categorized some crucial aspects of quantum information technologies, including quantum hardware, quantum networks, quantum cryptography, and quantum machine learning (ML).

\subsubsection{Others}
There are several surveys that focus on high-dimensional quantum communication~\cite{cozzolino2019high}, quantum microwave technology~\cite{WOS:000958092100001}, and quantum information processing technology~\cite{Flamini2018,arXivHanzo}. 
Specifically, Cozzolino \textit{et al.}~\cite{cozzolino2019high} discussed the generation, propagation, and measurement of high-dimensional quantum states and the benefits of using high-dimensional quantum states in communication.
Casariego \textit{et al.}~\cite{ WOS:000958092100001} reviewed the state-of-the-art quantum microwave technology, discussed some of the most recent results, such as quantum teleportation of an unknown microwave coherent state, quantum radar and imaging, and quantum illumination (QI).
Flamini \textit{et al.}~\cite{Flamini2018} provided the developments of quantum repeaters and distributed blind quantum computing protocols for photonic quantum networks.
Recently, Hanzo~\textit{et al.}~\cite{ arXivHanzo} summarized the latest advances in quantum information processing, sensing, and communication with an emphasis on quantum error mitigation. Moreover, the relevant techniques of quantum ML and quantum radar were discussed.

\subsection{Quantum Computing}

Quantum computing relies on algorithms and circuits designed for quantum computers. A significant amount of research has focused on quantum algorithms~\cite{HeLiangHuang2022NearTermQC, 8972916, WOS:001137396000001,WOS:001045556300001, 8540839}, quantum ML/AI~\cite{Dunjko2018, WOS:000939178900001, 2019Quantum,  9870532}, and quantum computers~\cite{9384317, GYONGYOSI201951, WOS:001054330900001 }. In addition, there are some review studies involving quantum Internet of Things (IoT)~\cite{ CHAWLA20232191, WOS:001114162000001} and post-quantum cryptographic~\cite{20202Towards}.

\subsubsection{Quantum Algorithms}
Numerous researchers are working on developing quantum algorithms~\cite{HeLiangHuang2022NearTermQC, 8972916, WOS:001137396000001} and its applications of the quantum search algorithm in classical communication system~\cite{WOS:001045556300001, 8540839}.
Huang \textit{et al.}~\cite{ HeLiangHuang2022NearTermQC } provided a comprehensive overview of near-term quantum computing techniques, covering variational quantum algorithms, quantum error mitigation (QEM), quantum circuit compilation, benchmarking protocols, and classical simulation. 
Li \textit{et al.}~\cite{8972916} summarized the typical quantum intelligent algorithms in quantum optimization and learning, such as the quantum evolutionary algorithm, quantum particle swarm algorithm, and quantum immune clonal algorithm. They also provided related experiments to verify quantum intelligent algorithms compared with traditional intelligent algorithms.
Kou \textit{et al.} \cite{ WOS:001137396000001} provided a review of dynamic optimization based on quantum computing. The principles of different quantum optimization algorithms were discussed for dynamic optimization problems and their improved forms of quantum optimization algorithms.

Some authors have reviewed the applications of quantum computing in the fields of telecommunications and power engineering.
Phillipson \cite{ WOS:001045556300001} explored the transformative impact of quantum computing on telecommunications, and investigated its applications to solve computationally intensive problems.
Botsinis \textit{et al.}~\cite{8540839} summarized the benefits of quantum search algorithms in the context of classical communication systems for solving a suite of optimization problems, 
such as diverse multi-user transmission and detection, joint channel estimation and data detection, and optimal multi-objective routing.
Quantum computing has broad and critical implications for power systems, including increasing the uptake of renewable energy and boosting computer efficiency. 
Moreover, Liu \textit{et al.} \cite{ WOS:001034626400001} applied quantum computing to power engineering, described an important quantum algorithms, and provided quantum computing applications in power systems.
}

\subsubsection{{\color{black}Quantum ML/AI}}
{\color{black}
In recent years, researchers have found that novel ML algorithms that incorporate quantum computing features can achieve the acceleration of traditional algorithms. 
Dunjko \textit{et al.}~\cite{Dunjko2018} studied the interactions between quantum information processing, ML, and AI, including a quantum generalization of ML-type tasks, quantum enhancements for ML, quantum learning agents, and quantum AI.
Zeguendry \textit{et al.} \cite{ WOS:000939178900001} provided the research path from quantum basic theory to quantum ML algorithm and discussed fundamental algorithms for quantum ML, including quantum 
support vector machines (SVMs), variational quantum circuits, and quantum neural networks (QNNs).
Nawaz \textit{et al.}~\cite{2019Quantum} proposed a novel quantum computing-assisted, quantum ML-based framework for NG communication networks while clarifying its potential enabling technologies.
Duong \textit{et al.}~\cite{9870532} reviewed the quantum-inspired ML in NG wireless networks and discussed how it provides NG networks with defense mechanisms against potential cyber-attacks.}

\subsubsection{{\color{black}Quantum Computers}}
Computationally intensive applications, including ML, chemical simulations, and financial modeling, may be well-suited for noisy intermediate-scale quantum (NISQ)  computers~\cite{9384317}. A major challenge is mapping quantum circuits onto NISQ hardware while also considering the physical constraints of the underlying quantum architecture. Li \textit{et al.}~\cite{9384317} presented a survey of AI and heuristic-based quantum circuit compilation methods, such as genetic algorithms, genetic programming, ant colony optimization, AI planning, and heuristics methods employing greedy algorithms, dynamic and graph optimization techniques.

Researchers have also discussed the quantum hardware and software architecture of quantum computing systems.
For example, Gyongyosi \textit{et al.}~\cite{GYONGYOSI201951} reviewed the basic quantum hardware blocks of quantum computers, the conditions of large-scale quantum computers, and progress on the implementation of quantum devices, computers, and algorithms.
Khan \textit{et al.} \cite{ WOS:001054330900001} reviewed the software architectures of quantum computing systems, including application domains of quantum software architecture, architectural process, modeling notations, design patterns, tools, and frameworks.

\subsubsection{{\color{black}Quantum IoT}}
As the number of IoT devices grows, so does the volume of data, and traditional computing methods cannot keep up. In contrast, quantum computing technology can help process and analyze enormous amounts of data more efficiently. 
Chawla \textit{et al.}~\cite{ CHAWLA20232191} analyzed quantum-secured structures for IoT, including attacks on IoT applications, quantum-resistant solutions for securing IoT communication, and quantum authentication methods.
Peelam \textit{et al.} \cite{ WOS:001114162000001} explored various aspects of the application of quantum computing in the field of IoT to improve accuracy, speed, and security, such as IoT optimization using quantum computing, faster computation at IoT terminals, quantum-based encryption and authentication methods in IoT, and quantum sensors for~IoT.

\subsubsection{{\color{black}Post-Quantum Cryptographic}}

Post-quantum cryptography is the new generation of cryptographic algorithms designed to withstand attacks on existing cryptographic algorithms by quantum computers, it will have a far-reaching impact on the existing public-key cryptosystems.
Tiago \textit{et al.}~\cite{20202Towards } analyzed the impact of quantum-computing attacks (based on Grover's and Shor's algorithms) on blockchain and studied how to apply post-quantum cryptosystems to mitigate such attacks. In addition, they provided extensive comparisons of the main types of post-quantum public-key and digital signature schemes, followed by an overview of post-quantum cryptographic IoT systems designed for protecting IoT systems from known quantum computing attacks~\cite{9420742}.

\subsection{Key Differentiators of This Survey }

The surveys discussed above have provided insights into a range of specific aspects, but they do not consider NG wireless systems. However, the employment of quantum technology in NG communications has the potential to revolutionize the way data is transmitted and processed. It may lead to significant advances in a variety of fields. Further study of these topics is necessary to fully understand the potential of quantum technology and develop effective strategies for incorporating them into NG communication systems.

We examine how different quantum communication techniques can be applied at different communication interfaces in the different layers of NG systems. For example, we summarize the latest research on multi-band quantum access~\cite{PhysRevResearch.3.043014, 9220903, app9245285, 9003426, refId0, e23091223, 2020Hybrid, 9492803, 8976167, 8761168, 8108494 }, multi-user quantum communications~\cite{bahrani2015orthogonal, 6545781, 2021Quantum, 9013407, 8756986, 8648086}, quantum MIMO~\cite{9739032, 2005Quantum, zhou2019shot}, quantum resource allocation~\cite{9552894, WOS:000685891700006, wang2019protection, QKD_ON}, and quantum routing~\cite{yangquantum}.  We also highlight the practical considerations of implementing these technologies in practice. Overall, this review intends to offer a comprehensive overview of the potential of quantum communication technology in the context of NG communication systems and to identify areas for further research development.

%% file: Chapter_1.tex
\section{Channel modeling and capacity}

The study of quantum channel theory can provide the theoretical basis for the application of quantum information in communication and signal processing. In particular, the establishment of mathematical models of quantum channels is essential in deriving their capacities, which can be distinctively different from those of classic channels. This section exemplifies several quantum channel models and discusses the quantum channel capacity.

{\color{black}\subsection{Quantum Channel Capacity} }

{\color{black}In classical information theory, there is a maximum information transmission rate for any given classical channel, which is the channel capacity of the channel. Different from} classical channels, quantum channels have the ability to transmit not only classical information but quantum information as well~\cite{A_brief_tutorial_on_quantum_information}. As a result, quantum channels possess both classical and quantum capacities. The classical capacity of a quantum channel represents the maximum amount of classical information that can be transmitted by the channel. It can be calculated by considering the upper bound on the amount of information accessible at the receiver through quantum measurement operators~\cite{Holevo1973BoundsFT,8998224}.
This upper bound, {\color{black} referred to as the Holevo bound,} is expressed as:
\begin{equation}
\label{I}
I\left( {X;Y} \right) \le S\left( {{\rho _{_X}}} \right) - \sum\limits_k {{p_{_k}}S\left( {{\rho_{_k}}} \right)}  = \chi ,
\end{equation}
where {\color{black} $I\left( {X;Y} \right)$ is the mutual information with $X$ being the set of source information symbols at the transmitter and $Y$ being the measurement results at the receiver; $S\left(  \cdot  \right)$ is the von Neumann entropy; ${\rho_{_k}}$ is the density operator corresponding to the input quantum state with probability ${p_{_k}}$, $1 \le k \le n$; $ \rho_{_X}$ is the average density operator with ${\rho _{_X}} = \sum\limits_k {{p_{_k}}{\rho _{_k}}} $. 
Here, $\chi$ is called the Holevo quantity.} 

The von Neumann entropy, $S\left( \rho_{_X} \right)$, is a measure of information contained in a quantum state~\cite{bookchina}, as given by
\begin{equation}
S\left( \rho_{_X}  \right) =  - {\rm{Tr}}\left[ {\rho_{_X} \log \left({ \rho _{_X}}  \right)} \right],
\end{equation}
where ${\rm{Tr}}\left[  \cdot  \right]$ represents the trace operation.

{\color{black}Consider the spectral decomposition, ${\rho_{_X}} = \sum\limits_x {{\lambda _\kappa}} \left| x \right\rangle \left\langle x \right|$, of ${\rho_{_X}}$, where  $\left\{ {\left| x \right\rangle } \right\}$ forms a complete set of orthonormal basis and the eigenspectrum ${\lambda _\kappa}$ forms a set of probability distributions. The random variable corresponding to the eigenvalue is $X$. It is easy to find that $S\left( {{\rho_{_X}}} \right) = H\left( X \right)$, where $H(X)$ is the Shannon entropy of the random variable $X$. Since the source is a quantum system, the von Neumann entropy is represented by the density operator, while Shannon entropy is represented by the probability density. Using the properties of Shannon entropy, any quantum state of a $d$-dimensional quantum system, $S\left( \rho  \right) \ge 0$, where eigenstates can form nice orthonormal basis with the identical probability density~\cite{A_brief_tutorial_on_quantum_information}.}
In classical information theory, we assume that the source $X$ sends classical messages ${x_\omega }\left( {\omega  = 1,2, \cdots,\Omega } \right)$ with corresponding probabilities $p\left( {{x_\omega }} \right)$. The classical Shannon entropy is $H\left( X \right) =  - \sum\limits_\omega  {p\left( {{x_\omega }} \right)} \log p\left( {{x_\omega }} \right)$. Here, $S\left( {{\rho _{_X}}} \right)$ can be expressed by the entropy of the probability distribution of the eigenvalues, i.e.,  
\begin{equation}
S\left( {{\rho _{_X}}} \right) = H\left( {\left\{ {{\lambda _\kappa }} \right\}} \right) =  - \sum\limits_\kappa  {{\lambda _\kappa }\log \left( {{\lambda _\kappa }} \right)},
\end{equation}
where ${{\lambda _\kappa }}$ denotes an eigenvalue of the density operator ${{\rho _{_X}}}$.

In classical communication systems, the source entropy $H\left( X \right)$  is the bound of the mutual information $I\left( {X;Y} \right)$, hence $I\left( {X;Y} \right) \le H\left( X \right)$.  In quantum communication systems, the maximum of accessible information is obtained when the states are orthogonal pure states, $\chi  = H\left( X \right)$, as visualized as the Bloch sphere in Fig.~\ref{Bloch_sphere}. For mixed states inside the Bloch sphere or non-orthogonal pure states, $\chi  < H\left( X \right)$. Therefore,
\begin{equation}
I\left( {X;Y} \right) \le \chi  \le H\left( X \right).
\end{equation}

In addition to the classical capacity, the quantum capacity of a quantum channel represents the maximum volume of quantum information that may be transmitted over the channel. This is measured by the maximum amount of quantum coherent information~\cite{Ramakrishnan_2021}, denoted by $C$, as given by
\begin{equation}
\begin{aligned}
C &= \max I\left[ {{{\rho }_{_X}};{\cal N}\left( {{{ \rho }_{_X}}} \right)} \right]\\
&= \max \left[ {S\left( {{\cal N}\left( {{{ \rho }_{_X}}} \right)} \right) - S\left( {{{ \rho }_{_X}};{\cal N}\left( {{{ \rho }_{_X}}} \right)} \right)} \right],
\end{aligned}
\end{equation}
where $I\left[ { \rho_{_X} ;{\cal N}\left( \rho_{_X}  \right)} \right]$ is the quantum coherent information, and ${\cal N}$ denotes the channel's mapping action of the input density operator to the output density operator gleaned from the channel.

\begin{figure}
\centering
\includegraphics[width=3.7 in]{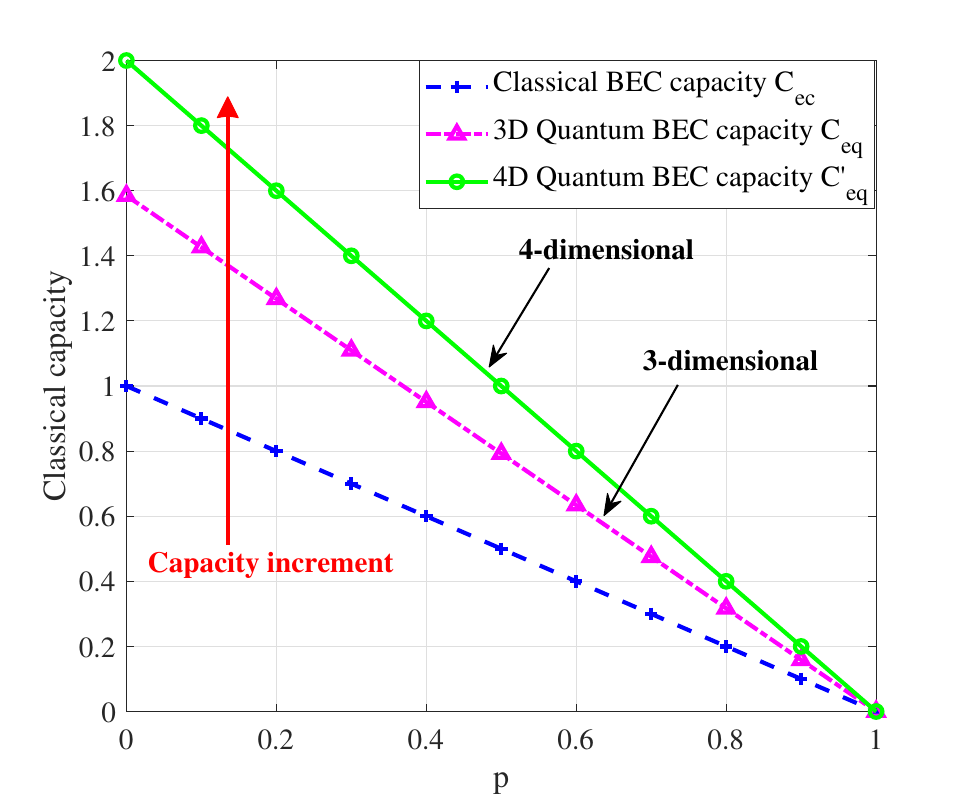}
\caption{Comparison of the classical BEC capacity and of the quantum BEC capacity vs. the erasure probability $p$. Compared to current classical communications, quantum communications have the powerful capability to transmit more information due to the increased Hilbert space dimension of the quantum channel~\cite{gyongyosi2018survey}.}
\label{Channel_capacity}
\end{figure}

Compared to classical communications, quantum communications have the potential to transmit more information due to the increased Hilbert space dimension $d$ of the quantum channel~\cite{cozzolino2019high, gyongyosi2018survey}. 
More explicitly, as shown in Fig.~\ref{Channel_capacity}, the classical capacity of a quantum channel exceeds that of a classical channel, where ${C_{ec}}$  is the classical binary erasure channel (BEC) capacity, ${C_{eq}}$  is the classical capacity of a quantum BEC, and $p$ is the erasure probability~\cite{gyongyosi2018survey,1999Entanglement}. 
${C_{ec}} = 1 - p$, and ${C_{eq}} = \left( {1 - p} \right)\log d$.
The classical channel can be viewed as a special case of the quantum channel associated with $d = 2$. Indeed, a quantum superposition state, which is the linear combination of different quantum states, can help the quantum channel transmit more classical information bits than the classical channel~\cite{gyongyosi2018survey}.

In the $d$-dimensional Hilbert space, there are at most $d$ pairwise distinguishable quantum states~\cite{gyongyosi2018survey}.
The options of increasing the number of photonic degrees of freedom have been reviewed in~\cite{cozzolino2019high} with the objective of generating high-dimensional quantum states. In this context, the authors of~\cite{cozzolino2019high} discussed the representation of quantum information in the increased Hilbert spaces of high-dimensional quantum systems relying on qudits. Briefly, a qudit is a quantum system that is not constrained to a 2D space. It can have any integer number $d$ of levels. A physical embodiment of $d$-dimensional systems is constituted by a photon.
The authors of~\cite{cozzolino2019high} also pointed out that different numbers of photonic degrees of freedom or their combinations can help expand the Hilbert space. For example, the orbital angular momentum (OAM) of light is one of the most frequently exploited photonic properties harnessed for generating high-dimensional quantum states. An arbitrarily large Hilbert space can be created since the number of dimensions can be any positive integer.

\subsection{\color{black}Quantum Channel Models} 

\begin{table*}
\centering
\caption{\color{black}A summary of quantum channels}
\begin{tabular}{|m{2.25cm}<{\raggedright}| m{4cm}<{\raggedright}| m{5cm}<{\raggedright}| m{5cm}<{\raggedright}|} \hline
\textbf{\makecell[c]{Quantum Channel}} & \textbf{\makecell[c]{Application Scenario}} & \textbf{\makecell[c]{Strength}} & \textbf{\makecell[c]{Limitation}} \\ \hline

\textbf{\makecell[c]{Depolarizing \\channel }} & 
$\bullet$ Optical fiber communication; 

$\bullet$ QKD~\cite{qcdepolar};

$\bullet$ Quantum computing and communication \cite{akyildiz20206g,sharma2020entanglement,2021The}. &  
$\bullet$ Compensated polarization loss; 

$\bullet$ Increased transmission distance; 

$\bullet$ Improved security. & 
$\bullet$ Classical capacity subject to the Holevo information of the channel;

$\bullet$ Quantum effects limit the transmission rate. \\ \hline

\textbf{\makecell[c]{Bosonic Gaussian \\channel}} & 
$\bullet$ Optical fiber communication; 

$\bullet$ QKD; 

$\bullet$ Quantum relay; 

$\bullet$ Quantum measurement and detection \cite{qcampbos,2019Tight}. & 
$\bullet$ The optical fiber quantum channel can be approximated as a Bosonic Gaussian channel for CV quantum information transmission. & 
$\bullet$ Cannot transmit non-Gaussian states; 

$\bullet$ Security threat by information leaks and eavesdropping. \\ \hline

\textbf{\makecell[c]{Free-space \\quantum channel}} 
& $\bullet$ Quantum communication between satellites~\cite{zhou2019shot};

$\bullet$ Satellite-ground quantum communication~\cite{9099546}; 

$\bullet$ Quantum ranging;

$\bullet$ QKD~\cite{djordjevic2020joint,9013407,vu2020design}; 

$\bullet$ Quantum teleportation.
& $\bullet$ High-speed data transmission;

$\bullet$ Long-distance quantum communication;

$\bullet$ Cost-effective deployment and maintenance. 
& $\bullet$ Susceptible to atmospheric effects such as absorption, aerosol scattering, and turbulence. \\ \hline

\end{tabular}
\label{tab.channel}
\end{table*}


{\color{black}Tab.~\ref{tab.channel} summarizes the features of specific types of quantum channel models, including the depolarizing channel, Bosonic Gaussian channel, and free-space quantum channel.}

\subsubsection{Depolarizing Channel}
{\color{black}A transmitted qubit can be depolarized during transmission; in other words, the polarization state of the qubit is mixed or lost.}
A depolarizing channel can have three basic forms: bit-flip, phase-flip, and bit-phase-flip channels. A depolarization channel in the Hilbert space ${\cal H}$ is characterized by
\begin{align}
{\rho _{out}} = {{\cal N}_d}\left( {{\rho _{in}}} \right) = \frac{{p{I_{\cal H}}}}{2} + \left( {1 - p} \right){\rho _{in}},
\end{align}
where ${\rho _{out}}$ and ${\rho _{in}}$ are the density operators of the output and input quantum states, respectively. ${{\cal N}_d}$ refers to the channel's mapping action that transforms the input density operator to the output density operator in the depolarizing channel. Here $p$ is the depolarizing probability, while ${I_{\cal H}}$ is defined as 
\begin{align}
{I_{\cal H}} = \frac{1}{2}\left( {{ \rho _{in}} + {\sigma _x}{ \rho _{in}}{\sigma _x} + {\sigma _y}{ \rho _{in}}{\sigma _y} + {\sigma _z}{\rho _{in}}{\sigma _z}} \right),
\end{align}
where ${\sigma _x} = \left( {\begin{array}{*{20}{c}}
0&1\\
1&0
\end{array}} \right),{\sigma _y} = \left( {\begin{array}{*{20}{c}}
0&{ - i}\\
i&0
\end{array}} \right)$, and ${\sigma _z} = \left( {\begin{array}{*{20}{c}}
1&0\\
0&{ - 1}
\end{array}} \right)$ are the Pauli matrices, representing the above-mentioned bit-flip, phase-flip, and bit-phase-flip, respectively.

\subsubsection{Bosonic Gaussian Channel}
{\color{black}The Bosonic Gaussian channel is a widely used quantum channel, where a transmitted quantum state is the Gaussian state after processing by linear optical elements,}
as given by~\cite{2006One}: 
\begin{align}\label{9}
{\rho _{out}'} = {{\cal N}_b}\left( {{{ \rho }_{in}'}} \right) = \int_{{{\cal N}_b}} {D\left( \xi  \right){{\rho }_{in}'}{{ D}^\dag }\left( \xi  \right)p\left( \xi  \right){d^2}\xi },
\end{align}
where $ \rho _{out}'$ stands for the density operator of the output quantum state, ${{\cal N}_b}$ denotes the channel's mapping action of the input density operator to the output density operator in Bosonic Gaussian channels, $\xi$ is the complex amplitude, $p\left( \xi  \right) = \frac{1}{{\pi N_c}}{e^{ - \frac{{{{\left| \xi  \right|}^2}}}{N_c}}}$, and ${\rho _{in}'}$ is the density operator of the input quantum state,
\begin{align}
{ \rho _{in}'} = \frac{1}{{\pi N_c}}\int_{{{\cal N}_b}} {{e^{ - \frac{{\left| \psi  \right|}}{N_c}}}} \left| \psi  \right\rangle \left\langle \psi  \right|{d^2}\psi .
\end{align}
In \eqref{9}, $ D\left( \xi  \right) = {e^{\xi {{ a}^\dag } - {\xi ^*} a}}$ is the displacement operator, while ${\xi ^ * }$ is the conjugate of $\xi$. Here, ${{{ a}^\dag }}$ and ${ a}$ are the creation operator and the annihilation operator of the input state $\left| \psi  \right\rangle $, respectively. ${N_c}$ is the variance of channel noise (i.e. the average photon number).

\subsubsection{Free-Space Quantum Channel}
{\color{black}This channel can transmit quantum states through free space.} In free-space quantum communication systems, atmospheric turbulence can affect the quantum states and be modeled using various probability distributions, e.g., a log-normal, Gamma-Gamma, or negative exponential model~\cite{8439931}. Atmospheric turbulence is due to temperature and pressure fluctuations. The Gamma-Gamma model is widely used to describe turbulence accurately. It has the following probability distribution~\cite{2006Error}:
\begin{align}
f\left( \eta  \right) = \frac{{2{{\left( {uv} \right)}^{^{\frac{{u + v}}{2}}}}}}{{\Gamma \left( u \right) \cdot \Gamma \left( v \right)}} \cdot {\eta ^{\frac{{u + v}}{2} - 1}} \cdot {K_{u - v}}\left( {2\sqrt {uv\eta } } \right),
\end{align}
where $\Gamma \left(  \cdot  \right)$ denotes the Gamma function, $u$ and $v$ are scintillation parameters, $\eta$ stands for the intensity of the turbulence strength, and ${K_{u - v}}\left(  \cdot  \right)$ stands for the modified Bessel function of the second kind of order $u - v$.

\subsection{Summary}

{\color{black}Channel modeling is indispensable for wireless communication system design, performance evaluation, and network planning and deployment. More efforts have to be invested into developing versatile channel models for all quantum communication channels, such as THz and visible light quantum channels, as well as ISAC quantum channels. 
}

%% file: Chapter_2.tex
\section{Native Quantum PHY Layer for Future Communications}

This section reviews technologies that might find their way into NG quantum-enabled PHY layers, such as multi-band quantum access, multi-user quantum communications, quantum MIMO, and quantum relays. We extend the point-to-point quantum communication philosophy to multipoint-to-multipoint scenarios suitable for ultra-dense NG scenarios.

\begin{table*}
\caption{A summary of different quantum access technologies}
\centering
\begin{tabular}{| m{2.35cm}<{\raggedright}| m{4.5cm}<{\raggedright}| m{4.5cm}<{\raggedright}| m{4.5cm}<{\raggedright}|} 	\hline

\makecell[c]{\textbf{Quantum Access}} & \makecell[c]{\textbf{Optical Quantum Access}} & \makecell[c]{\textbf{THz Quantum Access}} & \makecell[c]{\textbf{mmWave Quantum Access}}  \\ \hline

\textbf{\makecell[c]{Frequence range}} & $\bullet$  Visible light: 430 - 790 THz

$\bullet$ Optical fiber: 180 - 240 THz

& $\bullet$ 0.3 - 10 THz

& $\bullet$ 30 - 300 GHz

\\ \hline

\textbf{\makecell[c]{Bandwidth}} & $\bullet$ MHz to several GHz

& $\bullet$ 10 - 100 GHz

& $\bullet$ MHz to several GHz

\\ \hline

\textbf{\makecell[c]{Application scenario}} & $\bullet$ Remote communication~\cite{9220903}; 

$\bullet$ Fiber optic broadband access.

&  $\bullet$ Wireless and covert communication~\cite{9492803,8976167,8761168}; 

$\bullet$ Wireless local area network and sensor network.

&  $\bullet$ Wireless and covert communication~\cite{8108494}; 

$\bullet$ Wireless personal local area network; 

$\bullet$ Wireless metropolitan area network

\\ \hline

\textbf{\makecell[c]{Strength}} &  $\bullet$ Wide available spectrum transmission over long distances;  

$\bullet$ High security.

&  $\bullet$ Medium bandwidth transmission over a medium distance; 

$\bullet$ High security. 

&  $\bullet$ High bandwidth transmission at close range; 

$\bullet$ Better transmission quality and reliability. 

\\ \hline

\textbf{\makecell[c]{Limitation}} & $\bullet$ Higher cost; 

$\bullet$ Sensitive to atmospheric turbulence and optical noise;

$\bullet$ Easily affected by the attenuation and dispersion of optical fiber.

&  $\bullet$ Short transmission distance; 

$\bullet$ Easily affected by atmospheric turbulence and noise;

$\bullet$ Relatively low technology maturity.

&  $\bullet$ Short transmission distance; 

$\bullet$ Relatively low technology maturity; 

$\bullet$ High equipment cost. 

\\ \hline

\textbf{\makecell[c]{Key techniques}} & $\bullet$ QKD:  \cite{PhysRevResearch.3.043014,9220903,app9245285}

$\bullet$ Tracking system in wireless quantum communication networks:  \cite{refId0}

$\bullet$ Multiple-quantum-well photodetector: \cite{9094211, 9547724}

&  $\bullet$ Quantum secret sharing:\cite{e23091223}

$\bullet$ QKD:\cite{2020Hybrid,9492803,8976167,8761168}

$\bullet$ Quantum-dash laser diode:  \cite{9514429}
 
$\bullet$ Waveguide:
\cite{ nano10122436}

$\bullet$ Quantum cascade laser:
\cite{ https://doi.org/10.1002/pssa.202000407,9370720,2019Dispersion,doi:10.1021/acsphotonics.9b00411,2020Ultrafast,2020Unveiling }

$\bullet$ Photodetectors:
\cite{ doi:10.1021/acs.nanolett.9b05207,9566893,Liu_2019, doi:10.1021/acs.nanolett.1c02022,Yachmenev_2022,2022103, shao2021research, Thermoelectricgraphenephotodetectors }

$\bullet$ Broadband THz half-wave plate:
\cite{9409611 }

&  $\bullet$  Distribution of quantum entanglement over communication channels: \cite{8108494}

$\bullet$ Quantum-dash laser source:  \cite{9654335,Tareq:20,9542190}

$\bullet$ Quasi-tunable mmWave frequency synthesizer:  \cite{app112210742}

$\bullet$ Beamformer: \cite{8918104}

$\bullet$ Quantum-dash laser diode:  \cite{9514429}

$\bullet$ Quantum-dash laser comb source:\cite{TAREQ2021102553,9409932}

$\bullet$ Quantum dot coherent comb laser:  \cite{9435040}

\\ \hline

\end{tabular}
\label{tab.QuantumAccessTechnology}
\end{table*}

\subsection{Optical, THz, and mmWave Quantum Access}

{\color{black}In high-frequency spectra, quantum phenomena have  been observed in optical band~\cite{9220903}, THz band~\cite{6953069}, as well as mmWave band~\cite{8108494}.}
The carrier frequency of THz waves has attracted attention as a benefit of their abundance of bandwidth~\cite{2011Present}.  
Studies on mmWave and THz quantum communication typically focus on the distribution of quantum keys in indoor environments and satellite-to-satellite wireless communications. Moreover, the research of new materials in support of THz communications is also an active subject area~\cite{9895675, 9999512}. 
{\color{black}Tab.~\ref{tab.QuantumAccessTechnology} summarizes multi-band access technologies conceived for quantum communications, including optical, THz and mmWave quantum access.}

\subsubsection{\textbf{Optical Quantum Access}}
{\color{black}This access technology is suitable for optical fiber communication systems and can be applied to long-distance communication scenarios, such as optical fiber broadband access and optical fiber to the home. Generally, optical frequencies have been preferred for the implementation of quantum communication protocols, such as quantum teleportation~\cite{9023997} and QKD~\cite{8597918}. This is partly due to their low background black-body radiation~\cite{8108494}. 
However, optical signals are susceptible to attenuation and dispersion of the fiber, reducing transmission distance and quality.}
Existing QKD systems typically use photons to transmit information through optical fiber or free-space optical (FSO) channels~\cite{9492803}. Compared to optical fiber, FSO links can be more cost-effective. Given the lack of global standards, it is important to conceive QKD protocol converters for long-distance relaying across heterogeneous QKD networks~\cite{9964020}. 
{\color{black}For example, Pirandola \textit{et al.}~\cite{PhysRevResearch.3.043014} studied the fundamental ingredients of CVQKD systems, and Zhao \textit{et al.}~\cite{9220903} characterized QKD over FSO links in different scenarios. }

In addition to QKD, researchers are also exploring other applications of the optical frequency band, such as tracking systems~\cite{refId0} and multiple-quantum-well-based photodetectors~\cite{9094211,9547724}. Alkhazragi \textit{et al.}~\cite{9094211} achieved a record data speed of 7.4-Gbits/s using a selective-wavelength semi-polar InGaN/GaN-based multiple-quantum-well micro-photodetector in VLC. In~\cite{9547724}, the fabrication of a multi-functional ultraviolet-C micro-LED having a monolithically integrated photodetector was characterized.

\subsubsection{\textbf{THz Quantum Access}}
{\color{black}This technology is appropriate for close-range communication systems, e.g., wireless local area networks.}
Compared with optical bands, THz frequency bands have the advantage of less delicate pointing, acquisition and tracking. They are also less affected by ambient light, atmospheric turbulence, scintillation, clouds, and dust~\cite{2020Terahertz,2019Wireless}. In particular, a non-zero secret key rate can be achieved at room temperature as a benefit of the relatively low thermal noise at THz frequencies~\cite{9507495,8976167}.

There have been many studies on QKD in THz bands~\cite{8976167,8761168,WOS:001141997100001,9492803,2020Hybrid,e23091223}. A promising physical hardware architecture was designed for implementing a QKD system in THz bands~\cite{8976167}. A feasibility study of LEO-based QKD between micro-satellites relying on THz links was carried out in~\cite{8761168}.
For the continuous-variable quantum communication systems, Kundu \textit{et al.} \cite{ WOS:001141997100001} summarized recent advances in the design and analysis of CVQKD systems operating at THz frequencies, and discussed how MIMO transmission can be used in THz CVQKD to improve secret key rates and increase coverage distances. 
In THz bands, there are studies about the use of multi-carrier multiplexing techniques for CVQKD~\cite{9492803}, hybrid linear amplifier-based detection of thermal-state CVQKD~\cite{2020Hybrid}, and satellite-to-satellite communication system involving CV quantum secret sharing~\cite{e23091223}.
In addition, Mallick \textit{et al.}~\cite{9514429} described the application of an OFDM-based 5G fronthaul in both the THz and mmWave bands. In that system, THz and mmWave signals were generated with the assistance of L and C band quantum dash laser diodes.

The research community of materials and manufacturing processes invented for THz devices used in quantum-related applications, such as waveguides, quantum cascade lasers (QCLs), and photodetectors, has also made significant progress in recent years~\cite{7393447,7735237,9397381,nano10122436}.
QCLs constitute a crucial component of nanotechnology and have compelling applications in the THz and mid-infrared bands~\cite{https://doi.org/10.1002/pssa.202000407}. The use of quantum-compatible materials has also been studied at THz, mmWave, and other radio frequencies~\cite{9370720,2019Dispersion,doi:10.1021/acsphotonics.9b00411,2020Ultrafast,2020Unveiling}, such as chiral metamaterial arrays designed for modulating QCL polarization~\cite{doi:10.1021/acsphotonics.9b00411}, harmonic injection and mode-locking for modulating round-trip frequencies~\cite{2020Ultrafast}, and scanning near-field optical microscopy designed for the simultaneous measurement of the associated transport and scattering properties~\cite{2020Unveiling}.

Photodetectors constitute another crucial component of quantum communication systems and have become a rapidly evolving research area during the past few decades~\cite{doi:10.1021/acs.nanolett.9b05207,9566893,doi:10.1021/acs.nanolett.1c02022,Liu_2019,Yachmenev_2022,2022103,shao2021research,Thermoelectricgraphenephotodetectors}. Sophisticated techniques have been developed for uncooled THz photodetectors, 
such as quantum-dot single-electron transistor-based photodetectors~\cite{9566893,doi:10.1021/acs.nanolett.1c02022}, 
and sub-THz photodiodes~\cite{Liu_2019}. Recent studies have analyzed popular implementations of THz quantum-well photodetectors designed for high-speed detection, imaging, and communications~\cite{shao2021research}.
In addition, the popular quantum-domain (QD) composite pulse control technique has been harnessed for the design of a THz half-wave plate, using multi-layered metamaterials~\cite{9409611}.

\subsubsection{\textbf{mmWave Quantum Access}}
{\color{black}This access technology is suitable for communication systems, including wireless personal local area networks and wireless metropolitan area networks. Moreover, mmWave is less affected by atmospheric turbulence and noise during transmission. However, it can be easily affected by atmospheric absorption and scattering. }

The feasibility of direct QKD in the mmWave regime was also considered~\cite{2011Continuous} with the ambitious objective of supporting quantum entanglement~\cite{8108494} with no need for integrating additional optical hardware into the transceiver.

Significant progress has been made in quantum entanglement distribution~\cite{8108494} and quantum-dash laser sources~\cite{9654335,Tareq:20,9542190}.
In this context, Hosseinidehaj \textit{et al.}~\cite{8108494} analyzed quantum entanglement distribution for transmission over mmWave channels and quantified the transmission requirements. In~\cite{9654335}, a wireless quadrature phase shift keying (QPSK) transmission scheme relying on quantum-dash lasers operating in the 28 GHz mmWave was examined. An L-band quantum-dash laser suitable for mmWave signals was presented in~\cite{Tareq:20}.
Tareq \textit{et al.}~\cite{9542190} reported a 2 Gbps QPSK wireless transmission system using a 28 GHz mmWave carrier relying on a so-called injection-locked quantum-dash technique. {\color{black}Liu \textit{et al.}~\cite{app112210742} proposed a low-cost tunable ultra-wideband electrooptic frequency comb-based mmWave synthesizer. They used a quasi-tunable frequency in their experiments to generate mmWave signals.}

The authors of~\cite{8918104} demonstrated that a concatenated fiber-wireless QPSK transmission system relying on a dual-wavelength InAs/InP quantum-dash laser source is capable of transmitting 28 and 30 GHz mmWave signals. 
{\color{black}Furthermore, quantum-dash-based laser comb source~\cite{TAREQ2021102553,9409932} and quantum dot-based coherent comb laser source~\cite{9435040} that support fiber-wireless transmission systems in the access network were developed.} 

\begin{figure} 
\centering
\includegraphics[width=3.7 in]{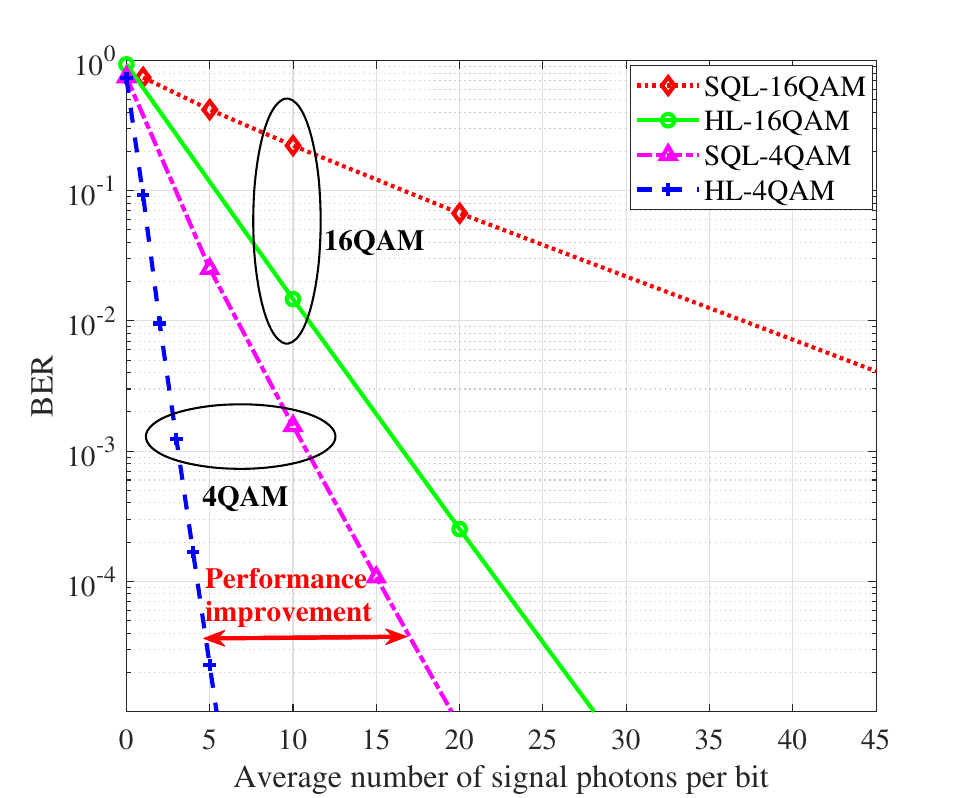}
\caption{Comparison of SQL~\cite{SQL20896} and HL~\cite{Helstrom1969}. HL of quantum systems represents better error probability than the SQL of classical systems. Hence, the quantum systems approaching the HL can potentially support new NG applications~\cite{8281456}.}
\label{Comparison_of_SQL_and_HL}
\end{figure}

\begin{figure}[htbp]
\centering
\includegraphics[width=3.7 in]{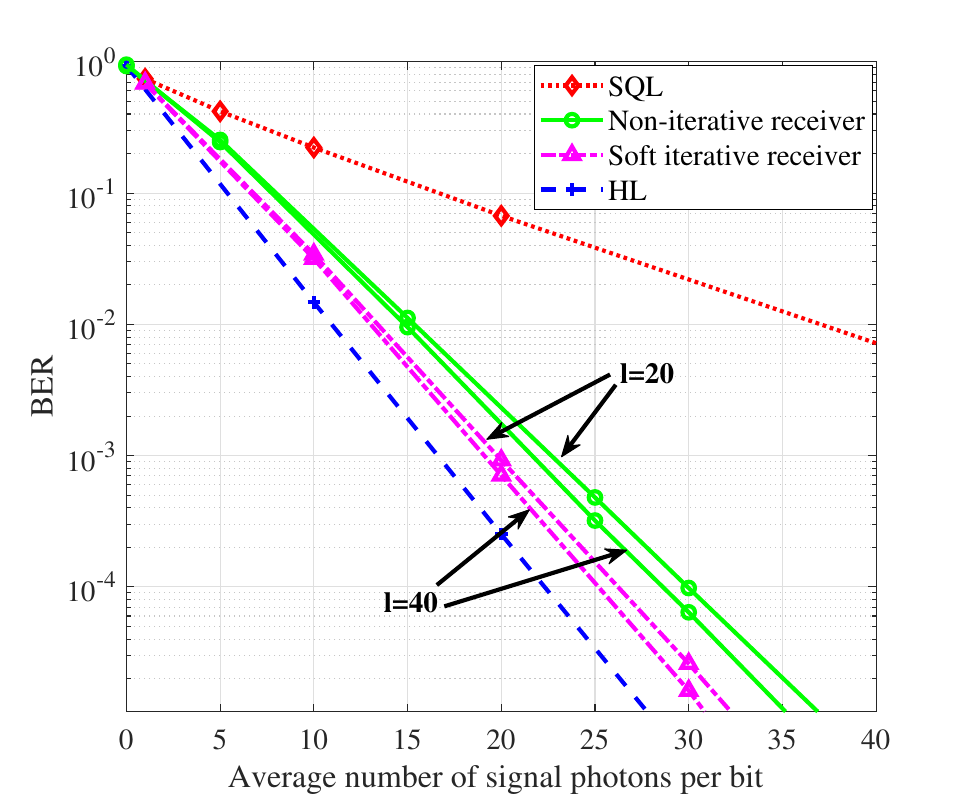}
\caption{Comparation of the error probabilities of the soft iterative receiver and non-iterative receiver for 16QAM. The soft iterative receiver achieves a significant performance improvement over its non-iterative counterpart and is close to the HL~\cite{8281456}.}
\label{Error_probability_of_soft_iterative_receiver_and_non-iterative_receiver_for_16QAM}
\end{figure}

\subsubsection{\textbf{Other Technical Enhancements}}

In order to harmonize the heterogeneous frequency bands and technologies of the electromagnetic waves and the particle-oriented photonic carriers, Xu \emph{et al.}~\cite{2021Petahertz } introduced a Petahertz communication framework. They unified the existing infrared, visible light, and ultraviolet subbands, and explored the potential of the PHz band for extremely bandwidth-thirsty applications. Furthermore, they discussed scenario-dependent channel models, modulation schemes, the expected system performance, and a suite of multiple access and networking techniques. 

The Helstrom limit (HL)~\cite{Helstrom1969} represents the ultimate error probability limit of quantum communication systems. The quantum systems approaching the HL have to exhibit ultra-high receiver sensitivity. Hence, approaching the HL in practice is challenging. Fig.~\ref{Comparison_of_SQL_and_HL} compares the HL of quantum receivers to the standard quantum limit (SQL) of classical receivers~\cite{SQL20896}. It is observed that the HL shows about three orders of magnitude bit error rate (BER) improvement over the SQL at five signal photons/bit for a 4QAM scheme.
Wei \textit{et al.}~\cite{8281456} developed a method for distinguishing QD signals with the radical objective of approaching the performance of the HL~\cite{Helstrom1969}. Their method utilizes beam splitters and iterative receiver techniques. As a benefit, it outperforms its non-iterative counterpart at the cost of an increased receiver complexity~\cite{6600857, 2014arXiv1412.4486Z}. 
Specifically, the received 16QAM QD signals are divided into $l$ identical quantum states by $\left( {l - 1} \right)$ beam splitters. In the iterative quantum state observation process of the receiver, the optimal measurement operator and the decoding results are updated to gradually improve the soft log-likelihood ratio (LLR) of each information bit.

Fig.~\ref{Error_probability_of_soft_iterative_receiver_and_non-iterative_receiver_for_16QAM} shows that, when $l = 40$ iterations are used, the proposed 16QAM soft iterative receiver achieves a significant performance improvement over its non-iterative counterpart and approaches the HL. In fact, the iterative quantum receiver exhibits an excellent performance even for fewer iterative measurement steps (e.g., when $l = 20$).
Polarization states~\cite{ 9130765 } and polarization entanglement of photons~\cite{ yang2020quantum } are utilized to create QSDC schemes. 

By contrast, the authors of~\cite{ 8281456 } focused on CV Gaussian coherent states and, therefore, enjoyed high information capacity, signal strength, flexible high-order modulation, and convenient implementation. The employment of CV Gaussian coherent states provides inherent compatibility with existing classical optical communication networks because CV systems can be readily implemented with off-the-shelf lasers and homodyne or heterodyne detectors~\cite{8439931}.
Moreover, the CV Gaussian coherent states discussed in~\cite{ 8281456 } can also be incorporated into the QSDC schemes developed in~\cite{9130765, yang2020quantum } to bring the above benefits into QSDC systems.

Sharma \textit{et al.}~\cite{sharma2020entanglement} investigated the behavior of entangled states under various noise influences and observed the sudden disappearance and reappearance of entanglement in the presence of specific noise types and parameter values. 
Harraz \textit{et al.}~\cite{harraz2018two, harraz2020n} proposed techniques for protecting two-qubit states from amplitude damping to ensure high-accuracy restoration of the original states. This can apply to any mixed/entangled state, and is robust to significant amounts of noise.
Rahman \textit{et al.}~\cite{Rahman_2022} explored the impact of channel impairments on the similarity between the initial and final quantum states, while the authors of~\cite{2021Complexity} compared different channels in the context of single qubit message transmission and identified amplitude damping and depolarization noise as the main impairments~\cite{2021Complexity}.
The authors of~\cite{8967070} evaluated different communication systems in the face of white quantum noise and related their QD metrics -- such as the number of conveyed bits per photon pair -- to conventional measures like spectral efficiency and error correction performance. The results indicated a capacity of approximately 20 bits per photon pair (10 bits per photon) for Shannon's capacity, while single photon transmission per channel only reached 3 bits per photon pair.
Chitambar \textit{et al.}~\cite{2021The} provided a mathematical expression for both the family of Werner-Holevo channels defined in~\cite{2002Counterexample, 2014On} and for arbitrary qubit channels. They revealed substantial differences in the channel's ability to establish and sustain entanglement. These differences manifested themselves in the probability of successfully transmitting a randomly selected classical message over the channel. 

\subsection{Multi-User Quantum Communications}

\begin{figure} 
\centering
\includegraphics[width=3.5 in]{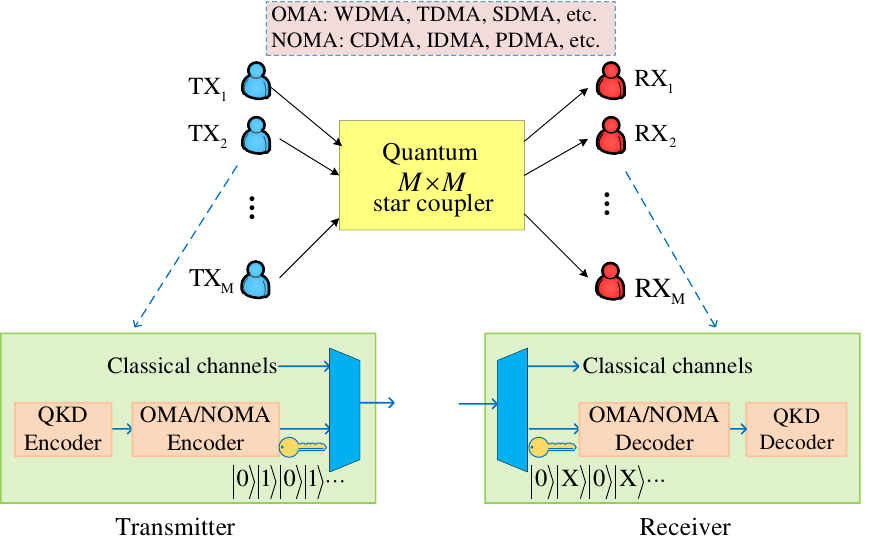}
\caption{ QD multiple access systems, including WDMA, TDMA, CDMA, IDMA, etc. All of the quantum users can be connected to each other via a quantum star coupler, which is modeled as a sequence of beam splitters that broadcasts all transmitters’ QD signals to all receivers~\cite{6253218}.}
\label{Quantum_multiple_access}
\end{figure}

{\color{black}In recent years, a great deal of research has been conducted on point-to-point quantum communications. There has been an increase in the deployment of point-to-point quantum communication systems due to its growing commercial flexibility~\cite{9684555}.
Heim \textit{et al.} \cite{ heim2014atmospheric } reported on a quantum communication experiment carried out in an urban setting across a 1.6 km point-to-point free-space link. 
}

{\color{black}Some preliminary research has been conducted on the combined point-to-point transmission of quantum and classical signals~\cite{ WOS:000492996000051}.
Noting previous experiments on measurement-device-independent QKD (MDI-QKD) networks were limited to point-to-point communication, Tang \textit{et al.} \cite{ WOS:000475395200002 } demonstrated a plug-and-play MDI-QKD network that enables point-to-multipoint communications among three parties.}
The next milestone will be the conception of multi-user quantum communication to provide quantum-secured communications over NG systems~\cite{QuantumComputingBasedChannelandSignalModeling}, as shown in Fig.~\ref{Quantum_multiple_access}. A diverse range of QD multiple-access methods has been proposed, including quantum time division multiple access (TDMA)~\cite{8756986, 6253218}, quantum orthogonal frequency division multiple access (OFDMA)~\cite{bahrani2015orthogonal, 6545781}, quantum code division multiple access (CDMA)~\cite{6253218, rezai2021quantum, sharma2020quantum}, and quantum interleave division multiple access (IDMA)~\cite{9013407, 8648086}.

\subsubsection{\textbf{Quantum TDMA}}

In TDMA systems, each user communicates with the base station in different time slots.
Yang \textit{et al.}~\cite{8756986} aimed to achieve max-min fairness in the downlink of a single-cell QKD system with a base station supporting several users in the presence of an eavesdropper. They combined TDMA QKD with the NOMA philosophy in order to increase the system's throughput without jeopardizing its security. A secret key was first generated and negotiated between the base station and users as part of the QKD protocol, followed by NOMA-based CD data transmission. They also proposed an iterative method for solving the resultant non-convex max-min problem that maximizes the minimal secure rate in the set of all users.
Razavi \textit{et al.}~\cite{6253218} proposed a multi-user QKD network that allows any pair of users to exchange a secret key without trusting any other nodes in the network. They considered a passive star-topology QKD network in which TDMA was used to enable multiple access among quantum users, as shown in Fig.~\ref{Quantum_multiple_access}.
All users can securely communicate with each other using TDMA QKD, enabling them to distinguish between the QD signals of different users. 

\subsubsection{\textbf{Quantum OFDMA}}
OFDMA maps multi-user data streams to different time- and frequency-domain resource slots of OFDM wareforms.
QKD may be integrated with OFDM if an all-optical OFDM transmitter and receiver pair is used in quantum communication channels. In this context, 
Bahrani \textit{et al.}~\cite{bahrani2015orthogonal} characterized QKD over all-optical OFDM links. The QKD encoder generates QD signals in the form of pulses that carry information about the encoded key bits in each subchannel. 
A pair of OFDM-QKD schemes were considered in~\cite{bahrani2015orthogonal} for quantum-secured communication in the face of realistic system imperfections. One of the transmitter designs was based on frequency-offset-locked lasers. The other utilized a mode-locked laser relying on an optical splitter.

\subsubsection{\textbf{Quantum CDMA}}

\begin{figure} [t]
\centering
\includegraphics[width=3.5 in]{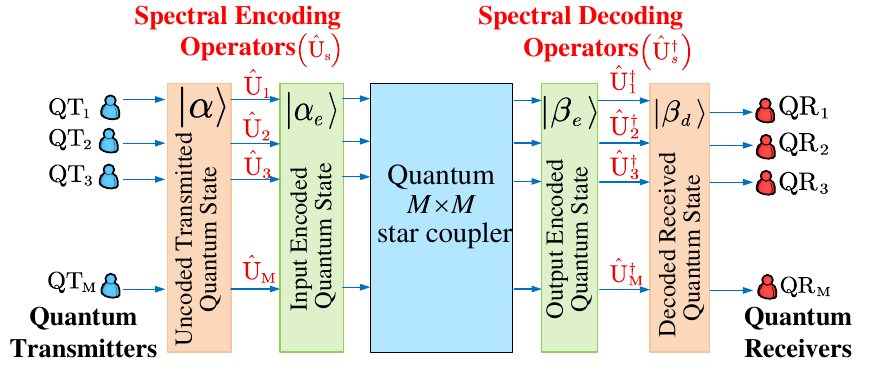}
\caption{Quantum CDMA. Each transmitter sends a pure quantum state, e.g., a coherent state or a Fock state, through a quantum CDMA spectral encoding unit~${\rm{\hat U}}$. The star coupler broadcasts each transmitter’s spectrally encoded QD signal to all the receivers. The quantum receiver, modeled by an ideal photodetector, performs the decoding operation through the spectral decoding unit~${{\rm{\hat U}}^\dag }$ to recover the original quantum state~\cite{rezai2021quantum}. }
\label{Quantum_CDMA}
\end{figure}

Razavi~\cite{6253218} presented a CDMA-based, multi-user, star-topology-based QKD network that is capable of mitigating the interference effects. They also found the average lower bounds of the secret key generation rates when decoy-state protocols were used. As shown in Fig.~\ref{Quantum_multiple_access}, all quantum users are connected to each other via a QD star coupler, which can be modeled as a sequence of beam splitters that broadcast all transmitters' QD signals to all receivers.
The transmitter includes a QKD encoder and a CDMA transmitter to support multiple quantum users.
The CD and QD signals are assigned two different bands and combined by a band multiplexer.
The receiver includes a band demultiplexer, which separates the quantum and the classical channels, and a CDMA receiver, followed by QKD measurements.
Rezai\textit{ et al.} \cite{rezai2021quantum} proposed a new quantum CDMA technique based on continuous-mode quantum light pulse spectral encoding and decoding, as depicted in Fig.~\ref{Quantum_CDMA}. Observe that in quantum CDMA, each transmitter sends a pure quantum state, e.g., a coherent state or a Fock state, through a quantum CDMA spectral encoding unit~${\rm{\hat U}}$. The star coupler broadcasts each transmitter’s spectrally encoded QD signal to all the receivers. 
The quantum receiver, e.g., a photodetector, can recover the original quantum state within the spectral decoding unit~${{\rm{\hat U}}^\dag }$. As a further advance, Sharma and Banerjee~\cite{sharma2020quantum} considered a single optical fiber as a transmission channel and studied spread spectrum techniques in which multiple users transmitted their qubits.
They improved the performance of CDMA-based QKD links relying on single photons without any amplifiers and modulators.
They also used receiver filters to mitigate the spectral overlaps, thereby improving the signal-to-noise ratio (SNR).

\begin{figure}[t]
\centering
\includegraphics[width= 3.5 in]{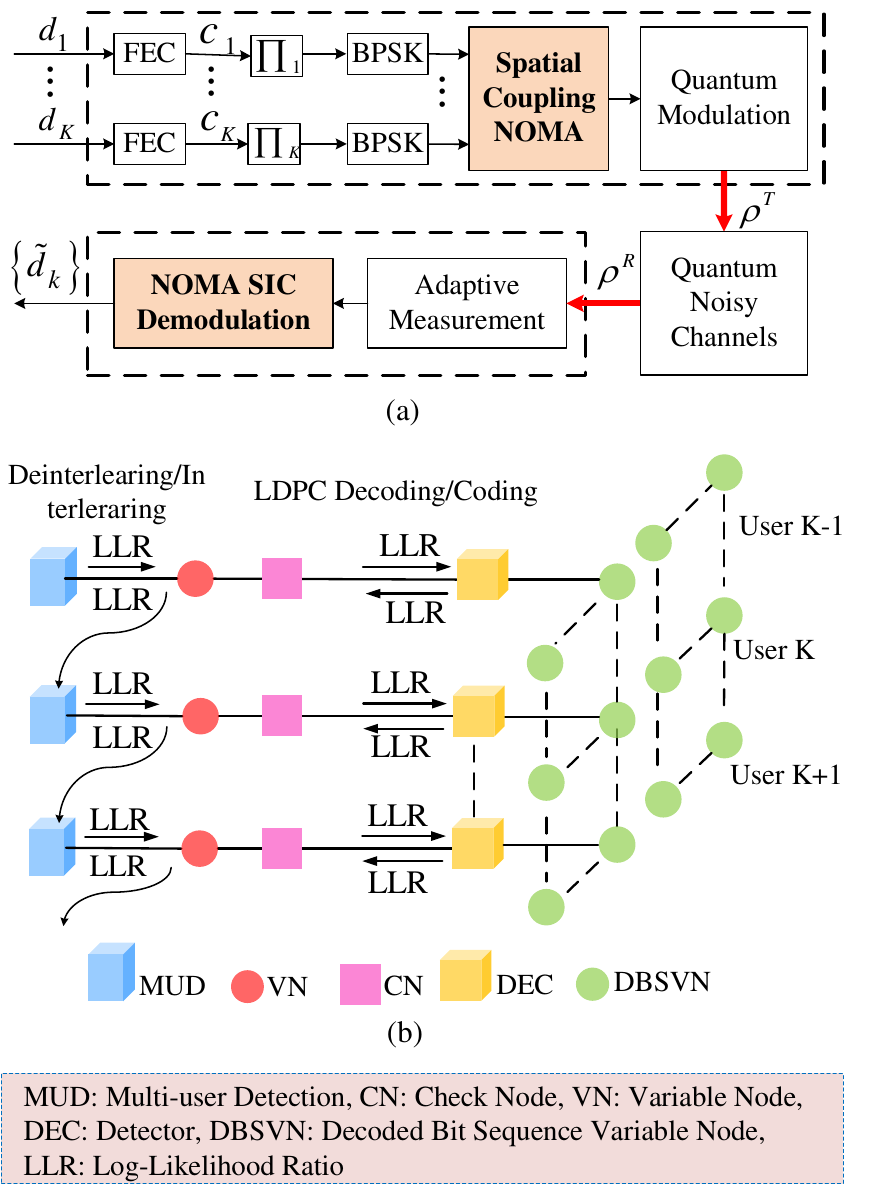}
\caption{The schematic of an IDMA-based iterative multi-user NOMA-QKD system. The system detects photons having different polarization states once the quantum measurement is completed. The \textit{a-priori} and \textit{a-posterior} LLRs of the QD signals are exchanged iteratively using an iterative multi-user detection method~\cite{9013407, 8648086}.}
\label{The_schematic_of_the_proposed_NOMA-MQKD_QKD_system}
\end{figure}

\subsubsection{\textbf{Quantum IDMA}}

IDMA may be viewed as a chip-interleaved CDMA system, where the users are distinguished by their unique, user-specific interleavers. An iterative IDMA-based NOMA-QKD multi-user system operating in the presence of Poisson shot noise was proposed for FSO communication in~\cite{9013407}. As depicted in Fig.~\ref{The_schematic_of_the_proposed_NOMA-MQKD_QKD_system}, to support multi-user information transmission, a quantum transmitter was designed to support NOMA. As shown in Fig.~\ref{The_schematic_of_the_proposed_NOMA-MQKD_QKD_system}(a), the IDMA-based iterative multi-user NOMA-QKD system harnesses an iterative detector that detects photons exhibiting different polarization states, once the quantum measurement was completed. The \textit{a-priori} and \textit{a-posterior} LLRs were processed iteratively for multi-user detection, see Fig.~\ref{The_schematic_of_the_proposed_NOMA-MQKD_QKD_system}(b). Despite the presence of atmospheric turbulence and polarization mismatch noise, the IDMA multi-user NOMA-QKD scheme exhibited rapid convergence and robust performance. It significantly improved the secure key rate compared to traditional wavelength division multiple access (WDMA) based QKD systems.
Shen \textit{et al.}~\cite{8648086} proposed a new iterative IDMA detection aided quantum spatial coupling NOMA scheme, which relied upon exploiting the intrinsic non-orthogonal characteristics of coherent state multi-user communications. This system exhibited high robustness against different practical QD imperfections, such as mode mismatch and dark counts.

\begin{figure}[t]
\centering
\includegraphics[width=3.5 in]{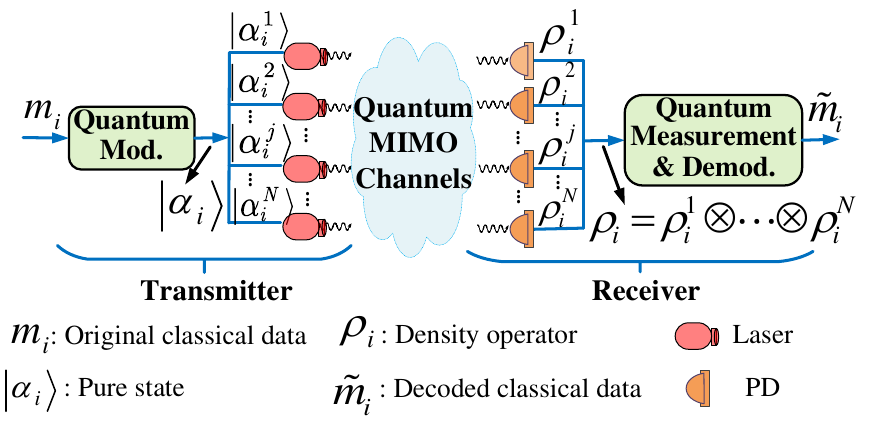}
\caption{Quantum MIMO. Quantum states can be transmitted over several quantum channels. The MIMO diversity effect can effectively reduce the impact of quantum states from quantum channel impairments, such as quantum noise, depolarization, and atmospheric turbulence.
Furthermore, quantum MIMO multichannel multiplexing can significantly increase the quantum system's capacity~\cite{9739032}.}
\label{Quantum_MIMO}
\end{figure}

\subsection{ Quantum MIMO}

Classical MIMO techniques aim to improve the BER and/or increase channel capacity by harnessing multiple transmission channels created by multiple antennas. 
Quantum communications can also benefit from the employment of MIMO techniques~\cite{9099546,9507495,9739032,2005Quantum,zhou2019shot} relying on the architecture shown in Fig.~\ref{Quantum_MIMO} to transmit quantum states over several quantum channels. The MIMO diversity can be beneficially exploited to effectively reduce the deleterious impact of quantum channel impairments, such as quantum noise, depolarization, and atmospheric turbulence imposed on the fragile quantum states. 
Furthermore, QD MIMO multichannel multiplexing has the potential to increase the quantum system capacity.
For quantum MIMO systems, the entanglement effect can provide a unique information channel, potentially increasing the degree of freedom in the quantum system. 

\begin{figure}[t]
\centering
\includegraphics[width=3.5 in]{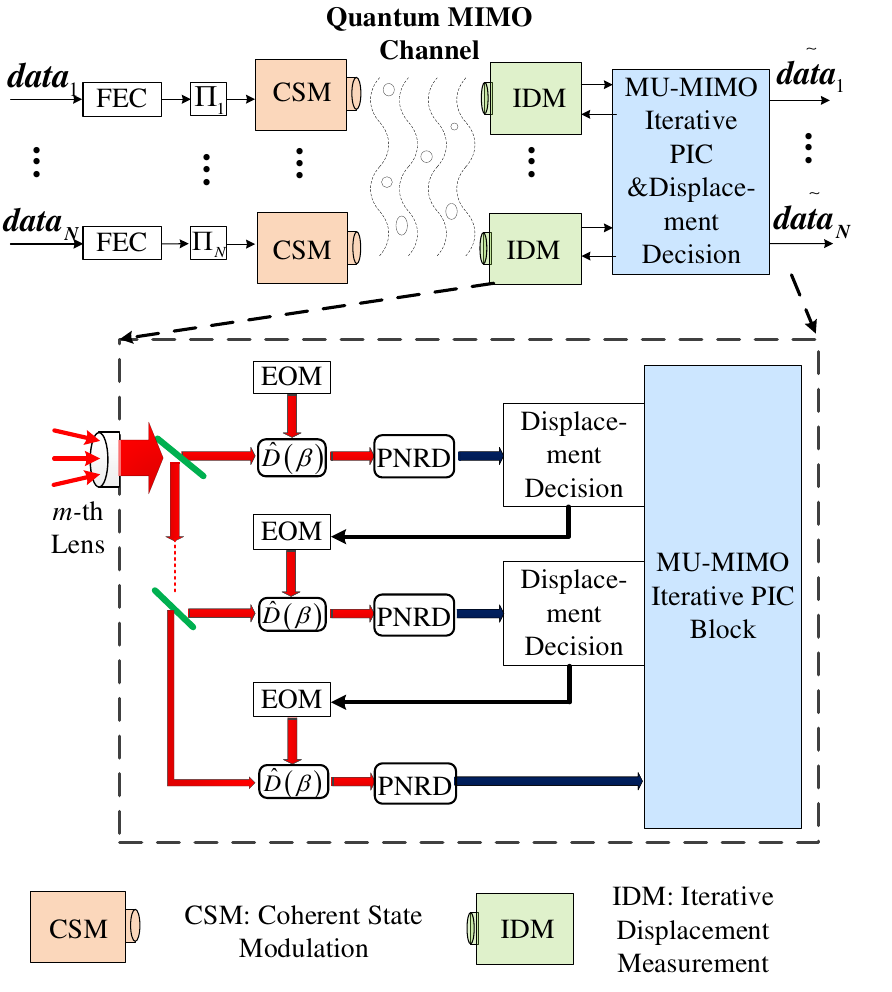}
\caption{A non-orthogonal coherent-state MIMO uplink system over noisy quantum atmospheric turbulence channel. At the transmitter, the information bit sequences are borne by the coherent state optical QD signals presented on the infinite-dimensional orthogonal basis of Hilbert space. The receiver's detection scheme relies on the iterative displacement quantum measurement and the iterative massive MIMO PIC detection~\cite{zhou2019shot}.}
\label{The_diagram_of_the_proposed_quantum_massive-MIMO_system}
\end{figure}

Yuan \textit{et al.}~\cite{9099546} proposed an FSO quantum communication system to leverage receiver diversity using an optical combining technique~\cite{2017Multi} and a generalized Kennedy receiver~\cite{2007Discrimination}. A threshold-detection-based Kennedy receiver
was harnessed to mitigate the influence of both turbulence
and thermal noise, and shown to outperform the SQL given by the homodyne receiver.
Kundu \textit{et al.}~\cite{9507495} employed multiple antennas in the THz band for a MIMO CVQKD scheme, proposed a transmit-receive beamforming scheme using singular value decomposition (SVD), and derived an analytical expression for the secret key rate of the system. Their simulations demonstrated that the MIMO CVQKD scheme outperforms a traditional single-input single-output system in secret key rate and transmission distance. Additionally, multiple antennas applied in the THz have the potential to mitigate path loss.
Kundu \textit{et al.}~\cite{9739032} conceived a channel estimation scheme for a THz MIMO CVQKD system. This system utilized optical-to-THz and THz-to-optical converters for QD signal transmission and detection. Again, an SVD-based beamforming method was employed at both the transmitter and receiver. During the key generation phase, the input-output relationship between Alice and Bob was characterized by incorporating the channel estimation overhead, the additional noise arising from the channel estimation error, and detector noise. The effect of both the channel estimation error and of the pilot overhead on the secret key rate of the system was analyzed.

A communication system relying on MIMO QKD based on the BB84 protocol~\cite{bennett2014quantum} was devised in~\cite{2005Quantum } for increasing the secret key rate.
This was achieved by harnessing multiple receivers to communicate concurrently with multiple transmitter elements. However, interference effects due to scattering and turbulence in atmospheric channels reduce the secret key rate of the system and increase the QBER. A new mathematical model was developed for analyzing the effects of crosstalk and interference on MIMO QKD systems in~\cite{2005Quantum }. 

\begin{figure}[t]
\centering
\includegraphics[width=3.7 in]{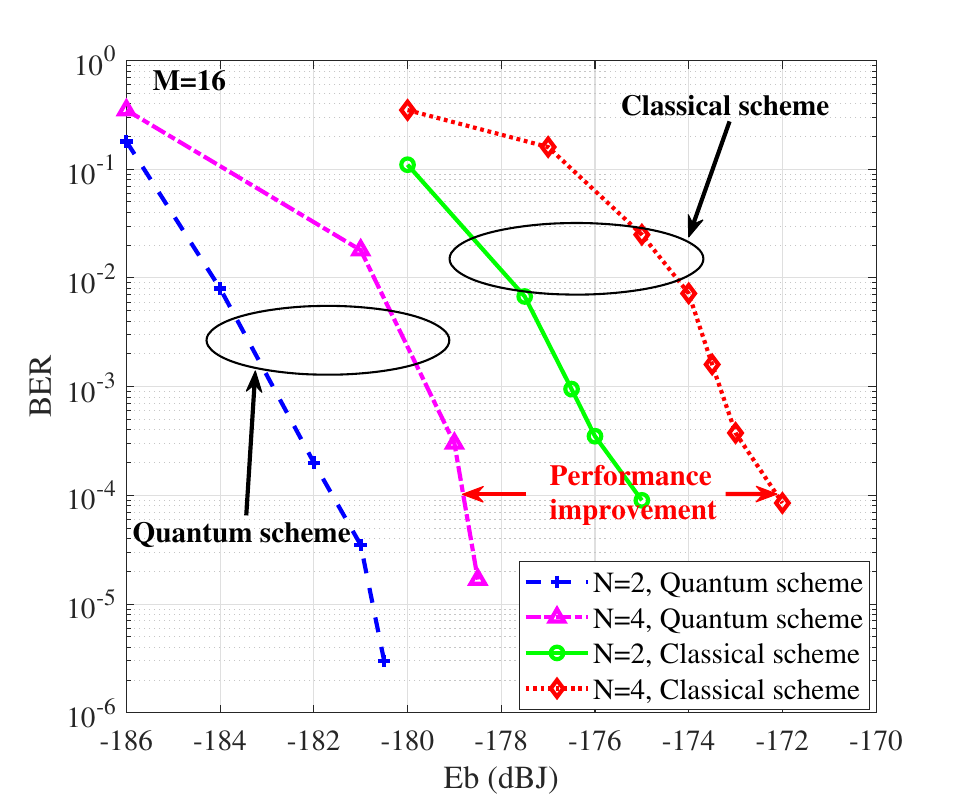}
\caption{Comparison of the quantum MIMO scheme and the classical scheme under strong turbulence fading. In the case of $N \times M = 4 \times 16$, the quantum MIMO is able to achieve an improvement of 6.25 dBJ per bit in performance at a BER of ${10^{-4}}$ for a Gamma-Gamma fading link~\cite{zhou2019shot}.}
\label{Comparison_of_proposed_quantum_MU-MIMO_scheme}
\end{figure}

Zhou \textit{et al.}~\cite{zhou2019shot} proposed a non-orthogonal uplink coherent-state MIMO system for transmission over noisy quantum atmospheric turbulence channels. As seen in Fig.~\ref{The_diagram_of_the_proposed_quantum_massive-MIMO_system}, this system comprises $N$ transmitters and a base station with $M$ receivers. The information bit sequences are conveyed by the coherent state optical QD signals spanning the infinite-dimensional orthogonal basis of Hilbert space from the transmitter to the receiver. The atmospheric turbulence fading channel encountered was modeled by the Gamma-Gamma distribution. The statistical model for the received QD signals was established. The receiver's detection scheme relies on the iterative displacement-based quantum measurement and massive MIMO-based parallel interference cancellation (PIC) aided detection. 
Fig.~\ref{Comparison_of_proposed_quantum_MU-MIMO_scheme} compares the BER performance between this quantum multi-user MIMO scheme and a classical scheme~\cite{article2012Zhou}. In the case of $N \times M = 4 \times 16$, this scheme can achieve an improvement of 6.25 dBJ per bit at a BER of ${10^{-4}}$ for a Gamma-Gamma fading link. This scheme exploits the multi-dimensional nature of the QD signals through multi-stage displacement measurements.

\subsection{Quantum Relays}

\begin{figure}[t]
\centering
\includegraphics[width=3.5 in]{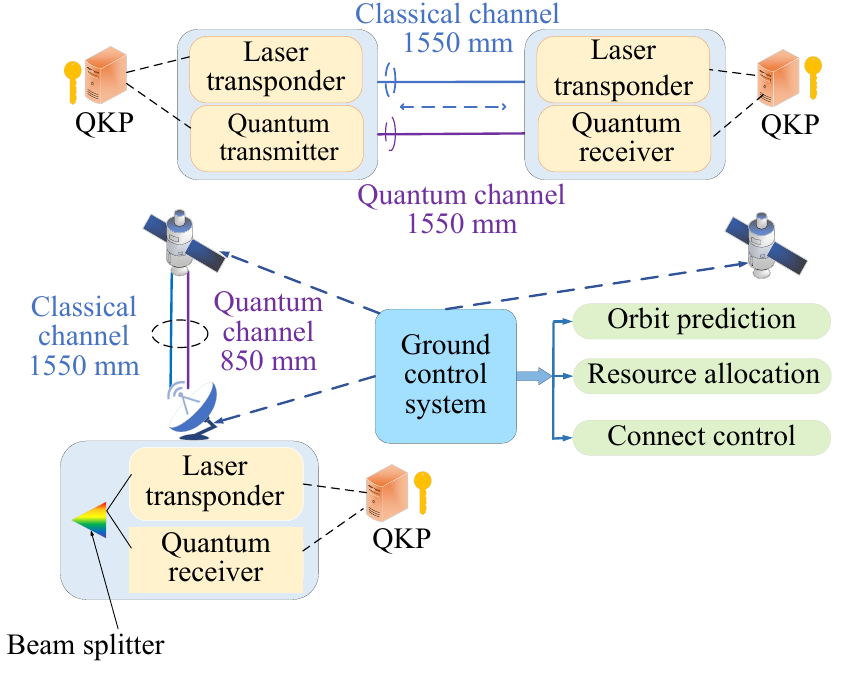}
\caption{The basic structure of the QKD satellite system enabled by the QKP. Each node has a quantum transceiver, a laser transceiver, and a QKP. Routing selection and key distribution are determined by the control system. Satellites and controllers can be connected via radio or optical links~\cite{huang2020quantum}.}
\label{Basic_structure_of_QKP_enabled_satellite_QKD_system}
\end{figure}

\begin{figure}[t]
\centering
\includegraphics[width=3.5 in]{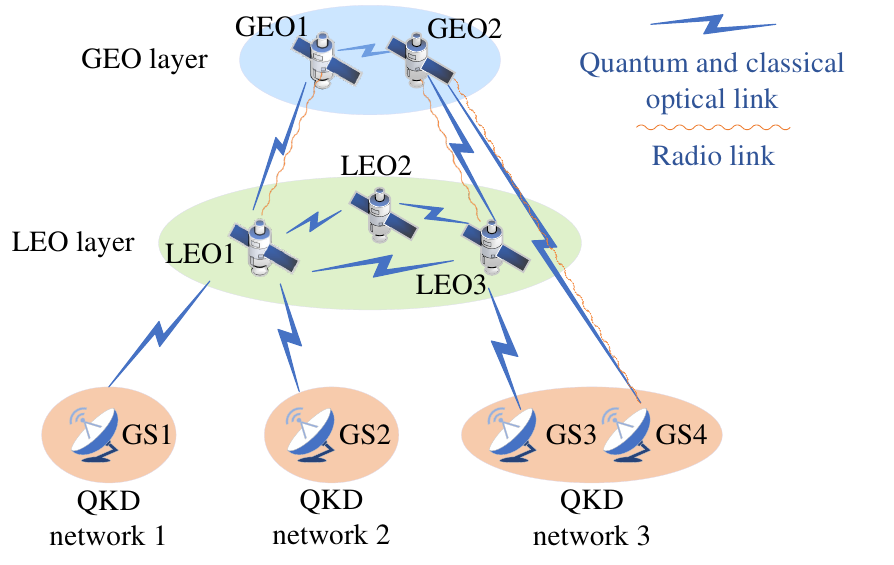}
\caption{Network architecture for double-layer quantum satellites. Quantum satellite nodes are composed of LEO and GEO satellite layers. Satellites in the same orbital layer are connected by inter-satellite links, and a satellite can communicate with another satellite in a different orbit layer via an inter-orbital link~\cite{huang2020quantum}.}
\label{Architecture_of_double-layer_quantum_satellite_networks}
\end{figure}

Long-distance QKD requires quantum relaying techniques. Quantum relays have been harnessed by LEO satellites, high-altitude platforms (HAPs), and terrestrial quantum network nodes on the ground \cite{9148563,vu2020design}. Relay-aided QKD networks tend to ensure secure encryption of each link by a dedicated secret key and facilitate end-to-end global key exchange between the source and destination. Routing algorithms meeting specific quality-of-service (QoS) considerations have been studied in QD relay networks \cite{yangquantum}. Experimental progress has shown the feasibility of QD relay techniques across multiple countries and continents~\cite{2018Satellite}.

Wang \textit{et al.} \cite{9148563} proposed a QKD network harnessing LEO satellites for relaying long-distance QKD requests and for achieving intercontinental QKD. Huang \textit{et al.} \cite{huang2020quantum} developed an architecture for global QD satellite networks, overcoming the limitations of single-layer satellite networks. Their architecture includes a trusted-repeater layer and a second layer that employs a quantum key pool (QKP) for supplying secret keys to ground stations. As shown in Fig.~\ref{Basic_structure_of_QKP_enabled_satellite_QKD_system}, the system consists of nodes having quantum transceivers. Each node is assigned a secret key for its routing selection. This architecture enables communication between satellites in different orbital layers, forming a global QKD network~\cite{huang2020quantum}; see Fig.~\ref{Architecture_of_double-layer_quantum_satellite_networks}.

{\color{black}Vu \textit{et al.}~\cite{ vu2020design} investigated the design and performance of relay-assisted satellite FSO/QKD systems used in secure vehicular networks. HAPs equipped with optical amplify-and-forward nodes serve as relay stations.  Secrecy performance in terms of QBER and ergodic secret-key rate was studied analytically under the influence of transceiver misalignment, receiver velocity variation, receiver noises, and atmospheric turbulence conditions.

A basic constraint that drastically limits the optimum rates achievable by two distant participants affects both quantum and private communications. Quantum repeaters can play an important role in addressing this issue.
Pirandola \textit{et al.}~\cite{PhysRevA.96.032318} studied the ultimate rates for transmitting quantum information, distributing entanglement, and distributing secret keys between a sender and a receiver with the assistance of unlimited two-way classical communication involving all participants.}
Additionally, in~\cite{ding2019quantum}, tailor-made coding schemes were designed for classical-quantum relay channels, including multihop, coherent multihop, decode-forward, and partial decode-forward relaying schemes.

\subsection{Quantum Error Correction and Mitigation}

The interaction of quantum systems with the environment results in decoherence, which can be mitigated with the aid of quantum error control techniques. The two main categories of quantum error control techniques are quantum error correction coding (QECC)~\cite{8423050,7950914,8019800,7515170,8218756,9581297, 8691729, 8299427} and QEM~\cite{ZhenyuCai2022QuantumEM,9294106,9638483,9684862}.

\subsubsection{QECC}
As for QECC, Babar \textit{et al.}~\cite{8423050} provided a detailed tutorial on the duality between classical and quantum coding theories and offered a guide on creating stabilizer-based QECCs. They also conceived design examples relying on the classical and quantum Bose-Chaudhuri-Hocquenghem codes and convolutional codes.
In addition, Cuvelier \textit{et al.}~\cite{9581297} proposed a general framework for harnessing QECCs in the context of classical non-coherent detection-aided space-time block codes.

Since quantum hardware is still in its infancy, Chandra and his colleagues~\cite{7950914,8019800,7515170,8218756,8691729} conducted a series of studies.
In particular, Chandra \textit{et al.}~\cite{7950914} explored the performance of low-complexity quantum codes having modest hardware requirements. They quantified quantum coding bounds and derived closed-form approximations of the trade-off between the minimum distance and coding rate of quantum codes. 
In another study, Nguyen~\textit{et al.}~\cite{8019800} investigated the hypothesis of whether it is feasible to harness network-coding principles in the quantum domain and conceived a relatively large-scale quantum network coding scheme for disseminating entangled qubits via quantum teleportation and QKD. Nguyen~\textit{et al.}~\cite{7515170} studied the Hashing bound of entanglement-based quantum channels and paved the way for the powerful design tool of extrinsic information transfer chart for improving the normalized throughput of realistic quantum devices.

Chandra~\textit{et al.}~\cite{8218756} revisited the classical topological error correction codes having codeword bits arranged in a lattice structure and exploited the so-called ``classical-to-quantum isomorphism''~\cite{8423050} to construct their quantum dual pairs. They also analyzed the QBER and fidelity of these codes when exposed to a quantum depolarizing channel, and quantified the specific depolarizing probability, below which the QECCs were capable of mitigating the QBER.
Since the depolarizing probability of quantum circuits is time-variant, Chandra \textit{et al.}~\cite{8691729} conceived computationally efficient variable-rate concatenated QECC designs, relying on combining short block codes and unity-rate codes as outer and inner codes, respectively. These designs are reconfigurable in coding rate, and can be conveniently reconfigured for different depolarizing probabilities. This offers relatively high quantum coding rates, since the unity-rate coding component clearly does not reduce the code rate and succeeds in approaching the hashing-bound under the idealized simplifying assumption of having flawless QECC hardware. 

However, the popular memoryless channel assumption routinely stipulated in QECC studies does not always hold true for realistic hardware. Previous error events tend to influence future errors. Hence, Izhar \textit{et al.}~\cite{8299427} examined quantum turbo codes (QTCs) in the face of depolarizing channels exhibiting memory and proposed a QTC scheme capable of exploiting error correlation during decoding. But again, new research is required for designing QECCs relying on realistic decoherence-infested hardware.

\subsubsection{QEM}
As a class of promising techniques for reducing the computational error of variational quantum algorithms, QEM can protect hybrid quantum-classical computation from decoherence but suffers from sampling overhead, which erodes the computational speed.
A thorough examination of the sampling overhead imposed by QEM was given in~\cite{ 9294106, 9638483, 9684862}. Out of a wide class of actual quantum channels with the same average fidelity, Pauli errors had the lowest sampling overhead~\cite{ 9294106}. 
In the presence of narrowband quantum noise, which corresponds to large-depth, high-error-rate quantum circuits, a general framework, known as permutation filters demonstrates reliable convergence to the global optimum~\cite{ 9638483}. 

A trade-off was revealed between the error scaling behavior and the computational overhead of quantum circuits protected by Monte Carlo-based QEM and those unprotected by QEM~\cite{ 9684862}. A channel inversion technique based on Monte Carlo sampling was employed, which adds extra computational error that, in turn, might be removed at the expense of an additional sample overhead.
In addition, Suguru \textit{et al.}~\cite{ ZhenyuCai2022QuantumEM } studied the framework of symmetry expansion, a kind of QEM that uses the intrinsic symmetries of the system and post-processing of measurement results to project the state into the noise-free symmetric subspace virtually. Although symmetry expansion has been limited to error mitigation of quantum states immediately before measurement, they successfully generalized symmetry expansion for state preparation.

\subsection{Summary}

In this section, we have reviewed the technologies that can be potentially utilized for quantum-enabled PHY layer in future native quantum systems, including multi-band quantum access, multi-user quantum communications, quantum MIMO, and quantum relays. By comparing the performance of quantum systems and the theoretical limits of classical systems, we observe that quantum communication is capable of outperforming classical communication in error probability. A community effort is required to extend point-to-point quantum communications to multipoint-to-multipoint scenarios suitable for future ultra-dense connections, such as multi-user quantum communication for quantum IoT and quantum MIMO systems resistant to atmospheric turbulence.

However, achieving QD multiple access in support of multiple users sharing entangled quantum states poses significant challenges. The fragility of entangled states and their susceptibility to environmental impairments and noise over long distances hinder the maintenance of coherence. Distributing entangled states among multiple users requires specialized equipment and presents substantial practical implementational challenges.
On the other hand, the computational complexity of quantum communication systems in the face of multi-user access presents another hurdle. The limited availability of quantum computing hardware is a practical barrier. This limitation restricts the performance of multi-user quantum communication systems and presents challenges in realizing such systems in real-world settings.

%% file: Chapter_3.tex
\begin{table*}
\caption{\color{black}A summary of quantum multiplexing}
\centering
\begin{tabular}{|m{2cm}<{\flushleft}| m{5cm}<{\flushleft}| m{4cm}<{\flushleft}| m{5cm}<{\flushleft}| }	
\hline
\textbf{\makecell*[c]{Method}} & \textbf{\makecell*[c]{Feature}} & \textbf{\makecell*[c]{Strength}} & \textbf{\makecell*[c]{Limitation}}  \\ \hline

\textbf{WDM QKD} &

$\bullet$ Can multiplex several independent quantum channels over a single fiber, enabling simultaneous secure communication between multiple pairs of users.

$\bullet$ Can be integrated into existing telecommunication networks.

& $\bullet$ Achieves high data rate, enhances network capacity \cite{2014Quantum,eriksson2019wavelength}

$\bullet$ Addresses challenge of resource allocation, improves network efficiency and security \cite{zhao2018resource}

$\bullet$ Ensures timely key provision \cite{cao2018time}

& $\bullet$ Requires sophisticated equipment and precise calibration \cite{2014Quantum}

$\bullet$Integration with a large number of classical channels may introduce complexities \cite{eriksson2019wavelength}

$\bullet$ Efficiency of RWTA algorithm may depend on network size and dynamics \cite{zhao2018resource}

$\bullet$ Complexity of time-scheduled QKD scheme may increase network overhead \cite{cao2018time}

\\ \hline

\textbf{SDM QKD}  & $\bullet$ Multiple spatial modes employed to transmit quantum states, enabling parallel quantum communication between multiple parties.

& $\bullet$ Offers flexibility in QKD deployment \cite{urena2019modeling}

$\bullet$ Provides insights into optimizing QKD-SDM compatibility~\cite{lin2020telecommunication}

& $\bullet$ Crosstalk between quantum and classical channels may limit performance \cite{urena2019modeling}

$\bullet$ Assumes idealized conditions~\cite{lin2020telecommunication}

\\ \hline

\textbf{WSDM QKD}& $\bullet$  Quantum signals transmitted in one outer core while classical signals transmitted in other cores. 
& $\bullet$ Enables simultaneous transmission of multiple quantum channels in MCF optical system

$\bullet$ Efficiently mitigates inter-core crosstalk~\cite{cai2019experimental}
& $\bullet$ Requires precise coordination of wavelength and core allocations

\\ \hline

\end{tabular}
\label{tab:quantum_multiplexing}
\end{table*}

\section{Quantum MAC and Upper Layer Issues} 

{\color{black}In this section, we investigate the key technologies of the quantum MAC and upper layer. Among them, the quantum MAC layer is responsible for managing and controlling the establishment, maintenance and release of quantum communication links, allocating and managing communication resources in the quantum communication system. The quantum network layer manages the network topology and selects routes, and the quantum application layer can realize various quantum applications, such as QKD, quantum teleportation, and other quantum security strategies.
}

{\color{black}\subsection{Quantum MAC Layer Issues}}

\subsubsection{Quantum Multiplexing} 

Wavelength division multiplexing (WDM) offers several benefits in addition to supporting clustered BS deployments, including the extension of radio access networks (RANs) to decentralized and heterogeneous RANs, while enhancing communication network flexibility and transparency. It enables the reuse of passive optical networks or metro rings for transporting various types of traffic, and facilitating multi-modal data transmission. WDM-based QKD networks can be employed for secure communications. The allocation of frequencies, time, power, and wavelengths is a key in WDM-based quantum networks. 

Tab.~\ref{tab:quantum_multiplexing} collates the salient examples of quantum multiplexing. Patel \textit{et al.} demonstrated the employment of dense WDM for QKD, while supporting 10 Gb/s bidirectional classical channels in a single fiber \cite{2014Quantum}. 
Eriksson \textit{et al.} \cite{eriksson2019wavelength} showcased the co-propagation of QKD signal with a large number of WDM-based coherent classical data channels, achieving a data rate of 18.3 Tbits/s when a CVQKD channel was jointly transmitted with 100 classical WDM channels. These experiments demonstrated significant improvements in the number of WDM channels and bit rate compared to previous co-propagation results.

Zhao \textit{et al.} \cite{zhao2018resource} studied resource allocation in QKD-secured WDM-based optical networks, and proposed a QKD-enabled network architecture, where a software-defined networking (SDN) controller can allocate different wavelengths to the classical data channel, the QD signal channel, and the public interaction channel using WDM. A new routing, wavelength, and time-slot allocation (RWTA) algorithm was designed to integrate classical routing algorithms, such as the $K$-shortest path and Dijkstra's algorithms, as well as classical channel allocation algorithms, such as first-fit wavelength allocation~\cite{zhao2018resource}. Their simulations revealed the blocking probability of the QD signal channel increases as the key updating periods decrease, while the security performance improves.

Cao \textit{et al.} \cite{cao2018time} proposed a QKD-over-WDM architecture for decoupling secret keys from the infrastructure and for ensuring timely provision of secret keys. They presented the first-ever time-scheduled QKD scheme specifically designed for WDM networks. They developed a computationally efficient heuristic  RWTA algorithm, including persistent or dynamic secret key consumption, even or uneven time-slot allocation, and continuous or discrete quantum key provision. 
Urena \textit{et al.} \cite{urena2019modeling} evaluated a multi-core fiber (MCF)-based QKD link constructed using spatial division multiplexing (SDM). Their analysis explored the feasibility of stand-alone QKD transmission for QKD-only channels and the simultaneous transmission of QKD and classical channels. They showed that a typical homogeneous MCF having low inter-core crosstalk can support multiple quantum channels. However, the presence of both quantum and classical channels can lead to excessive crosstalk between the cores, necessitating the use of a heterogeneous fiber to reduce crosstalk.

The compatibility of QKD with classical communication in SDM transmission was also investigated in~\cite{lin2020telecommunication}, who conceived analytical models of noise sources in QKD links for transmission over heterogeneous MCFs. Their analysis focused on intra-core crosstalk and intra-core spontaneous Raman scattering. Considering different core and wavelength allocations, they obtained lower bounds for the secret key rates and the QBER. In another study, Cai \textit{et al.}~\cite{cai2019experimental} demonstrated QKD using CD signals over a seven-core fiber and developed a wavelength-space division multiplexing (WSDM) strategy and mitigated crosstalk for simultaneous transmission of QD and CD signals in an MCF optical system.

\subsubsection{ Quantum Resource Allocation}

Compared to classical communications, quantum information transmission relies on more complex resource allocation. Researchers have dedicated significant efforts to developing resource allocation algorithms for QKD networks, such as the wavelength, fiber core, and time slot allocations within WDM-based QKD architectures. Furthermore, QKD has been integrated both with classical SDN~\cite{Yuan2019Multi} and network function virtualization (NFV)~\cite{wang2019end} to provide flexible supervision and uninterrupted secret key provision \cite{zhao2018resource,wang2019end}. The challenge is that the generation of secret keys is a painstakingly slow and tedious process, which must supply new keys as frequently as possible. In recent years, QD techniques, such as quantum games, have been shown to exhibit promise in resource allocation, supplementing traditional game theory and convex optimization methods \cite{QKD_ON}.

\begin{figure} 
\centering
\includegraphics[width=3.5 in]{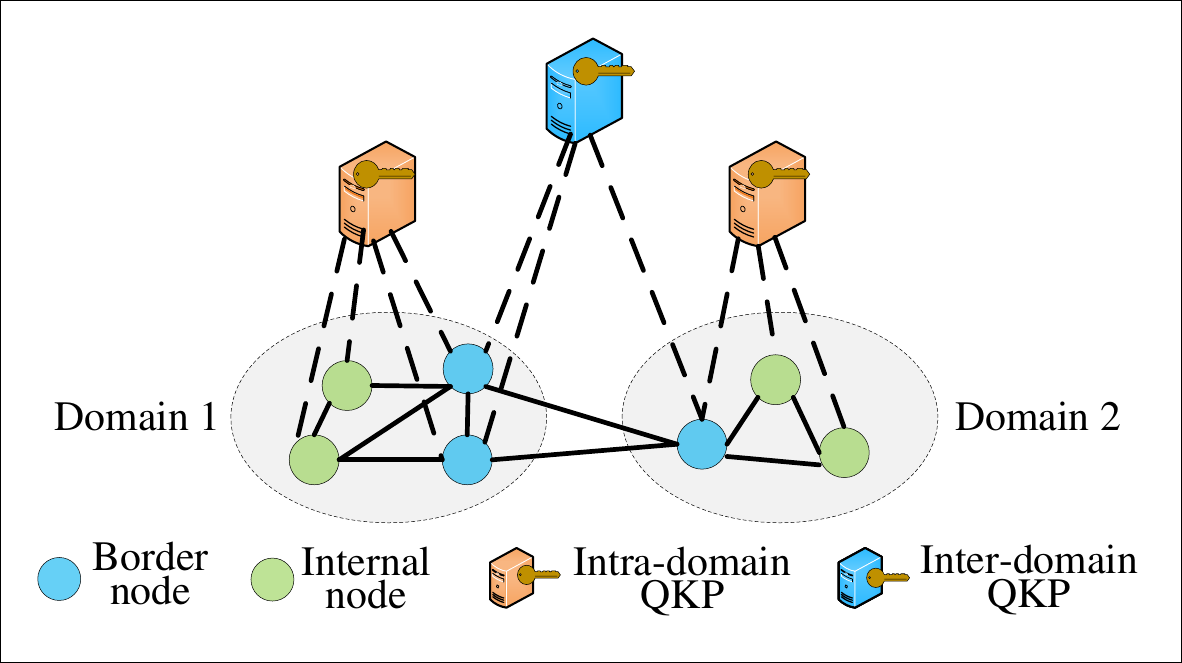}
\caption{Multi-domain QKD optical network. In the case of a multi-domain QKD optical network, the intra-domain QKP of every intra-domain QKP node and the inter-QKP of every boundary node is indispensable. Service allocation will be determined by how key resources generated on direct and non-direct links  are used to supplement key pools~\cite{QKD_ON}.}
\label{Multi-domain_QKD_optical_network}
\end{figure}

To achieve seamless global connectivity, future core networks are expected to be multi-domain networks where autonomous systems can collaborate. As the network scale and flawless connection requirements between different equipment manufacturers escalate, multi-domain optical networks become a practical area of study. In such networks, the adjacent nodes within a single domain must mutually endorse the negotiation and generation of key resources. However, having both intra-domain and inter-domain QKPs becomes indispensable in multi-domain QKD optical networks. Quantum key resources are valuable, and excessive service requests may quickly deplete the key pool. 
Wang \textit{et al.}~\cite{QKD_ON} proposed an adaptive scheme for multi-domain QKD in optical networks, which carefully balances the demand for the number of hops, improving both the service-blocking probability and the key resource utilization, as shown in Fig.~\ref{Multi-domain_QKD_optical_network}.
Wang \textit{et al.} \cite{wang2019protection} developed a pair of schemes for controlling the key volumes in optical networks using QKD. These schemes incorporate key-volume-dependent adaptive routing and a key update mechanism, offering a reliable solution for allocating and protecting resources. Their simulation results demonstrated that shared key-volume-dependent adaptive algorithms outperform their non-shared counterparts in terms of both blocking probability, time-slot utilization efficiency, and proportion of secret key ``consumption''.

In the context of multi-core optical networks, Yu and colleagues \cite{yu2021routing} proposed a set of four routing, core, and wavelength allocation (RCWA) algorithms. These algorithms, namely RCWA using spectrum perception (RCWA-SP), RCWA employing crosstalk perception (RCWA-XTP), RCWA operating without perception (RCWA-WP), and RCWA with core perception (RCWA-CP), rely on different core-allocation schemes. Specifically, RCWA-SP selects the channel having the least busy wavelength to optimize network resource utilization. RCWA-XTP preserves cores to prevent interference between the quantum core channel and other types of requests. RCWA-WP and RCWA-CP impose reduced complexity but may exhibit eroded performance. Numerical results demonstrated that RCWA-SP reduces the blocking probability, while RCWA-XTP minimizes inter-core crosstalk.

To accomplish resource-efficient QKD networking with QKP, Zhang~\textit{et al.}~\cite{RoutingChannelKeyRateandTimeSlotAssignment} put forward a new routing, channel, key-rate, and time-slot assignment (RCKTA) problem. Four distinct network configurations—allowing or prohibiting the usage of trusted relays and optical bypasses—were used to solve the problem. A scalable and nearly optimum heuristic approach was developed to cut down the execution time of building the QKD route.

{\color{black}\subsection{Quantum Network Layer Issues}}

The quantum network layer is a fundamental component of quantum communication systems, enabling quantum information transmission, entanglement distribution, and secure communication. Several challenges can hinder the reliability, scalability, and practical deployment of quantum networks.

First, scalability remains a significant hurdle. Most existing quantum networks are small-scale or limited to point-to-point connections; expanding them to metropolitan or global scales requires a vast number of quantum repeaters and memory nodes. However, the current state of these technologies is not mature; for instance, quantum memory devices face challenges in maintaining quantum states for extended durations and achieving efficient read/write operations~\cite{WOS:000400864700001}. 

Second, quantum noise affects the performance and scalability of quantum networks. It affects quantum gates by increasing error rates, reducing fidelity, and causing decoherence, degrading the success rate of quantum operations and hindering network expansion~\cite{belghachi2023quantum}. Long-distance entanglement is essential for scalable quantum networks, yet it deteriorates over distance due to decoherence~\cite{WOS:000590075700003, WOS:000590075700001}. Quantum repeaters can restore entanglement across long distances, but noise within repeater memory nodes can compromise the fidelity of entanglement between network nodes.

Moreover, resource constraints pose additional challenges. Quantum resources, e.g., entanglement pairs and quantum memory, are limited, necessitating efficient scheduling to facilitate multi-user sharing. Quantum networks must dynamically establish entangled connections between multiple nodes, requiring novel algorithms for routing, error correction, and security protocols~\cite{wehner2018quantum}. The management and control software for quantum networks must be re-engineered, incorporating complex system design and optimization to accommodate the unique demands of quantum communication.

\begin{figure} 
\centering
\includegraphics[width=3.5 in]{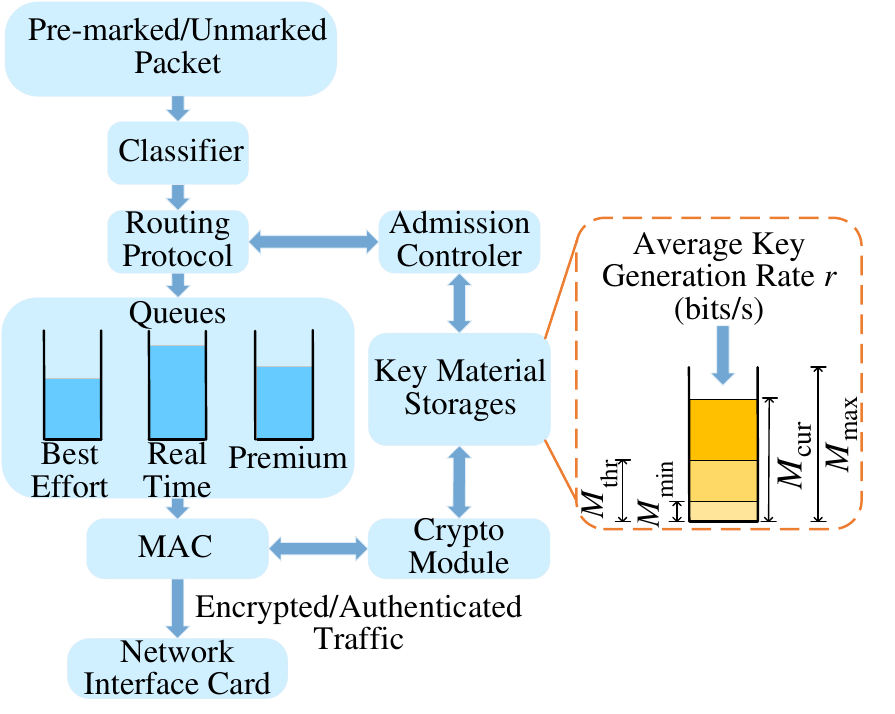}
\caption{Flexible QoS model for QKD and traffic processing. The flexible QoS model distinguishes three traffic classes with the corresponding differentiated services, namely, best-effort,  real-time, and premium classes~\cite{8935373}. }
\label{FQKD_model_and_traffic_processing}
\end{figure}

\subsubsection{Quantum Routing in QKD Network}

Quantum routing is the process of monitoring changes in network topology, exchanging routing information, establishing and maintaining routes, and forwarding information-carrying quantum states in a quantum network~\cite{ Quantumrouting }.
Recent advancements have focused on optimizing key generation, traffic prioritization, and efficient key distribution.
A cost function-based approach has been introduced to dynamically steer relays toward faster key generation links, enhancing key exchange success rates and ensuring balanced participation across all QKD links. Mehic~\textit{et al.}~\cite{8935373} addressed real-time communication challenges by developing a QoS-based routing protocol that classifies and prioritizes traffic at the ingress node using separate queues, as shown in Fig.~\ref{FQKD_model_and_traffic_processing}. This protocol incorporates caching and return-loop detection, minimizing key material consumption while maintaining efficient forwarding.
For secure multicast services, an optimized global key distribution scheme was proposed in~\cite{2020Tree}, leveraging a key-relay mechanism and an auxiliary topology that dynamically maps connections and weights between QKD nodes. By integrating a routing and key assignment algorithm, this scheme demonstrated improved security probabilities compared to traditional methods.
These innovations contribute to more efficient, scalable, and secure QKD networks.

\begin{table*}
\caption{\color{black}A summary of quantum network service and architecture}
\centering
\begin{tabular}{|m{3cm}<{\flushleft}| m{5cm}<{\flushleft}| m{4cm}<{\flushleft}| m{4cm}<{\flushleft}|}	
\hline
\textbf{\makecell*[c]{Deploment Scenario}} & \textbf{\makecell*[c]{Contribution}} & \textbf{\makecell*[c]{Strength}} & \textbf{\makecell*[c]{Limitation}}  \\ \hline

Multi-user quantum wireless communication network   &$\bullet$ A self-organized bilayer network architecture to ensure the rapid and steady transmission of quantum  information~\cite{7570198}  & $\bullet$ Reduced resource consumption &$\bullet$  High implementation complexity  \\ \hline

 SDN/VNF-based QKD-enabled network  
& $\bullet$ Support network  services dynamically  created by chaining VNFs over  multiple network domains~\cite{wang2019end} 

$\bullet$  Integration of application, SDN, and QKD~\cite{hugues2019monitoring}

$\bullet$ Multi-tenant QKD network architecture based on SDN and QKP~\cite{Yuan2019Multi}

& $\bullet$  Establish dynamic service \cite{wang2019end} 

$\bullet$ Real-time resource allocation \cite{hugues2019monitoring}
 
$\bullet$ Efficient secret-key assignment \cite{Yuan2019Multi}

& $\bullet$ High integration complexity \cite{wang2019end}

$\bullet$ High implementation complexity \cite{hugues2019monitoring}

$\bullet$ Scalability validation \cite{Yuan2019Multi} 
\\ \hline
QKD-enhanced end-to-end service & $\bullet$ Required control plane workflows and protocol extensions for the integration of QKD keys into network services~\cite{aguado2018virtual} & $\bullet$  Integration with major protocols & $\bullet$ Standardized protocols  \\ \hline

QKD over optical backbone network & $\bullet$  QKD network architecture for optical backbone networks with trusted and untrusted relays~\cite{9373434} & $\bullet$  Reduced cost & $\bullet$ Security model validation  \\ \hline

\end{tabular}
\label{tab:quantum_network_service_and_architecture}
\end{table*}

\subsubsection{Entanglement Routing}   
Entanglement routing, a technique for creating remote entanglement connections between quantum nodes, plays a crucial role in realizing the capabilities of quantum networks~\cite{FidelityGuaranteedEntanglementRouting}. 
Recent advancements in entanglement routing for quantum networks have focused on optimizing resource utilization, improving scalability, and enhancing routing efficiency. Various strategies have been proposed to address key challenges, such as congestion, multi-hop routing, and resource allocation. 
Li~\textit{et al.}~\cite{SwappingBasedEntanglementRouting} introduced a swapping-based entanglement routing scheme that reallocates idle entanglement resources to alleviate bottlenecks, significantly improving end-to-end entanglement distribution rates. Shaban~\textit{et al.}~\cite{SPARQ} explored a space–air–ground quantum network, integrating satellites, HAPs, and ground users while leveraging deep reinforcement learning (RL) for optimized entanglement distribution. Similarly, Shi~\textit{et al.}~\cite{ConcurrentEntanglementRouting} proposed a contention-free path selection algorithm tailored for multi-hop routing in quantum networks, introducing quantum-specific metrics beyond traditional hop count and physical distance.

Other works have focused on structured routing approaches and resource efficiency. Zeng~\textit{et al.}~\cite{EntanglementRoutingDesign} proposed a two-phase entanglement routing framework, where offline pre-computed routes are deployed in real time to optimize network throughput and scalability. Kim~\textit{et al.}~\cite{ResourceEfficientLinksetConfiguration} introduced a degree-aware shortest path selection algorithm and a resource-efficient link-set configuration technique, enabling optimal path selection and efficient link resource utilization in large-scale quantum networks.
Collectively, these advancements highlight the importance of intelligent entanglement routing strategies, resource optimization, and scalability enhancements for future quantum networks.

\subsubsection{Quantum Network Service and Architecture}

The rapid expansion of quantum user equipment and IoT devices necessitates diverse QoS requirements, such as data rate and latency, which traditional RANs struggle to meet~\cite{8766143}. Open RAN (O-RAN) has emerged as a solution by treating services separately, allowing some applications to connect directly to the core network while others pass through multiple radio control functions. O-RANs integrate technologies like NFV, mobile edge computing (MEC), and QKD-based encryption to enhance efficiency and security~\cite{8766143}.
Tab.~\ref{tab:quantum_network_service_and_architecture} summarizes the key technologies in quantum network service and architecture.

Quantum-secured architectures are being explored to strengthen network resilience. The use of Greenberger-Horne-Zeilinger (GHZ) states in quantum wireless networks has been shown to reduce computational complexity and resource consumption~\cite{7570198}. In optical networks, quantum-aware virtual network functions (VNFs) have been introduced, leveraging SDN-enabled optical control and NFV orchestration to allow dynamic network service reconfiguration~\cite{wang2019end}. Furthermore, integration of QKD with IPsec and control plane protocols, such as OpenFlow and NETCONF, enables secure end-to-end encryption for multi-domain communication~\cite{aguado2018virtual}.

To further enhance network security and reliability, real-time monitoring of quantum parameters within SDN frameworks has been proposed, enabling adaptive routing and protection against link-level attacks~\cite{hugues2019monitoring}. A multi-tenant QKD network architecture utilizing SDN and secret-key rate-sharing was developed to provide scalable key management for multiple users~\cite{Yuan2019Multi}. In addition, a QKD-enabled optical backbone network was introduced employing trusted and untrusted relays, integrating cost, security, and performance optimization models~\cite{9373434}. 
These advancements collectively drive the evolution of quantum-secured network infrastructures, ensuring scalability, dynamic adaptability, and enhanced security.

\subsection{Quantum Application Layer Issues}

\begin{table*}
\caption{\color{black}A Summary of Primary Quantum Security Strategies}
\centering
\begin{tabular}{|m{2.5cm}<{\raggedright}| m{5cm}<{\raggedright}| m{5cm}<{\raggedright}| m{4cm}<{\raggedright}|}
\hline
\textbf{Technique} & \textbf{Brief Description} & \textbf{Strength} & \textbf{Limitation} \\ 
\hline

\textbf{QKD} & 
\begin{itemize}[leftmargin=*]
    \item LDPC code-based rate-adaptive reconciliation schemes for QKD systems~\cite{ai2018quantum}. 
    \item Joint QKD-post-quantum cryptosystem~\cite{djordjevic2020joint}.
    \item Reliability of eavesdropping detection in BB84 QKD protocol~\cite{WOS:000866556800054}.
    \item Evaluation of eavesdropping error-rates in high-dimension QKD system~\cite{WOS:000730601600001}.
    \item Modified BB84 QKD protocol utilizing disentropy of autocorrelation function~\cite{WOS:000748319000005}.
\end{itemize} & 
\begin{itemize}[leftmargin=*]
    \item Improved secret key rates~\cite{ai2018quantum}.
    \item Extended transmission distance~\cite{djordjevic2020joint}.
    \item Eavesdropping detection and prevention~\cite{WOS:000866556800054}.
    \item Reduced eavesdropping error rate~\cite{WOS:000730601600001}.
    \item Increased security of the QKD protocol~\cite{WOS:000748319000005}.
\end{itemize} & 
\begin{itemize}[leftmargin=*]
    \item High implementation complexity~\cite{ai2018quantum,WOS:000730601600001, WOS:000748319000005}.
    \item High integration complexity~\cite{djordjevic2020joint}.
    \item Trade-off between detection accuracy and resource consumption~\cite{WOS:000866556800054}.
\end{itemize} \\ 
\hline
\textbf{QSDC} & 
\begin{itemize}[leftmargin=*]
    \item QSDC protocol utilizing entangled photon pairs~\cite{yang2020quantum}.
    \item Semi-QSDC protocol~\cite{rong2020semi}.
    \item Controlled bidirectional QSDC protocol~\cite{pan2021controlled}.
    \item Single-photon-memory QSDC scheme~\cite{9129730}.
    \item Three-party QSDC protocol~\cite{WOS:000763396900002}.
    \item Bidirectional multi-party DSQC scheme~\cite{WOS:000848596900003}.
\end{itemize} & 
\begin{itemize}[leftmargin=*]
    \item High SNR due to the heralding function of entanglement pairs~\cite{yang2020quantum}.
    \item Complete robustness~\cite{rong2020semi}.
    \item Avoidance of information leakage~\cite{pan2021controlled}.
    \item Resistance against multiple external attacks~\cite{9129730,WOS:000763396900002}.
    \item High-rate communication and eavesdropping detection~\cite{WOS:000848596900003}.
\end{itemize} & 
\begin{itemize}[leftmargin=*]
    \item Technical challenges in practical implementation~\cite{yang2020quantum,WOS:000848596900003,9129730}.
    \item Integration complexity of controlled protocols~\cite{pan2021controlled}.
    \item Integration with existing systems and infrastructure~\cite{WOS:000763396900002}.
\end{itemize} \\ 
\hline
\textbf{Quantum secret-sharing} & 
\begin{itemize}[leftmargin=*]
    \item Verifiable quantum secret-sharing protocol for three parties~\cite{WOS:000640429300001}.
\end{itemize} & 
\begin{itemize}[leftmargin=*]
    \item Ensured security against internal and external attacks~\cite{WOS:000640429300001}.
\end{itemize} & 
\begin{itemize}[leftmargin=*]
    \item Integration with existing systems and infrastructure~\cite{WOS:000640429300001}.
\end{itemize} \\ 
\hline
\textbf{Quantum security monitoring} & 
\begin{itemize}[leftmargin=*]
    \item In-service security monitoring using quantum states~\cite{WOS:000714702000040}.
\end{itemize} & 
\begin{itemize}[leftmargin=*]
    \item Improved compatibility and potentially higher data rates~\cite{WOS:000714702000040}.
\end{itemize} & 
\begin{itemize}[leftmargin=*]
    \item Practical implementation and compatibility challenges~\cite{WOS:000714702000040}.
\end{itemize} \\ 
\hline
\end{tabular}
\label{tab:quantum_application_layer_issues}
\end{table*}

\begin{figure}[t]
\centering
\includegraphics[width=3.5 in]{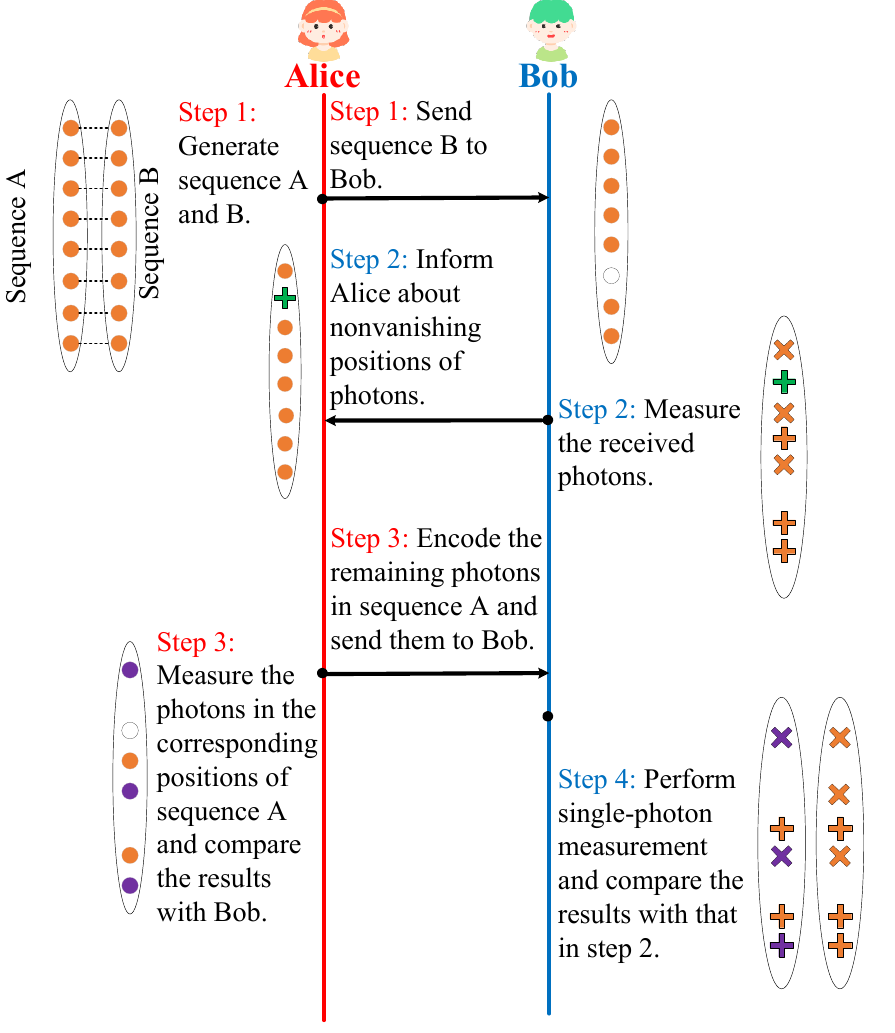}
\caption{A QSDC protocol utilizing entangled photon pairs. The orange circles connected by the dotted lines are entangled photons, and the white circles indicate the photons lost in transmission. The symbols + (Z-basis) and × (X-basis) indicate the single-photon measurement results. The remaining photons in step 3 represent Alice encoding the secret message through unitary operations $I$ (orange) and ${\sigma _y}$ (purple), respectively~\cite{yang2020quantum}.}
\label{QSDC_with_enhanced_eavesdropping_detection_probability}
\end{figure}

Tab.~\ref{tab:quantum_application_layer_issues} summarizes primary quantum security strategies in the quantum application layer. In space-based QKD systems utilizing entanglement-based protocols like the BB84 protocol, Ai  \textit{et al.} \cite{ai2018quantum} investigated LDPC code-based rate-adaptive reconciliation schemes. They compared adaptive-rate LDPC codes to fixed-rate non-adaptive LDPC codes when Alice and Bob are located at the ground stations 1000 km apart, and a LEO satellite generates and distributes entangled pairs of photons at an altitude of 500 km. The results show that adaptive-rate LDPC codes significantly improve the key rates for space-based QKD applications, offering higher flexibility than fixed-rate reconciliation schemes.

Djordjevic \textit{et al.}~\cite{djordjevic2020joint} proposed a joint QKD-post-quantum cryptosystem for enhancing the key rate vs. transmission distance trade-off. This cryptosystem combines a post-quantum cryptography-based subsystem with QKD to improve the reconciliation of information. By incorporating the QKD subsystem, the cryptosystem reliably prevents information leakage and detects eavesdroppers. Additionally, Djordjevic \textit{et al.}~\cite{djordjevic2020joint} also developed a scalable spatially coupled LDPC code that can be configured in real time. A twin-field-QKD-McEliece cryptosystem was simulated to achieve a distance of 1238 km over ultra-low-loss fiber, assuming an eavesdropper employing a quantum information set decoding attack.

{\color{black}To elaborate on the anti-eavesdropping capability of QKD protocols, Lee \textit{et al.}~\cite{WOS:000866556800054} analyzed the reliability of eavesdropping detection in the BB84 QKD protocol using Hoeffding's inequality. A trade-off was found between the accuracy of eavesdropping detection and the frugality of the quantum resources used by the BB84 protocol. }
{\color{black}Kamran \textit{et al.}~\cite{WOS:000730601600001} demonstrated the advantages of using dynamic spatial modes in QKD systems through the design and implementation of the KMB09 protocol using Laguerre-Gaussian OAM-based techniques. Moreover, Castro \textit{et al.}~\cite{WOS:000748319000005} developed a modified BB84 QKD protocol that relies on the disentropy of the autocorrelation function for detecting eavesdropping. This protocol offers an additional layer of security to the already secure quantum communication systems, making it a beneficial design option for secure data communication.}

{\color{black}Many researchers have proposed various QSDC protocols that can resist multiple external attacks to improve security~\cite{yang2020quantum,rong2020semi,pan2021controlled, WOS:000763396900002, WOS:000848596900003, WOS:000613003400134, 9129730}. For instance, }Yang \textit{et al.} \cite{yang2020quantum} proposed a QSDC protocol utilizing entangled photon pairs, as shown in Fig.~\ref{QSDC_with_enhanced_eavesdropping_detection_probability}. Unlike earlier protocols, this method involves measuring one photon of the entangled pair after the entanglement distribution. Alice generates EPR pairs and collects particles for transmission while keeping a pair for herself. Bob measures the received photons and informs Alice about their non-vanishing positions. Through measurements and comparisons, Alice ensures the secrecy of the communication. The remaining photons in sequence A are encoded by Alice, and the encoded photons are sent to Bob after inserting check bits. Bob performs single-photon measurements, and the results are compared with previous measurements. Simulation results demonstrated increased secrecy capacity due to the heralding function of the entangled photons.

Rong \textit{et al.} \cite{rong2020semi} presented a semi-QSDC protocol enabling bi-directional information transfer between a quantum-capable Alice and a classical Bob using single-photon transmission and Bell states. The protocol allows the legitimate user to read the message directly upon receiving the quantum state and prevents encoded messages from being leaked to eavesdroppers, even if they control the channel. By employing GHZ states and quantum networks, the protocol shows potential for secure and efficient communication in various applications, including private query systems involving third parties. Furthermore, a controlled bidirectional QSDC protocol was proposed in \cite{pan2021controlled} utilizing six-qubit entangled states, where Bell basis measurements and single-particle measurements were performed. The protocol is controlled by a third party named Charlie, who generates the six-qubit states and ensures security by inserting decoy states into the sequence sent to Bob.

{\color{black}Zhou \textit{et al.}~\cite{WOS:000763396900002} proposed a three-party QSDC protocol associated with hyperentanglement under polarization and two spatial longitudinal momentum degrees of freedom. Their analysis indicates that inserting decoy photons for eavesdropping detection improves security and can resist multiple external attacks. Deterministic secure quantum communication (DSQC) is another branch of QSDC that uses classical bits to transmit secret messages. In this context, Guo \textit{et al.}~\cite{WOS:000848596900003} proposed a bidirectional multi-party DSQC scheme based on $d$-dimensional $k$-particle GHZ states. They added $d$-dimensional single photons for eavesdropping detection and extended the scheme to a network in support of high-rate communication of secret messages across regions.}

QSDC is featured by theoretically perfect information security. However, in practice, the imperfections of optical devices can leave loopholes for quantum attacks.
To address this, Pan \textit{et al.} \cite{9129730} proposed a new single-photon-memory QSDC scheme based on a two-step QSDC protocol. They also carried out the security analysis of the scheme in the face of direct measurement attacks, intercept-resend attacks, and opaque attack strategies. None of these turned out to be a threat.

Apart from the QKD and QSDC protocols, studies have been conducted on other quantum security protocols.
For example, Jiang \textit{et al.}\cite{WOS:000640429300001} proposed a verifiable quantum secret-sharing protocol using eight entangled quantum states, ensuring security against internal and external attacks using decoy states and message randomness. Meanwhile, Gong \textit{et al.}\cite{WOS:000714702000040} introduced a method for optical, physical layer security monitoring, utilizing a QD signal with continuously variable quantum states for in-service protection.

\subsection{Summary}

{\color{black}Future networks may be expected to integrate multiple domains, necessitating efficient resource allocation algorithms in the MAC and upper layer for quantum information transmission. Some studies have extended mobile ad-hoc networking concepts to address the QoS of QKD networks. Quantum communications can leverage WDM, SDM, and WSDM to enable security. Real-time monitoring of key rates and QBERs can be achieved with the aid of SDN and NFV.

However, it is important to consider the limitations of quantum protocols. Quantum information transmission is susceptible to losses and noise, which can restrict the distance over which quantum communication is viable. Furthermore, the creation and manipulation of quantum states require specialized equipment, presenting challenges in practical implementations of quantum protocols. Moreover, quantum protocols tend to be more intricate than classical protocols, leading to analytical and conceptual challenges. These limitations contribute to the complexities associated with developing and implementing quantum protocols in real-world scenarios, e.g., Qinternet.
}

%% file: Chapter_4.tex
\section{Quantum AI and Computing for Future Communications}

Quantum information and AI constitute a pair of pivotal technologies in future communications. Integrating quantum technology with AI offers significant advantages, capitalizing on the rapid advancements in both fields. A compelling application is found in quantum edge computing, which can provide an integrated and secure edge cloud platform for various industries, such as banking, insurance, public security, e-commerce, and mobile communications \cite{9665527}.

{\color{black}Quantum computers can use algorithms for factorization (Shor's algorithm) and database searching (Grover's algorithm), and
solve optimization and search problems exponentially faster than classical computers~\cite{10.1007/978-981-16-5640-8_46}.
Quantum computers also leverage the principles of superposition and entanglement, and perform computations on multiple possibilities simultaneously to speed up pattern recognition~\cite{10.1007/978-981-16-4863-2_13}.
On the other hand, by combining AI techniques, developing and optimizing quantum algorithms becomes more efficient. Methods such as generative model and RL provide a new way to solve the difficult problem of quantum computing. For example, the use of generative models to generate optimized sequences of quantum gate operations and the use of RL algorithms to training optimize entanglement processes can improve the performance of algorithms. Combining AI technology to optimize the quantum gate operation and entanglement process in Grover algorithm can improve the search efficiency~\cite{https://cloud.tencent.com/developer/article/2387355}. Additionally, AI can accelerate quantum computing in quantum circuit optimization, hardware design, error correction, and noise mitigation~\cite{WOS:001004939000007}. In the future, with the further development of quantum computing and AI, the mutual promotion of the two will open up more possibilities for scientists and developers.
}

\subsection{Quantum Search Algorithms}

Quantum search algorithms are a class of algorithms used in quantum computing to search an unsorted database or solve related problems.  Quantum computers can process multiple inputs simultaneously due to the principles of superposition.  This allows quantum search algorithms to explore multiple paths in parallel, significantly speeding up the search. The applications of quantum search algorithms include searching through large databases, finding a specific item in an unsorted list, and solving optimization problems. 
Grover's algorithm, proposed in 1996, is one of the most known quantum search algorithms that uses quantum parallelism to search for an item in an unordered database~\cite{Grover1996AFQ}. It can search through an unsorted database of $N$ elements in $\mathcal{O}(\sqrt{N})$ time, which is exponentially faster than classical algorithms  that require $\mathcal{O}(N)$ time~\cite{10318071}. Grover's algorithm uses quantum parallelism and interference to increase the probability of the correct answer while reducing the probability of incorrect answers. 

{\color{black}In~\cite{8688404}, the complexity of
sparse code-based multiple access was reduced by an efficient
quantum-search algorithm harnessed for multi-user detection.}
The study presented in~\cite{8540839} investigated the potential of quantum algorithms to solve large-scale search problems encountered in the physical and network layers of wireless communication systems. They considered a suite of optimization problems, such as multi-user detection, channel estimation, and finding optimal transmit precoding weights and routes.  
As a further development, Chandra \textit{et al.}~\cite{10.36244ICJ.2022.3.1} surveyed wireless problems, where quantum search algorithms can contribute substantially to reduce the number of cost-function evaluations compared to their classical counterparts, and improve speed, security, and efficiency.
Mondal \emph{et al.}~\cite{9340593} proposed a semi-quantum framework to reduce the query complexity of both the Durr-Hoyer algorithm~\cite{1996A} and of the Boyer-Brassard-Høyer-Tapp~\cite{1996Tight} algorithm. Their simulation results characterizing a MIMO-OFDM system showed that the symbol error rate of the proposed algorithm was close to that of the classical maximum likelihood search. 

{\color{black}Quantum random-access lookup of a string of classical bits is a necessary ingredient in several important quantum algorithms. In some cases, the cost of quantum random-access memory (qRAM) limites the implementation of quantum algorithm.
Matteo \emph{et al.}~\cite{8962352} studied the cost of fault-tolerantly  implementing a qRAM. They analyzed a number of circuit families that function as qRAM, discussed trade-offs in the number of qubits and run-time, and estimated resource costs when embedded in a surface code.}

\subsection{QD Optimization}

{\color{black}QD optimization merges principles from quantum computing and optimization theory to tackle complex computational problems. These algorithms often exploit quantum parallelism to simultaneously evaluate multiple potential solutions, with potentially faster convergence and more efficient exploration of solution spaces than classical optimization algorithms. 

QD optimization has significant applications in solving combinatorial optimization problems, such as the traveling salesman problem, graph coloring, and vehicle routing. 
Kim \textit{et al.}~\cite{9548770}
developed a quantum optimization algorithm to minimize the overlap
between observation satellite constellations in the mmWave band.
Kakaraparty \textit{et al.}~\cite{9666048} discussed AI and quantum
computing in mmWave communication systems designed for
UAVs. 
Alanis \emph{et al.}~\cite{8329527} proposed an evolutionary QD Pareto-optimization algorithm for finding ``almost all'' of the best paths at near-polynomial complexity in wireless multi-hop networks.
{\color{black}Quantum walks are the quantum analog of classical random walks,
and constitute a popular quantum computing architecture exhibiting reliable cryptographic characteristics,
which have been harnessed for 5G-IoT applications.
For instance, El-Latif \textit{et al.}~\cite{8972594} proposed new
symmetric key distribution algorithms for secure encryption by
exploiting the properties of quantum walks.

An active development of quantum optimization algorithms is the
quantum-behaved particle swarm optimization (PSO) algorithm,
which leverages the principles of quantum mechanics to enhance the classical PSO algorithm.
By representing particle positions probabilistically and using the mean best position for updates,
quantum-behaved PSO improves exploration capabilities, avoids local optima, and accelerates convergence.
This algorithm has been applied in resource scheduling~\cite{9112145},
wavelength assignment in a time and WDM system~\cite{zhang2020qos},
and energy efficient vehicular networks~\cite{9204821}.

Quantum annealing can potentially offer speedups for complex and high-dimensional problems
by leveraging quantum superposition and tunneling to explore multiple solutions simultaneously
with an increased likelihood of finding the global optima.
This algorithm has been applied in large-scale MIMO-aided wireless communication
for an improved SNR and BER~\cite{DBLP:conf/sigcomm/KimVJ19},
and for solving the downlink vector perturbation-based transmit precoding problem~\cite{9500557}.

Quantum genetic algorithms, including quantum bacterial foraging optimization (BFO)
and quantum approximate optimization,
can explore solution spaces more efficiently and achieve faster convergence than classical genetic algorithms,
by leveraging superposition, entanglement, and quantum gates~\cite{8790134, Frank2020Quantum}.
In quantum BFO, qubits are employed to represent the position of each bacterium.
Superposition allows bacteria to exist in multiple states at once, increasing the diversity of the search.
Entanglement helps in sharing information between bacteria, improving the collective behavior and convergence.
Quantum rotation gates are applied to update the positions of bacteria, guiding their movement in the search space.
In quantum approximate optimization,
the problem is represented as a Hamiltonian-based cost function,
encoding the constraints and objectives.
Quantum gates are applied and optimized to minimize the cost function.
Quantum genetic algorithms have been applied to MIMO communication~\cite{9225711}
and combinatorial problems under hard constraints~\cite{9222273}.
}

Quantum game theory is a branch of game theory that extends the philosophy of classical game theory by combining the principles of quantum mechanics and quantum computation with conventional games. In quantum game theory, players may have access to quantum strategies, which can involve using superposition, entanglement, and other quantum phenomena to make decisions. This allows for a more nuanced analysis of strategic interactions, especially in scenarios where classical strategies may not be optimal or where quantum effects play a significant role.}
Xu \textit{et al.}~\cite{xu2020quantum} designed a pair of quantum game models to deal with the mission assignment and quality optimization of existing crowdsourcing platforms relying on homogeneous workers. A quota-oriented crowdsourcing game was first constructed, only to reveal that no optimal strategy exists to maximize the payoffs of the workers. Then, a quantum game model was proposed to address the issue. Furthermore, the authors of~\cite{xu2020quantum} studied a quantum-based quality-aware crowdsourcing game using a density matrix-based method. 
Their analysis showed that -- as expected -- entanglement can enhance all participants' performances.

\subsection{Quantum ML}

Quantum ML models are a class of ML algorithms and models that leverage principles from quantum computing to perform various tasks more efficiently or enable new capabilities not achievable with classical ML methods. These models incorporate quantum computing concepts, such as superposition, entanglement, and quantum parallelism, to process and manipulate data in ways that classical computers cannot.
Quantum computation techniques can enhance both deep learning
and RL models, as well as generative and
discriminative learning models often used for data classification.
Havlicek \textit{et al.}~\cite{havlivcek2019supervised} conceived and experimentally implemented a pair of QD techniques on a
superconductive processor for supervised learning, which involved
constructing either a classifier or an estimator. Their first
algorithm, which uses a variational quantum circuit, categorizes data similarly to traditional SVMs by constructing a classifier function in the feature space. The second algorithm, namely a quantum kernel approximator, approximates the kernel function on a quantum computer and achieves the same results as the best classical SVM. Both of these quantum algorithms rely on using the quantum state space as their feature space, which can only be recognized by a quantum computer.

Chen \textit{et al.}~\cite{ChenQuantumMachineIntelligence2021} attempted to classify data stored in the form of quantum states by
training quantum circuits. This is an extension of the
well-established quantum state discrimination technique designed to
address the specific challenge that it is impossible to discriminate
against non-orthogonal quantum states deterministically. 
A quantum circuit was designed to perform arbitrary unitary transformations, which was trained to attain maximum generalization capability using the Adam optimizer.
It was shown that quantum circuits can be
trained, and quantum data can be discriminated
by balancing the errors and the inconclusiveness of classification
tasks. 

{\color{black}Xu and Ren~\cite{9343852} conceived a pilot encoding method based on quantum learning, which is capable of operating in the face of malicious attacks.  Specifically, the multi-user pilots are encoded as so-called diacritical subcarrier activation patterns. The base station then decodes the pilots from the observed subcarrier activation patterns by relying on the joint detection of the attack mode and the user activity with the help of a quantum learning network.} Huggins \textit{et al.}~\cite{huggins2019towards} designed a quantum tensor
network circuit for processing large numbers of samples and components in real-world data.  The geometry of the tensor-network-equivalent quantum circuits was specified, and the parameters of unitary operations were optimized individually for specific ML tasks. By relying on their circuits, quantum ML, including discriminative and generative learning, can be implemented in a unified theoretical and algorithmic framework that harnesses both classical and quantum computing. Tests of discriminative and generative tensor network models were carried out based on the MNIST dataset through supervised learning.

To model the probability distribution of any natural data, Gao
\textit{et al.}~\cite{doi:10.1126/sciadv.aat9004} proposed a quantum
generative model (QGM), which uses the probability of entangling
many-body states for modeling correlations within the data. QGMs are
more generalizable than classical factor graphs since they constitute quantum realizations of factor graphs. Statistical function divergences between the generative model and the given sample data set can be minimized with the help of the Kullback-Leibler divergence~\cite{Harremoes2014Renyi}. The QGM offers substantial improvements in representational power in the runtime of learning and the runtime of inference.

{\color{black}
Duong \textit{et al.} \cite{9870532} presented the latest techniques in quantum computing and gave a comprehensive overview of its potential through ML. In addition, quantum-inspired ML applications in resource allocation and network security are introduced.
Ansere \textit{et al.} \cite{10342775} proposed a quantum deep RL algorithm to exponentially improve the cache learning speed and content cache delivery efficiency in multi-dimensional continuous large action spaces. 
They also proposed a new quantum-empowered deep RL approach that combines quantum computing theory with ML to significantly achieve the trade-off between exploration and development through quantum parallelism~\cite{10318071}. 
On the other hand, in order to provide the privacy of semantic data, quantum anonymous transmission of quantum semantic states seeks to hide the identities of communication parties. Khalid \textit{et al.} \cite{10251843} investigated the potential benefits of combining quantum anonymous teleportation and variational quantum computing to develop quantum semantic communication, thereby enabling Metaverse users to interact with virtual environments more reliably and securely.}

\begin{table*}[t]
\caption{Summary of Quantum ML and Quantum Neural Network Techniques}
\centering
\begin{tabular}{|>{\raggedright\arraybackslash}m{6cm}|>{\raggedright\arraybackslash}m{7cm}|>{\centering\arraybackslash}m{2cm}|}
\hline
\textbf{Quantum Learning Technique} & \textbf{Application} & \textbf{Examples} \\ \hline
Variational quantum unsampling & Deep learning & \cite{tacchino2021variational} \\ 
Variational quantum circuit & Deep RL & \cite{chen2020variational} \\ 
Continuous-variable quantum neural network & Convolutional, recurrent, and residual networks & \cite{killoran19} \\ 
Quantum transfer learning & Image recognition & \cite{mari2020transfer} \\ 
Quantum circuit learning & Neural networks in near-term, low-depth quantum circuits & \cite{PhysRevA.98.032309} \\ 
Quantum circuit for arbitrary unitary transformation & Data classification in quantum states & \cite{ChenQuantumMachineIntelligence2021} \\ 
Quantum tensor network circuit & Discriminative and generative learning & \cite{huggins2019towards} \\ 
Quantum generative model & Data correlation characterization & \cite{doi:10.1126/sciadv.aat9004} \\ \hline
\end{tabular}
\label{tab.quantumML}
\end{table*}

\subsection{Quantum Neural Network }

The state-of-the-art ML methods rely on the classical von Neumann
computation frameworks and are already extensively employed~\cite{zhao20tc}. However, the existing quantum computing
architectures are unsuitable for classical deep learning frameworks. A more viable solution is to develop deep-learning algorithms for quantum circuits, including QNNs.

{\color{black}QNNs are a class of neural network models that use quantum computing principles to represent and process data. They replace classical neurons with quantum analogs and can potentially excel in tasks, e.g., pattern recognition, classification, and regression. 
QNNs are often implemented as quantum circuits, where qubits are initialized in certain states and manipulated through a series of quantum gates.
These qubits can exist in superposition states, enabling the network to explore multiple paths simultaneously during computation.
Quantum gates can perform operations, such as superposition, entanglement, and phase shifts.}
Tab.~\ref{tab.quantumML} summarizes the key technologies in QNNs and their applications at a glance.

Tacchino \emph{et al.}~\cite{tacchino2021variational} implemented a QNN and proposed an approach in which separate quantum nodes rely on the novel technique of variational unsampling policies. The family of unsampling techniques aims to extract as much information as possible about the unitary operation of a circuit using a polynomial, increasing the number of outputs while relying on a given initial state. 
Chen \emph{et al.}~\cite{chen2020variational} reshaped classical deep RL algorithms, including the functions of
experience replay and training networks into variational quantum
circuits.
This type of quantum circuit has tunable parameters that can be optimized iteratively by a classical computer. The robustness to noise and circuit depth adaptations was demonstrated for variational quantum circuits.

Killoran \emph{et al.}~\cite{killoran19} developed a QNN with layered structures having CV parameters.  
The QNN employing both Gaussian and non-Gaussian gates constitutes a pair of affine transformations and nonlinear activation functions.  Based on the CV framework, the QNN can encrypt
strongly nonlinear manipulations and maintain strict unitarity.
They demonstrated how to integrate a classical system into a quantum architecture and developed quantum-based models for several training frameworks, including convolutional, recurrent, and residual models~\cite{killoran19}.  The Strawberry Fields
software platform~\cite{2018Strawberry} was used for constructing training experiments. By implementing a fraud detection classifier
and a classical quantum auto-encryption system, the experiments
demonstrated the generalizability of CV-QNN architectures.

Mari \emph{et al.}~\cite{mari2020transfer} applied the popular
transfer learning technique in a mixed neural network having both
classical and quantum components. A pre-learned classical system was revised and enhanced by a variational quantum circuit. This facilitates the processing of high-dimensional data (e.g., pictures) in advance by a classical method and integrates an appropriately selected group of critical characteristics with a quantum operator. A system-oriented software library, namely PennyLane~\cite{2018PennyLane}, was used for examining a high-definition image classifier using two distinguishable quantum
computing machines provided by IBM~\cite{zulehner19} and
Rigetti~\cite{rigetti22}.

Inspired by the recent developments in quantum data analyzers, Cong
\emph{et al.}~\cite{cong2019quantum} examined a quantum circuit
inspired by the philosophy of convolutional neural networks (CNNs).
The extension of CNNs to the QD is constituted by quantum
CNNs (QCNNs). By using on the order of $\mathcal{O}(\log (N))$ variational coefficients on $N$ qubits, the QCNN can achieve effective application of real-world quantum equipment.  To illustrate the functionalities of QCNN, Cong~\emph{et al.}~\cite{cong2019quantum} applied them to recognize quantum
states in a group of one-dimensional many-body systems and
demonstrated that QCNNs can confidently distinguish quantum states. They demonstrated that quantum error control
methods can be designed using the QCNNs to outperform known QD methods.
Mitarai \textit{et al.}~\cite{PhysRevA.98.032309} proposed a new
classical-quantum hybrid ML framework, termed quantum circuit
learning, for implementing neural networks in near-term, realizable,
low-depth quantum circuits. Their solution iteratively refines the
circuit parameters based on the input data until the circuits provide the expected outputs.

{\color{black} 
\subsection{Summary}
One of the advantages of quantum computing is its ability to provide exponential or even super-exponential acceleration in solving complex computing problems. However, quantum computing and communication research has been following relatively independent tracks. Only limited efforts have been invested in their joint design. The lack of research on the joint design of quantum computing and quantum communication stifles further progress in the quantum information field.  One of the main limitations of quantum AI is the unavailability of quantum computing hardware. Quantum computers are still in the early stages of development and are not widely available. As a result, it is expensive to perform large-scale quantum computing, thereby limiting the capabilities of quantum AI.
}

%% file: Chapter_5.tex
\section{Quantum Sensing and Timing for Future Communications}

\begin{figure}[t]
\centering
\includegraphics[width=3.5 in]{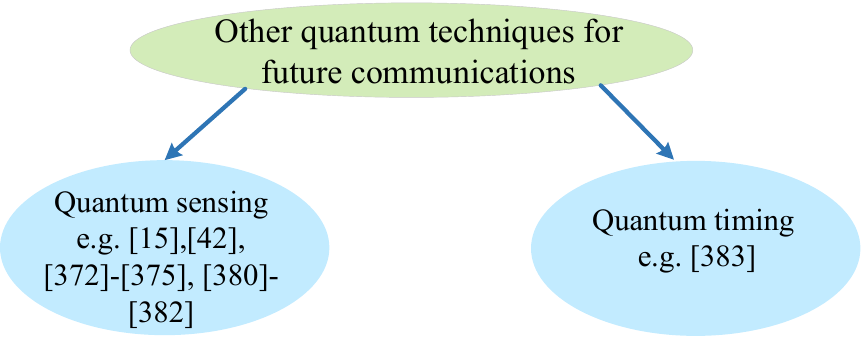}
\caption{Other quantum techniques for future communications. The integration of accurate quantum sensing and quantum communication can realize high-quality holographic communication. Quantum timing can provide high timing accuracy and high-precision positioning services for future networks. The combination of quantum sensing and quantum communication can significantly enhance information security and hardware efficiency, and offer an integration gain.}
\label{Other_quantum_techniques_for_6G}
\end{figure}

Quantum sensing and quantum timing use the principles of quantum mechanics to improve the accuracy and sensitivity of measurement and timing. Common techniques used include quantum interference, quantum entanglement, and the evolution of quantum states. 
Factors such as the fidelity of qubits and the retention time of quantum entanglement play an important role in the effectiveness of quantum sensing and quantum timing.
This section briefly summarizes quantum sensing and
timing, as shown in Fig.~\ref{Other_quantum_techniques_for_6G}. In QD, sensing and communication can be beneficially combined, giving birth to numerous new quantum services and enabling future networks. For example, the amalgam of high-resolution quantum sensing and quantum communication is capable of supporting holographic communication to promote deep integration of the real physical world and the virtual digital world. Quantum timing is capable of providing excellent timing accuracy and high-precision positioning services for future networks, benefiting emerging ISAC.

\subsection{Quantum Sensing}
{\color{black}

The measurement accuracy of traditional sensors is limited by thermal noise, quantum noise, and quantum randomness. 
Quantum sensing can significantly increase the accuracy of traditional sensing and detection~\cite{9904994}. Quantum sensors~\cite{4367523} are based on atoms~\cite{QuantumsensingIItechnologiesandtypicalexamples} and entanglement~\cite{QuantumsensingIIIapplicationsandprospects}, 
and exploit the unique characteristics of quantum mechanics, including quantum entanglement~\cite{8750548}, quantum coherence~\cite{9542465}, and quantum state squeezing~\cite{9908325}, to achieve high levels of precision.
These properties of quantum mechanics have been specifically optimized for use in quantum sensors.

An important application of quantum sensing is quantum inertial sensing, which is used to measure physical quantities, such as acceleration, angular velocity, and magnetic fields. 
Chen \textit{et al.} \cite{PhysRevApplied.12.044039} experimentally achieved high-precision multi-functional quantum sensing with a spatial resolution of 50 nm, demonstrating non-invasive subdiffraction electromagnetic field sensing using optical far-field super-resolution microscopy. Compared with traditional diffractive limited optical microscopy, the spatial resolution and sensing accuracy are improved.
Chen \textit{et al.} \cite{ PhysRevLett.128.180502} conducted a twin-field QKD experiment on 658 km fiber and tested a vibration sensing scheme with a positioning accuracy of about 1 km.
Sajeed \textit{et al.} \cite{10043372} proposed a novel approach for target detection based on quantum coherence.  They exploited a time-bin quantum interferometer to create sequences of coherent optical pulses toward a target. The experiment confirmed that coherence-based detection achieves a better performance than conventional intensity-based detection. This may open up new application avenues of coherence to quantum sensing, imaging, and communication.}

An emerging application is quantum remote sensing. The authors of~\cite{2020Quantum} examined the scenario,
where the client assigns a measurement mission to a distant server
relying on a quantum sensor, and the eavesdropper intercepts all the
classical information on the server.  The authors
of~\cite{2020Quantum} compared the information gleaned both by the
client and by an eavesdropper when considering the effects of both
dephasing and state preparation errors.
{\color{black}
Remote sensing can be compromised due to the measurement defect or forgery of sensing signals. Peng \textit{et al.} \cite{ PhysRevA.105.032615} developed a credible scheme for remote quantum sensing and a protocol to perform secure sensing of a parameter at the remote side. This protocol was applied to magnetic sensing, and showed that the client can reliably estimate the strength range of the magnetic field.}

\subsection{ Quantum Radar}
Quantum radar, as a branch of quantum sensing, exploits
the particle-based nature of light to achieve target detection. 
Quantum radars exploit entangled photons, allowing them to operate at extremely low powers~\cite{10027908}. They are useful for a range of applications, including biomedical~\cite{9528545} and security applications~\cite{9343595}, since they can effectively hide behind background noise as exemplified by the microwave-based system of~\cite{Barzanjeh_2020}. This technology has the potential of creating stealthy radar systems that emit minimal detectable electromagnetic radiation.

For example, Wilson \textit{et al.}~\cite{2020Progress} compared quantum radars and classical radars,
{\color{black} and discussed the fundamental concept, prerequisites, and potential gains of a quantum noise radar employing superconducting quantum circuits.}  
In~\cite{QTMS} and \cite{Quantumenhancednoiseradar}, they also discussed a new QI protocol, namely, quantum two-mode squeezing (QTMS), that has been successfully implemented in the microwave spectrum, described the physical realization of superconducting microwave circuits, and derived the predicted quantum enhancement observed experimentally. By alleviating the requirement for joint measurements, photodetectors, and delay lines, this new QI protocol is more practical than previous protocols.
As expected, their results demonstrated that the cost of a quantum radar is substantially higher than that of a classical radar. 

Nevertheless, the effective SNR of a quantum radar is expected to be better than that of a classical radar at the same transmit power and bandwidth. {\color{black} Quantum radar using entangled photons significantly improves the resolution than the non-entangled photons, but the entanglement is so fragile and unstable and usually difficult to create and preserve. Considering the above phenomena, Salmanogli \textit{et al.}~\cite{2019Entanglement } designed a quantum radar system to investigate the entanglement behavior in the process of generation, propagation, and reflection. The simulation results showed that the system can keep the detected photons entangled with the optical mode.
}

\subsection{Quantum Timing}

Traditional timing methods are often based on the principles of classical physics, e.g., atomic and optical clocks. Exploiting the evolution and interference effects of quantum states, quantum timing can achieve higher temporal resolution and measurement accuracy.  Quantum timing can apply to precise time synchronization, communication, and navigation. 

For instance, Dai \textit{et al.}~\cite{2020Towards} developed a satellite-based
quantum-secured time stamp transfer method relying on bidirectional
QKD techniques.  
Based on the Micius quantum satellite, they verified the feasibility of quantum-secured time stamp transfer of individual photons in sub-nanosecond precision over satellite-to-ground links. A QBER of less than 1\% is achieved with a time stamp transfer accuracy of 30 ps. 
Other applications include
space-air-ground network integration~\cite{2021Theshift}, joint communication, navigation, remote sensing~\cite{20206G}, and holographic interaction~\cite{9023997}.

Future aerial, terrestrial, and underwater network devices can have quantum-based clocks that are precisely synchronized to nanoseconds. This can enhance communication capacity and sensing accuracy.
ISAC~\cite{8999605,9162963} will be a key differentiating technique of future systems, which require ultra-reliable communication and high-resolution sensing capabilities unsatisfied by traditional communication and/or sensing techniques.

\subsection{Summary} 
{\color{black}Quantum sensing and timing are promising directions
in the field of QIT. These sensors have the
potential to improve the performance of current sensor
technologies. By jointly designing the timing and
sensing, future networks could benefit from precise localization and
synchronization~\cite{2020Quantum}. However, the relationships, theoretical boundaries, and performance trade-offs between quantum communication, quantum sensing, and quantum computing require substantial research.
}

%% file: Chapter_6.tex
\begin{table*}
\caption{A summary of main companies in the commercial phase working on quantum computing and quantum communications, where  \checkmark  indicates that the company is engaged in the corresponding research area, and a blank space indicates otherwise~\cite{companies}.}
\begin{center}
\begin{tabular}{p{1.5cm}<{\flushleft} p{5cm}<{\flushleft} ccccc} \hline
 \multirow{2}{*}{\textbf{\makecell{Headquarter \\ Location }}} &
\multirow{2}{*}{\textbf{Company}} & \multicolumn{5}{c}{\textbf{Key Research Area}} \\ 
\cline{3-7} 
&   &   \textbf{Algorithm} & \textbf{Crypto}  & \textbf{Software}  & \textbf{Chip}  & \textbf{Platform}\\ \hline

\multirow{8}{*}{\makecell*[c]{ Europe  }  }
 & ArQit (UK) \cite{arqit}  &  &\checkmark &\checkmark & &\checkmark    \\ 

 & KETS Quantum (UK) \cite{kets} &  & & &\checkmark &\checkmark    \\  
 & ID Quantique (Switzerland) \cite{idquantique}  &\checkmark  &\checkmark & & &\checkmark      \\ 
 &  VeriQloud (France) \cite{veriqloud}  &\checkmark  &\checkmark &\checkmark &    &       \\  
 & QuTech (Netherlands) \cite{QuTech}  &\checkmark  & & & & \checkmark    \\ \hline 
 
\multirow{4}{*}{\makecell*[c]{ US  }  }
&  Quantinuum~\cite{Quantinuum}  & \checkmark &\checkmark  &\checkmark  & &   \checkmark     \\ 
&  QuEra~\cite{QuEra}  & \checkmark &  &\checkmark  & &   \checkmark     \\ 
&  Qubitekk~\cite{Qubitekk}  &  &  \checkmark  &  & &   \checkmark    \\ 
&  IonQ~\cite{IonQ}  & \checkmark &  &\checkmark  & &   \checkmark     \\ 
&  Quantum Trilogy \cite{quantumtrilogy} &\checkmark  &\checkmark & & &      \\ \hline

\multirow{4}{*}{\makecell*[c]{ Canada  }  }
& D-Wave \cite{DWave}  & \checkmark & &\checkmark & \checkmark&     \\ 
& Xanadu \cite{Xanadu}  &  & &\checkmark & \checkmark & \checkmark    \\ 
& 1QBit \cite{1QBit}  & \checkmark & &\checkmark & &  \checkmark   \\ \hline

\multirow{4}{*}{\makecell*[c]{ China  }  }
 & QuantumCTek  \cite{QuantumCTek}  &  & \checkmark & \checkmark& &   \checkmark  \\ 
 & Qasky \cite{qasky}  &  & \checkmark  &\checkmark &\checkmark  &   \\  
& Origin Quantum \cite{OriginQuantumComputing}  &  & &\checkmark &\checkmark & \checkmark    \\ 
& Qudoor \cite{qudoor} &  & & &\checkmark  &\checkmark      \\ 
& Bose Quantum \cite{BoseQuantum} & \checkmark & &\checkmark &  &\checkmark      \\ \hline

\makecell*[c]{ Japan } 
& QunaSys \cite{QunaSys}  & \checkmark & &\checkmark &  & \checkmark  
\\ \hline

\makecell*[c]{ South Korea }  
& EYL \cite{EYL} & & \checkmark & &\checkmark &    \\ \hline

\makecell*[c]{ Singapore }  
& SpeQtral  \cite{SpeQtral } & & \checkmark & &\checkmark &  \checkmark  \\ \hline

\makecell*[c]{  Australia }  
& Q-CTRL  \cite{Q-CTRL } &\checkmark &  &\checkmark & &  \checkmark  \\ \hline

\end{tabular}
\end{center}
\label{tab.company}
\end{table*}

\begin{figure*}
\centering
\includegraphics[width=7.25 in]{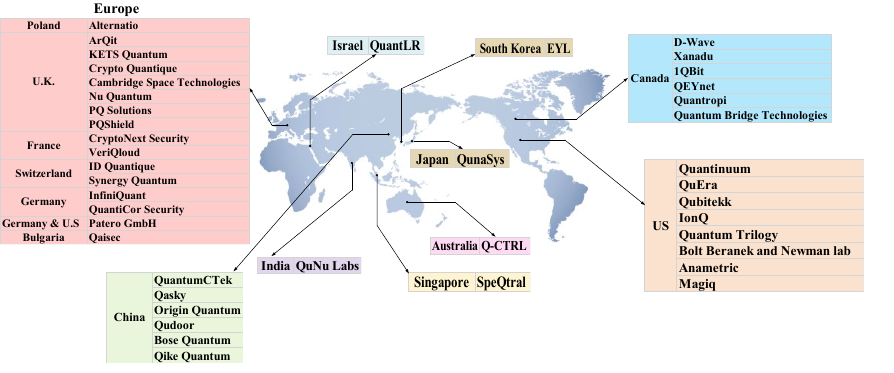}
\caption{{\color{black}Company location. Apart from academic and scientific demonstrations, significant progress is underway in developing and commercializing quantum technologies. The US, UK, and Canada are among the biggest and most active players in this field. Quantum communication technology companies have also been established in other countries around the world~\cite{companies}.}}
\label{Company_location}
\end{figure*}

\begin{figure}[t]
\centering
\includegraphics[width=3.5 in]{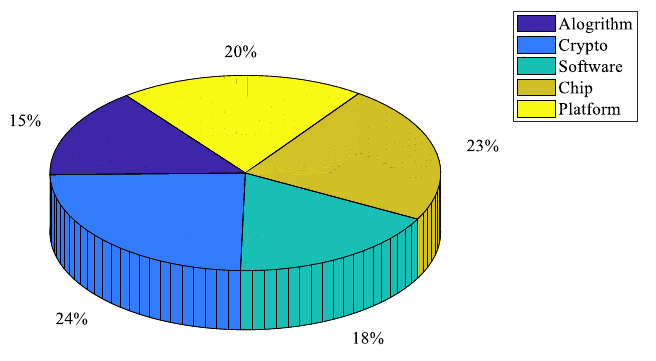}
\caption{Key research areas. The areas of research involved in quantum companies include algorithms, crypto, software, chips, and platforms. While most industrial progress has been made in encryption, attention has been increasingly drawn to the uptake of chip design for quantum systems.}
\label{key_research_areas}
\end{figure}

\section{Commercial Landscape and Case Study}

Apart from theoretical academic studies and scientific demonstrations, significant progress is underway in developing and commercializing quantum technologies. 
{\color{black} As summarized in Tab.~\ref{tab.company}, we highlight the primary research areas of some prominent quantum companies. The US, UK, Canada, and China are among the most active players. Globally, while significant industrial progress has been made in encryption, there is a growing focus on the chip design of quantum systems (see Figs. \ref{Company_location} and \ref{key_research_areas}).}

The US and China started early in the research of quantum communications. 
For example, Quantum Trilogy~\cite{quantumtrilogy} was founded in 2016 to commercialize the outcome of the research carried out since 1991 on quantum physics and cybersecurity.
Qasky~\cite{qasky} commercialized quantum cryptography research by the Chinese Academy of Sciences, including the quantum cryptography communication terminal equipment, network switching/routing equipment, core optoelectronic devices and modules, scientific instruments and network security control and application software, as well as integrated solutions for quantum information security systems.  

{\color{black}
\subsection{Europe}

In 2022, British company Arqit launched QuantumCloud, a cloud-based symmetric key protocol software platform, to provide flexible, lightweight, quantum-resistant encryption capabilities. QuantumCloud allows any device to download a lightweight software agent and then co-create an unlimited number of encryption keys with any number of other devices. The platform enables organizations to simplify and strengthen encryption on a global scale with no need for devices~\cite{arqit}.

In 2023, researchers of ID Quantique in Switzerland generated keys at rates of 64 Mbps and 3.0 Mbps over distances of 10.0 km and 102.4 km, respectively, by upgrading the high count rate of a single photon detector~\cite{grunenfelder2023fast}.
In the same year, researchers of this company developed a photon number resolution system with a 4-pixel parallel structure single photon detector, boasting 92\% single photon detection efficiency and capable of distinguishing 4-photon events~\cite{PhysRevApplied.19.064041}.  

\subsection{North America}

In 2022, researchers at Qubitekk, U.S., conducted research on QKD-based security authentication on actual smart grid systems. They made a demonstration on a deployed electric utility fiber network, and experiments showed that the authentication efficiency based on the Galois message authentication code is significantly higher than that of traditional schemes~\cite{alshowkan2022authentication}.
In 2023, the researchers used Quantinuum's H1-1 ion trap quantum processor to simulate phase transition with 20 qubits performing 18 steps of evolution. They devised a qubit reuse method that saved more than 73 qubits~\cite{WOS:001094561500001}.

In 2023, researchers at D-Wave, Canada, demonstrated the measurement of dynamics in 3D spin glass at 5000 superconducting qubits based on an annealing algorithm, showing significant advantages over slow stochastic dynamics simulations based on Monte Carlo algorithms~\cite{king2023quantum}.

In 2024, researchers at QuEra, U.S., built a quantum computing system featuring 280 physical qubits cooled by magneto-optical traps with rubidium atoms. This system prepared the largest logical qubit with significant code distance to date and achieved a GHZ state among four logical qubits with a direct measurement fidelity of 72\% and a post-selective fidelity of 99.8\%~\cite{bluvstein2024logical}.

\subsection{Asian-Pacific Region}

In 2018, the Hefei National Research Center for Physical Sciences at Microscale and the Department of Modern Physics used LEO satellites to establish a long-distance QKD network~\cite{2018Satellite}. Using the Micius satellite, QKD was performed among Xinglong and Nanshan in China and Graz in Europe, showcasing the potential for global secure communication networks, as shown in Fig.~\ref{Illustration_of_the_three_cooperating_ground_stations}.
In 2021, the Hefei National Research Center demonstrated an integrated space-to-ground quantum network, combining four metropolitan area networks, a long-distance fiber backbone, and a satellite-to-ground network with the Micius satellite~\cite{0An}, as shown in Fig.~\ref{Illustration_of_the_integrated_space-to-ground_quantum_network}. With over 700 fiber QKD links and two satellite-to-ground QKD links, this system achieved an average secret key rate of 47.8 kbps and a peak rate of 462 kbps, emphasizing its commercial viability for scalable quantum communication.
In 2023, QuantumCTek, China researchers performed classic-quantum cofiber on a few-mode fiber using a mode multiplexer/demultiplexer, combined with power wavelength allocation and other methods. This system achieved a QKD secret key rate of 2.7 kbps under the condition of 100.96 km fiber with a classical bandwidth of 1 Tbps~\cite{Dou:23}.
In the same year, QuantumCTek developed a time-bin phase encoding QKD system and demonstrated the field implementation of QKD over long-distance deployed aerial fiber. During the 70-day field test, this system achieved approximately a 1.0 kbps secret key rate with stable performance~\cite{Tang:23}.

In 2020, researchers at QunaSys, Japan, announced that its cloud service called Qamuy for performing quantum chemistry calculations on quantum computers was available in private beta. With this software, the performance comparison with quantum chemistry calculations and the performance comparison of various algorithms can be verified on a classical computer~\cite{ https://thequantumdaily.com/2020/10/23/japanese-company-reports-their-quantum-chemistry-cloud-service-is-in-private-beta/}.

In 2017, South Korea's EYL company launched a security new concept USB flash drive called TriGen in order to improve the security of USB storage devices. TriGen is a user's real name USB flash drive based on quantum cryptographic garble technology, loaded with a super-small quantum garble generator, which needs to be used when there is an Internet. During operation, the number authentication is performed every 3 seconds through the server and quantum gibberish, which will cause authentication errors and limit the use of TriGen in the case of disconnection~\cite{SouthKoreaCompany}.

In 2023, Singapore company SpeQtral and its collaborator announced the launch of Zenith, their first quantum random number generator designed specifically for the space sector, offering data transfer rates of 200 to 1,000 Mbps~\cite{SingaporeCompany}.

In 2024, Australian company Q-CTRL announced that they introduced a comprehensive quantum solver for binary packet optimization problems on a gate model quantum computer, and benchmarked this solver on an IBM qubit computer. Use of this new quantum solver increased the likelihood of finding the minimum energy by up to $\sim 1,500\times $ relative to published results using a D-Wave annealer, and it can find the correct solution when the annealer fails~\cite{AustralianCompany}.

\begin{figure}[t]
\centering
\includegraphics[width=3.5 in]{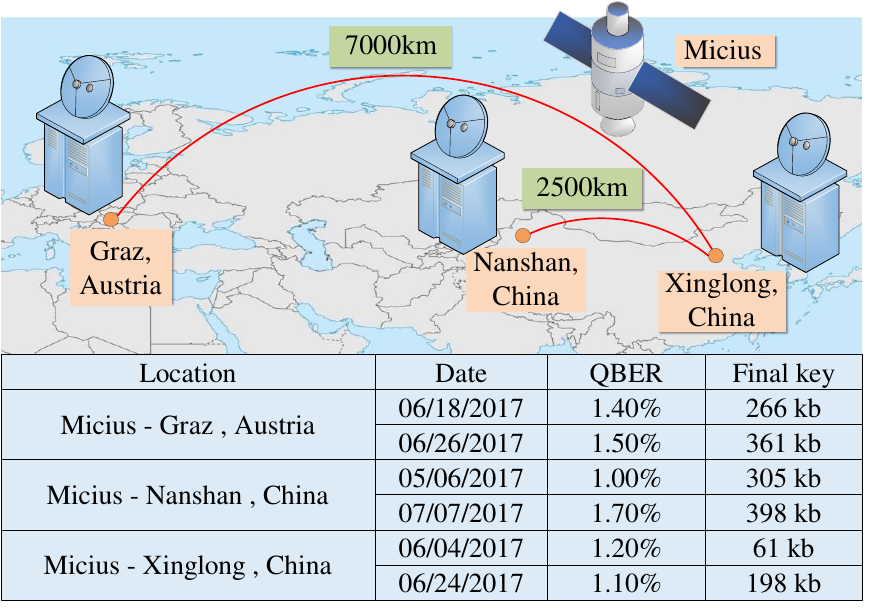}
\caption{Illustration of the three cooperating ground stations. Micius satellite 
is a trusted relay to perform QKD among three distant locations, 
i.e., Xinglong and Nanshan in China and Graz in Europe~\cite{2018Satellite}.}
\label{Illustration_of_the_three_cooperating_ground_stations}
\end{figure}

\begin{figure}[t] 
\centering
\includegraphics[width=3.5 in]{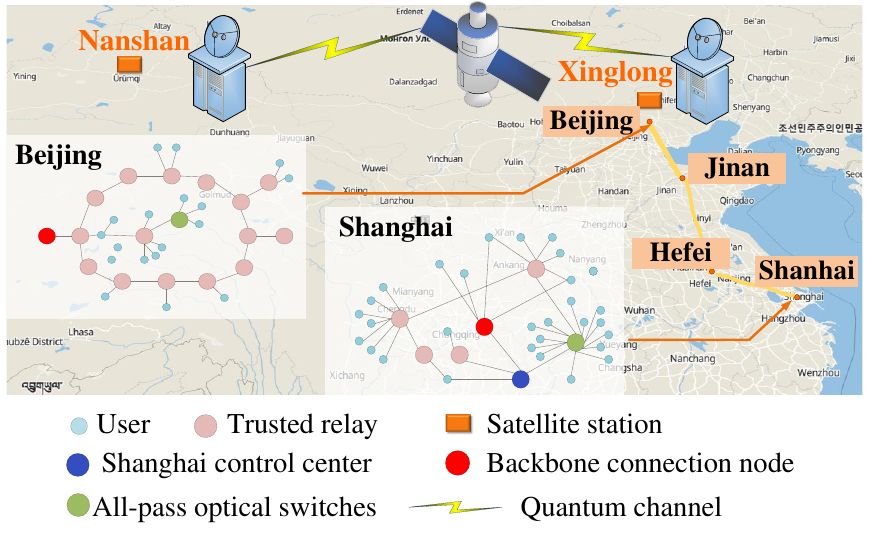}
\caption{An illustration of a space-to-ground quantum network. This experimental network consists of four QMANs in Hefei, Shanghai, Jinan, and Beijing. A backbone network communicates these four QMANs and a satellite-to-ground network with two ground stations and a quantum communication satellite–Micius satellite~\cite{0An}.}
\label{Illustration_of_the_integrated_space-to-ground_quantum_network}
\end{figure}

\subsection{Summary}
In this section, we have presented a comprehensive survey
of quantum technology companies worldwide, shedding light on their
diverse research endeavors. Furthermore, we have highlighted some of the
field experiments carried out by these companies.  Quantum technology
companies focus their research on algorithms,
cryptography, software development, chip design, and platform
development. Additionally, several companies have ventured into field
experiments involving quantum communication networks, leveraging
collaborative partnerships with scientific research institutions.
While quantum technology continues to experience rapid growth, there
remains a need to expedite the transformative application of quantum
technology. This necessitates the integration of quantum
communication, quantum computing, big data, cloud computing, and a
range of other cutting-edge technologies in the interest of
architecting the Qinternet.
}

%% file: Chapter_7.tex
\section{Lessons Learned and Open Challenges for Future Quantum Communications}

Quantum communications have witnessed remarkable advances, yet the integration of quantum technologies into commercial networks to ensure communication security and integrity presents a hitherto unexplored research frontier. Addressing this requires dedicated community efforts, with some of the interesting future directions as follows.

\subsection{Lessons Learned}

\subsubsection{Bottlenecks of Quantum Communication, Quantum Computing, and Quantum Sensing}
\begin{itemize}
\item Quantum Communication:
While QKD has found practical applications, reliable long-distance entanglement distribution remains a challenge. Current approaches rely on relays or satellites, necessitating the development of high-fidelity quantum repeaters and quantum memory~\cite{Pirandola:20}.
Moreover, the theoretical unconditional security of QKD is contingent on device reliability, so noise, signal loss, and error correction must be addressed~\cite{RevModPhys.92.025002}.
Additionally, the integration of quantum and classical systems is crucial for scalable quantum communication. Hybrid quantum-classical networks are explored, emphasizing the need for efficient allocation of classical and quantum resources, given the scarcity of quantum resources~\cite{9951308}.

\item Quantum Computing:
Quantum computers are susceptible to environmental noise, errors, and decoherence, necessitating resource-intensive error correction and fault-tolerant protocols~\cite{WOS:000457936400001}. While quantum computing shows promise, quantum ML and optimization remain in early stages, with hybrid quantum-classical approaches emerging for specialized tasks~\cite{WOS:000914622800003}. Advances in quantum teleportation could enable distributed quantum computing, but its scalability depends on robust quantum communication and entanglement~\cite{WOS:001415135100001}.

\item Quantum Sensing:
Quantum sensors offer high-precision measurements but are highly sensitive to environmental noise, requiring stringent control of conditions~\cite{10713118}. While quantum-enhanced measurements hold promise in fields like geophysics and medical diagnostics, challenges such as noise susceptibility remain~\cite{RevModPhys.89.035002}. Devices like atomic magnetometers and quantum gyroscopes are prone to interference from temperature and electromagnetic fields, demanding complex shielding~\cite{10399937}. In addition, the integration of laboratory-grade sensors, such as cold atom interferometers, into mobile platforms faces significant challenges in miniaturization~\cite{10.1117/12.2647923}.

\end{itemize}

\subsubsection{Lessons about Entanglement Generation and Quantum Error Correction}
\begin{itemize}
\item Generation and Distribution of Entangled Qubits:
The efficient generation and distribution of entangled qubits directly impact communication
distance. Factors, such as entanglement coherence
time~\cite{lin2020quantum} and entanglement consumption
rate~\cite{hu2021long}, play significant roles in determining the
effective quantum communication distance. Both a shorter entanglement
coherence time and higher entanglement consumption rate result in a
shorter communication distance~\cite{0Quantum,
  akyildiz20206g}. Developing novel technologies to extend
the entanglement coherence time and reduce the entanglement
consumption rate is paramount for increasing the quantum
communication distance.

\item Quantum Error Correction:
Enhancing the quantum error correction capabilities is crucial for mitigating the deleterious effects of decoherence~\cite{8766143,2019Quantum}. Existing quantum error correction methods are unable to clean up all errors inflicted by the low-fidelity quantum circuits~\cite{9762723,9720880}. Therefore, developing new techniques for efficient quantum error correction is essential.

One of the most popular quantum code design principles hinges on the so-called symplectic condition, which allows the creation of the quantum coding counterpart of classical codes. These codes independently treat the bit-flip and phase-flip qubit errors, resulting in a relatively low coding rate. Future research can create new codes that dispense with the symplectic condition to achieve high coding rates and, hence, require fewer qubits. These codes should also be intrinsically amalgamated with QEM techniques.

Moreover, integrating QD solutions with AI has the potential to revolutionize error correction and enhance the fidelity of quantum computing~\cite{9687466, 9861650}. This synergy can create more powerful ML algorithms in support of quantum cryptography~\cite{9687466, 9861650}.

\end{itemize}

\subsubsection{Realization of Large-Scale Quantum Networks}
\begin{itemize}
\item Massive Access for Quantum IoT:
Entanglement is a valuable resource in quantum systems and can play a critical role in multi-user systems. By combining these quantum
measurement methods with quantum multi-user detection~\cite{7163557} and QD diversity detection
techniques~\cite{8648086}, a combination of existing quantum reception
technologies can be realized to support the access of quantum users in
the turbulent outdoor channels of the quantum IoT~\cite{9099546}.
There are challenges both in realizing entanglement-based QD multi-access and in coping with the complexity of densely connected quantum communication systems~\cite{0Quantum}. Accurate detection of quantum states at the receiver requires the study of new quantum multi-user algorithms, such as quantum POVM~\cite{2010Performance} and square-root measurement~\cite{2002On}.

\item Scalability of Quantum Networks:
This scalability is primarily constrained by the collaboration of quantum nodes, with limited memory capacity being a key limiting factor. Insufficient memory affects both the storage and forwarding of quantum information, restricting network expansion. For example, in reconfigurable quantum entanglement distribution networks, limited memory can lead to conflicts or delays in resource allocation, when nodes need to share entanglement resources across multiple users~\cite{Howe2024}. Moreover, quantum error correction, which requires extra memory for redundant information, further strains node capacity. For instance, surface code error correction necessitates each logical qubit to correspond to multiple physical qubits, making memory limitations a significant bottleneck~\cite{PhysRevA.86.032324}.

Another major obstacle is the fidelity constraint, which worsens as network expands due to noise, decoherence, and imprecise quantum processes~\cite{WOS:000855773200002}. As the fidelity of entanglement decreases, sustaining high-quality entanglement over long distances becomes more challenging, reducing the network's reliability. This impacts the network's communication capacity, as fidelity constraints demand costly error correction~\cite{WOS:000795340400001, 10556066}.

Entanglement purification can mitigate fidelity loss, but comes with trade-offs between resource consumption and entanglement quality~\cite{WOS:001069297500003}. The probabilistic nature of purification also limits scalability, as multiple attempts are often required, reducing the success rate and hindering network growth~\cite{10628208, WOS:001315068200001}. Moreover, long-distance entanglement involving repeaters and entanglement swapping faces additional challenges, as failures in swapping increase the time and resources needed to establish stable communications~\cite{WOS:000486622300005, WOS:000792585200005}.

\item Economic and Policy Implications:
As a cutting-edge technology, the development of quantum communication gives rise to new industrial chains and business opportunities, including the manufacturing of quantum devices, the construction and operation of quantum networks, and related quantum software and services. 
 
In order to promote the development of quantum technologies, governments around the globe invest in the research and development of related technologies, including quantum devices, quantum networks, and quantum security. The governments can also encourage enterprises and research institutions to innovate and collaborate through financial support, tax incentives, and innovation policies. Furthermore, governments can formulate relevant laws and regulations to clarify technical standards, privacy protection measures, and security audit requirements to ensure the security and credibility of quantum communications. In general, quantum communication is a global challenge and opportunity that requires in-depth global cooperation and coordination. 

\end{itemize}

\subsection{Open Challenges}
\subsubsection{Design and Development of Quantum Devices}
\begin{itemize}
\item Quantum Device Design for THz Communications: 
The integration of THz technology with quantum communications holds
great potential for enhancing the connectivity and coverage of future
networks~\cite{Xiaohu2021Towards}. The performance of THz-CVQKD systems is constrained by hostile
impairments~\cite{2020Hybrid}. For example, due to a low SNR, the vacuum noise resulting from channel
loss reduces the correlation between the communicating parties. To overcome this, a potential
THz-CVQKD system requires frontier research in noise suppression
techniques, high-gain antennas, efficient detectors, and accurate
parameter estimation algorithms~\cite{9492803}. 

Moreover, high-quality
THz components, such as QCLs, modulators, and quantum cascade
detectors, are not yet widely available for quantum
communications~\cite{8761168}.
There have been developments in high-performance THz detectors utilizing quantum wells~\cite{2010Growth}, carbon nanotubes~\cite{2018Sensitivity}, and two-dimensional nanomaterials~\cite{2016THz}. However, these detectors
face limitations of their dark current performance and
absorption efficiency~\cite{2022103,shao2021research}. Furthermore, there have been only theoretical studies on quantum well infrared detectors, but experimental validations are scarce.  Addressing these challenges will be crucial for unlocking the full potential of THz technology in quantum communication.

\item Miniaturization of Quantum Terminals: 
The miniaturization of quantum devices, e.g., quantum communication
chips and quantum antennas~\cite{https://doi.org/10.1002/qute.201900120}, is in the focus of intensive current
research. Quantum communication chips provide quantum security that
can be used in compact devices, e.g., smartphones, tablets, and
smartwatches~\cite{Quantumcommunicationchips}. High-gain quantum antennas are required in far-field sensing~\cite{9204821,
  WOS:000532490000088}, quantum
communication~\cite{WOS:000564218701172}, quantum
imaging~\cite{WOS:000399354500007}, and energy
harvesting~\cite{WOS:000466596500003}.
These radical advances require new metamaterials and nanotechnology. The manufacturing and feeding of antennas require delicate processing due to the fragility of entangled states~\cite{2020Quantum}. Transferring the knowledge gathered in the design and fabrication of classical antennas to antennas employed in quantum systems is challenging. Solving problems, such as designing the antenna geometry and the initial quantum states, is a promising future research item.

\item Acquisition, Tracking, and Pointing for Mobile Quantum Communications: 
The expansion of both air- and space-based networks results in new communication nodes with high mobility, which poses challenges for quantum communications, owing to a plethora of hostile propagation effects~\cite{2021Drone}.  Ensuring accurate alignment between the transmitter and receiver devices is challenging due to the non-uniform distribution of electromagnetic fields and geographical variations~\cite{2016Free}. Building a space-based quantum communication network harnessing multiple satellites as relays
requires advanced acquisition, tracking, and pointing technology~\cite{8288586, Jono1999AcquisitionTA}.

\end{itemize}

\subsubsection{Integration of Quantum Communication Networks}
\begin{itemize}
\item Integration of Quantum Communication with Future Communication Infrastructure: 
The existing communication infrastructure is based on the principles of classical physics, while quantum communication is based on the principles of quantum mechanics. Integrating quantum networks with existing communications infrastructure requires addressing compatibility issues. This includes designing and implementing new protocols and standards to ensure interoperability between quantum and classical networks. Qubits in quantum communication are susceptible to ambient noise and loss, limiting the transmission distance of quantum signals. Meanwhile, existing communication infrastructures have often been designed for reasonably long distances. Therefore, integrating quantum and classical communication networks requires an enhancement in the transmission distance, stability, and fidelity of quantum signals. This may involve new devices, protocols, and network architectures and, consequently, increase costs and system complexity. 

\item Integration of Quantum Computing and Communications: 
Integrating quantum computing and communications into a joint design
gives birth to quantum-native computing and communication systems. This fusion has the potential of
revolutionizing secure communication networks, distributed computing,
and other domains~\cite{WOS:000473005200003,
  WOS:000400486300013}. Further research is required for
exploring its benefits as well as limitations, plus the development of
the hardware and software necessary  for practical
implementations~\cite{ WOS:000473005200003,
  WOS:000400486300013}. Additionally, integrating quantum computing
and communications with other technologies, such as MEC, presents
further opportunities for secure and efficient information
systems~\cite{8030322}.

\item Integrated Quantum Communications and Sensing:

By making use of quantum phenomena like superposition and entanglement, quantum sensing networks combine quantum communication with sensing to provide Heisenberg accuracy and unconditional security.  However, technological limitations, device flaws, quantum noise, and dynamic sensing situations pose challenges to quantum sensing networks' scalability and sensing capabilities in the NISQ era~\cite{QSN}. In addition,
a particular challenge in integrated quantum sensing and communication is the lack of unified evaluation criteria. The sensing accuracy may be affected by measurement errors caused by non-ideal hardware
factors~\cite{0Quantum}. Unfortunately, the existing communication-oriented channel
models do not distinguish between sensing and non-sensing targets. As
part of the future integrated quantum sensing and communication
systems~\cite{9232174}, it is necessary to distinguish the
multi-path echos of perceived targets and clutters, where the existing
communication-oriented channel models cannot apply~\cite{9687468}.  Furthermore, the relationships, theoretical
boundaries, and performance trade-offs between communication, sensing,
and quantum computing have to be further investigated.

\end{itemize}

\section{Closing Remarks}

{\color{black}This survey has explored the potential benefits of incorporating quantum communications into future systems. By proposing a quantum-native or quantum-by-design architecture, we examined technologies, such as QD MIMO, QD relaying, QD resource allocation, QD routing, QD multiplexing, as well as quantum AI and computing. We critically appraised a whole suite of recent research breakthroughs in these areas and adopted a comprehensive system-oriented view to evaluate the feasibility of using quantum communications in future applications.

The survey also identified research gaps and future directions, including the need to extend the entanglement's coherence time, reduce the entanglement consumption rate, explore new approaches for
efficient quantum error correction, develop THz quantum devices, and address challenges in the miniaturization of quantum terminals.

{\em Valued colleagues, our hope is that you would feel inspired to
  join this community effort in exploring this fascinating
  multi-disciplinary frontier in research, where a plethora of open
  questions may be solved by scientists from the entire electronics,
  mathematics, physics, computer science and material science
  community.}
}